\documentstyle[aps,prd,twocolumn,epsf]{revtex}

\tighten
\begin{document}


\firstfigfalse
\newcommand{\incpsfig}[1]{\centerline{\epsfxsize=3.0in \epsfbox{#1}}}

\wideabs{
\title{Three Dimensional Numerical Relativity with a Hyperbolic
  Formulation}

\author{Carles Bona${}^{(1)}$, Joan Mass\'o${}^{(1,2)}$, 
Edward Seidel${}^{(2,3,4)}$, and Paul Walker${}^{(2,3)}$}

\address{
  ${}^{(1)}$ Departament de Fisica, Universitat de les Illes Balears, SPAIN\\
  ${}^{(2)}$ Max-Planck-Institut f\"ur Gravitationsphysik,
  Albert-Einstein-Institut, Schlaatzweg 1, 14473 Potsdam GERMANY \\
  ${}^{(3)}$ Department of Physics,
  University of Illinois, Urbana, Illinois 61801 \\
  ${}^{(4)}$ Department of Astronomy,
  University of Illinois, Urbana, Illinois 61801 \\
  }

\date{Mar 13, 1998} \maketitle
\begin{abstract}
We discuss a successful three-dimensional cartesian implementation of
the Bona-Mass\'o hyperbolic formulation of the 3+1 Einstein evolution
equations in numerical relativity. The numerical code, which we call
``Cactus,'' provides a general framework for 3D numerical relativity,
and can include various formulations of the evolution equations,
initial data sets, and analysis modules.  We show important code
tests, including dynamically sliced flat space, wave spacetimes, and
black hole spacetimes.  We discuss the numerical convergence of each
spacetime, and also compare results with previously tested codes based
on other formalisms, including the traditional ADM formalism.
This is the first time that a hyperbolic reformulation of Einstein's
equations has been shown appropriate for three-dimensional numerical
relativity in a wide variety of spacetimes.
\end{abstract}
}


\section{Introduction and Overview}

The young field of three-dimensional (3D) numerical relativity has 
entered an exciting era.  As we review below at some length, strong 
theoretical and astrophysical motivations have led to increased 
activity and collaborations among many research groups, and general 3D 
relativistic problems are being attacked with increasing success.  In 
this paper we present results from a new and very general 3D advanced 
computer code, which we call ``Cactus''\cite{Masso98b}, designed to 
study these general problems in a collaborative environment.  This is 
the first in a series of papers on this code and its applications in 
numerical relativity and relativistic astrophysics.  At the same time, 
the paper is also a follow-up of previous papers in our continuing 
exploration of hyperbolic formulations of the Einstein equations for 
numerical relativity\cite{Bona92,Bona94b,Bona97a}.

\subsection{Motivation}
The imminent arrival of data from of the long awaited gravitational wave
detectors (LIGO, VIRGO, GEO600, TAMA; see, e.g., Ref.~\cite{Flanagan97b}
and references therein) has provided a sense of urgency in producing
realistic simulations of very strong sources of gravitational waves,
which can only be done through the full machinery of numerical
relativity.  One of the best candidates for early detection by the
laser interferometer network is increasingly considered to be black
hole mergers\cite{Flanagan97a,Flanagan97b}.  However, the signals are
likely to be weak enough by the time they reach the detectors that
reliable detection may be difficult without prior knowledge of the
merger waveform.  These are among the reasons that the NSF-funded
Binary Black Hole Grand Challenge Alliance has focused the efforts of
numerous US and international groups on developing codes for solving
the problem of 3D coalescing black holes (see, e.g, the latest round
of papers of the Alliance\cite{Abrahams97a,Cook97a,Gomez98a}).

Another important process in astrophysics that requires fully 
relativistic simulation is neutron star mergers (see, e.g.  
\cite{Mathews97}), which will produce a possibly detectable burst of 
gravitational waves\cite{Ruffert96a}.  These are 
sometimes considered as sources of gamma-ray bursts~\cite{ruffert95}, 
and the final state (e.g., a neutron star or black hole) is highly 
uncertain.  Most studies of this process have been Newtonian, but even 
post-Newtonian correction terms, which are still inadequate to 
describe the possible formation of a black hole, produce significant 
changes in the evolution\cite{Oohara96}.  More relativistic 
approximations to the Einstein equations produce still quite different 
(and controversial) results, indicating that the neutron stars may 
actually form black holes {\em before} the merger\cite{Mathews97}.  
The point we wish to make is that the merger process clearly requires 
a fully consistent relativistic treatment, which provides another 
motivation for development of powerful and general numerical codes to 
solve the full set of Einstein equations, in this case coupled to the 
relativistic fluid equations.  This research area is a particular 
application for which Cactus is being developed, although we will 
present only tests of the vacuum part of the evolution system in this 
paper.

Astrophysics aside, there are of course purely theoretical reasons to
develop robust 3D solvers to Einstein's equations.  As general
relativity is one the fundamental theories of physics, it needs to be
better understood in its most nonlinear regimes, which are usually the
most difficult to probe.  Again, numerical treatment of the full set
of Einstein equations is one of the main tools for studying the theory
in such regimes, and has already led to the discovery of unexpected
phenomena, such as critical phenomena in black hole
formation\cite{Choptuik93} (see Ref.\cite{Gundlach97d} for a recent
review), which has now been seen in spacetimes containing scalar
fields, fluids, and even in pure vacuum, gravitational wave
spacetimes.  Most of these studies have been carried out in 1D or in
rare cases in 2D\cite{Abrahams93a}, but little is known about the 3D
behavior\cite{Gundlach97b}.

Unfortunately, despite all of these motivations for 3D numerical
relativity, and the best efforts of many groups around the world,
progress has been slower than hoped and expected.  One of the reasons
for this is the the sheer complexity of the Einstein equations in 3D,
coupled with the immense computational needs for solving them.  For
example, an enormous amount of memory and time on the order of one CPU
day on a Teraflop computer will be required to produce a single,
highly resolved simulation of 3D black hole spiraling coalescence (see
Ref.~\cite{Anninos96c} for a review).  Developing well tested software
that simultaneously solves the Einstein equations, takes advantage of
high performance parallel computers, and can be effectively used by
the large number collaborators needed to develop algorithms is a
challenging software engineering problem in its own right.

However, the problems of 3D numerical relativity run far deeper than
computation and code development.  Given a sufficiently large computer
and perfectly debugged code, problems like coalescing black holes or
neutron stars would still not be solvable today, because of important
theoretical and algorithmic problems still to be addressed.  Perhaps
the best example to illustrate these problems is that of a spacetime
containing black holes.  

The presence of a singularity inside the black hole and the weak field
zone far from the hole gives rise to an extreme dynamic range.
Singularity avoiding slicings effectively keep time slices from
hitting the singularity, but lead to pathological time slices that
create huge gradients near the black hole horizon which cannot be
resolved, especially in 3D\cite{Anninos94c,Bruegmann96}.  Such
gradients lead to numerical instabilities with the standard
formulations of the equations, often causing codes to crash in the
interior well before the desired evolution can be carried out in the
radiation zone. Although the characteristic time necessary to obtain
accurate waveforms for the inspiral and merger of two black holes is
on the order of thousands of $M$ (see, e.g, Ref.\cite{Gomez98a}), even
state-of-the-art black hole collisions in axisymmetry
(2D)\cite{Anninos93b,Anninos94b} can only be evolved for hundreds of
$M$. (We will use units $c=G=1$ throughout this paper, so time and
spatial units for black hole simulations are in terms of the black
hole mass $M$.)

Success in evolving black holes in 3D has been mixed.  Partial
successes include colliding, equal mass black holes in 3D
\cite{Anninos96c}, and waveform extraction of distorted 3D black
holes\cite{Camarda97a,Camarda97b,Allen97a}. In both cases the
system is evolved successfully for tens of $M$ and although this
would be completely insufficient for the black hole coalescing
problem, it is enough time to study the waveforms produced in the
ring-down phase.  As verified by comparison with perturbation theory and 
axisymmetric simulations, these 3D simulations can produce highly 
accurate waveforms, but they ultimately crash both due to the ``grid
stretching''\cite{Shapiro86} effects created by singularity avoiding
slicings and due to poor outer boundary conditions.

One approach to understanding the expected waveforms that avoids these
problems is to solve the linearized equations describing black hole
gravitational wave interactions.  This approach has proven to be
remarkably robust in comparisons with a range of presently feasible
fully nonlinear simulations of distorted and colliding black hole
spacetimes~\cite{Allen97a,Price94a,Price94b,Abrahams95c,Baker96a,Gleiser96b},
but it does not solve the general coalescence problem.  Related
studies of a direct 3D integration of the perturbation equations show
that even such a simple linear problem is very demanding, having inner
and outer boundary difficulties \cite{Rezzolla97a} that can be
overcome through the machinery of adaptive mesh refinement
\cite{Berger84,Wild96,Papadapoulos98a}.

The problem of dealing with singularities, grid stretching, and inner
boundaries may be ultimately solved by the so-called AHBC (apparent
horizon boundary
conditions)\cite{Seidel92a,Anninos94e,Scheel94,Marsa96,Daues96a,Cook97a},
which are basically ingoing conditions on appropriate quantities
evolving near the black hole horizon coupled with appropriate gauge
conditions.  But other gauge problems may still lead to large
gradients as coordinates are sheared and squashed during the
evolution.  Hence much research into appropriate gauge conditions for
such dynamic spacetimes is needed.  Even in very weak wave spacetimes,
gauge problems can cause numerical codes to develop pathologies and
crash as coordinates evolve out of
control\cite{Anninos94d,Anninos96b,Balakrishna96a}. Recent
developments shed new light into the mathematical understanding of
these coordinate problems and gauge pathologies in
general\cite{Alcubierre97a,Alcubierre97b}. Furthermore, in order to
resolve the inner, strong field region near the black holes, the outer
boundary is generally placed uncomfortably close to the hole, where spurious
signals or reflections which propagate inward may be generated due to
inappropriate boundary conditions, masking the true physics taking
place in the interior (for a recent discussion, see
Ref.\cite{Abrahams97a}).

Despite these difficulties, there has been considerable progress in
evolving dynamical black hole spacetimes in the last year.
Br\"ugmann\cite{Bruegmann97} recently demonstrated that it is possible
to see some form of gravitational radiation from numerically
constructed true 3D black holes with spin and momentum. Unfortunately,
these feasibility studies seem to indicate that current techniques
have face more severe difficulties with these highly dynamical systems, and cannot yet
provide useful information for realistic gravitational wave
astronomy \cite{Bruegmann97}.  The Grand Challenge Alliance has
developed outer boundary conditions which appear to allow accurate
outgoing wave boundary conditions in three dimensional numerical
relativity \cite{Abrahams97a}. Moreover, using causal differencing and
a careful inner boundary treatment, the Alliance has been able to
transport a black hole several black hole radii across a grid
\cite{Cook97a}. Work by Daues and collaborators has allowed single
black holes to be evolved beyond 100M using dynamically determined
gauges \cite{Daues96a}. However, none of these treatments have shown,
to date, the ability to produce a long time stable evolution for
colliding or highly distorted black holes in three dimensions, and
many difficult problems remain to be solved.

Another very recent approach to 3D black hole evolution that
completely avoids the problems of grid stretching is characteristic
evolution, which has successfully evolved 3D rotating and distorted
black holes for essentially unlimited time periods ($t \approx
60,000M$\cite{Gomez97a,Gomez98a}).  These spectacular results are
achieved by using an ingoing characteristic foliation of the black
hole spacetime, using the horizon as an inner boundary.  However, it
is not clear yet if this method will be viable for evolution of very
highly distorted or colliding black holes, where focusing of ingoing
light rays may create caustics, leading to a breakdown of the
foliation.  Also, ironically, the method is presently most successful
when a black hole {\em is} present, creating an $S^{2} \times R$
topology; dealing with the so-called $r=0$ problem is difficult for
any formulation of the Einstein Equations, and is avoided by using
cartesian grids in the standard 3+1 formulations, but the
characteristic method cannot use cartesian grids, and would therefore
have to face this problem in the absence of a black hole (e.g., for
the coalescence of neutron stars).  Nonetheless, the possibility of
very long time evolutions demonstrated with the characteristic
evolution scheme is an exceptionally significant achievement that
seems likely to provide an alternate and superior approach for an
interesting class of 3D black hole spacetimes.

\subsection{Hyperbolic Numerical Relativity}
In recent years, much renewed research into theoretical foundations of
numerical relativity has led to the development of hyperbolic
formulations of the Einstein equations for numerical relativity, which
have numerous advantages over the standard ADM
formulation\cite{York79}. We have
addressed in detail this issue in a previous publication in this
series\cite{Bona97a}. In summary, they ({\em a}) provide a much better
starting point for the mathematical analysis of well-posedness and
existence of solutions\cite{Friedrich96,Reula98a}, ({\em b}) are
better suited than the standard ADM formulation to modern
numerical methods developed for computational fluid
dynamics\cite{Leveque92} and promise to handle large
gradients\cite{Bona94b,Bona97a}, ({\em c}) are more adapted to
providing natural boundary conditions either on the black hole horizon
or at the outer edge of the simulation, and({\em d}) still allow a very
general class of gauge conditions (many of which are yet to be
developed) that will be needed to control coordinate motion (although
see Ref.\cite{Alcubierre97b} for caveats of hyperbolic choices in the
gauge conditions).

Reula has recently reviewed, from the mathematical point of view, most
of the recent hyperbolic formulations of the Einstein
equations\cite{Reula98a} (This article, in the online journal ``Living Reviews
in Relativity'', will be periodically updated).
It is important to realize that the mathematical relativity field has
been interested in hyperbolic formulations of the Einstein equations
for many years and some systems that could have been suitable for
numerical relativity were already published in the
1980's\cite{Choquet83,Friedrich85}.  However, these
developments were not recognized by the numerical relativity community
until recently.

Choquet-Bruhat and Ruggeri already commented in 1983\cite{Choquet83}
on the possible importance of stable hyperbolic systems for numerical
applications. Following this suggestion, Bona and Mass\'o studied the
numerical relativity implications of the harmonic slicing
condition\cite{Bona88} and the advantages of systems of balance laws
from the numerical point of view\cite{Bona89}. In 1992 they proceeded
to develop the first hyperbolic formulation of the 3D Einstein
equations with numerical relativity in mind\cite{Bona92}. Special
emphasis was put on the idea of borrowing from the huge arsenal of
numerical methods available from the computational fluid dynamics
community.  

A complete 3D code was developed with this
formulation\cite{Masso92,Bona92b}, leading to an advanced parallel
version developed at NCSA called the ``H'' code.  Different variations
on this code were used in numerous applications in relativity, where
it was extensively tested on pure wave spacetimes\cite{Anninos94d},
and in computational science (see, e.g., \cite{Gjertsen96,Kuo97}).
This code forms the basis for some of the tests presented here, and
furthermore the computational science experience gained from
developing this code was essential in developing the more powerful
Cactus code, described below.  However, this formulation was
hyperbolic only for harmonic slicing (which amounts to a simple
algebraic condition on the lapse: $\alpha \propto \sqrt{g}$,
where $g$ is the determinant of the three--metric $g_{ij}$),
and it did not consider a shift, making it suitable only for a limited
range of problems in numerical relativity.

For these reasons, the system was generalized to apply to an arbitrary
shift and to an infinite family of lapse conditions, including maximal
slicing, in which case a mixed hyperbolic-elliptic system
results\cite{Bona94a,Bona97a}.  This system, currently known as the
``Bona-Mass\'o formulation'' (BM), takes the flux
conservative form, which already allows a wide class of modern
numerical methods not possible with the standard ADM formulation, for
{\em any} choice of lapse and shift.  But it has the additional
advantage of being hyperbolic (i.e., diagonalizable) if the lapse is
chosen from the particular infinite class of slicings defined below.
This formulation showed its superiority over the standard formulation
in spherical symmetry (1D) by evolving a black hole essentially
indefinitely, {\em without} apparent horizon boundary conditions.  Due
to the use of the eigenfields, the advanced numerical methods
available to such a formulation, and the improved outer boundary
treatment afforded by the formulation, it was able to handle the large
gradients that develop near a black hole with a singularity avoiding
slicing. Details of these numerical techniques and boundary treatments
are given in an accompanying paper in this series\cite{Arbona98a}. We
are presently working to carry these techniques into 3D, and this
paper takes the first step in addressing these issues. 

The BM system is now one among many hyperbolic systems, as other 
independent hyperbolic formulations of Einstein's equations were 
developed\cite{Fritelli94,Choquet95,Abrahams95a,Fritelli95,MVP95,Abrahams97b} at 
about the same time as Ref.\cite{Bona94a}.  To our knowledge, among 
these other formulations only the one originally devised in 
Ref.\cite{Abrahams95a} has been applied to spacetimes containing black 
holes\cite{Scheel97}, although still only in the spherically symmetry 1D 
case (a 3D version using full AHBC is under development\cite{CookScheelPrivateComm}.)

There is an additional important motivation for hyperbolic systems in
general relativity provided by the interest in relativistic
hydrodynamics, which will be needed to study systems like colliding
neutron stars.  Traditional approaches to relativistic hydrodynamics
treat the left and right hand sides of Einstein equations separately,
with different numerical methods, independent update routines, and so
forth.  However, relativistic hydrodynamics has a {\em single} set of
equations, mathematically and philosophically.  If the entire set of
Einstein equations, including the fluid equations (which should be
considered as a subset of the Einstein equations) could be formulated
as a single hyperbolic system, a unified numerical treatment of the
entire system would be possible.

\subsection{Goals of this Paper}
For all of these reasons, it is essential to develop robust and
general 3D numerical codes to attack the many problems in general
relativity and astrophysics waiting to be solved, testing and comparing
the different formulations of the Einstein equations.  With these strong
motivations, this paper has a two-fold purpose:

First, as follow-up of our previous work on the theoretical basis of 
our formulation\cite{Bona97a}, we present the first detailed testing 
of a hyperbolic formulation of Einstein's equations in 3D on a variety 
of spacetimes that have become established benchmarks for numerical 
relativity, including black hole and gravitational wave spacetimes.  
In this paper we will not try to advance the results of previous 3D 
codes but we show for the first time that with standard numerical 
methods for balance law systems (MacCormack and Lax-Wendroff schemes, 
discussed below), the BM formalism compares well with the
traditional ADM formulation.  In this paper we present results on the 
formulation in its most general form, allowing arbitrary slicings and 
shifts.  This form does not allow for advanced numerical methods based 
on the eigenfields of a hyperbolic system, or advanced boundary 
treatments.  Such methods are subject to further research and work is
in progress to apply them to this system of equations.  We also report 
on how to establish a set of techniques for rigorous verification and 
self-convergence testing.

Second, we present a code, called ``Cactus'', that provides a general,
high performance framework for 3D numerical relativity in a
collaborative environment, allowing for a number of formulations of
the equations, general gauge and initial conditions, different
numerical methods, analysis tools, etc. This code is being developed
as a general tool to be used for many different problems in 3D
numerical relativity, such as those described above. The philosophy
behind this approach is described in an accompanying paper\cite{Masso98b}. The performance and
parallelization aspects are described in accompanying
papers\cite{Clune98a,Walker98b}. Other tests of the code, including
matter tests, horizon finders, waveform extraction, etc. will be
published in future papers in this series, as a growing number of
international collaborators are extending the capabilities of the
current version. 

We proceed as follows: In Sec.~\ref{sec:basic} we discuss basic 
concepts of our code, including the systems of equations, coordinate 
systems, gauge choices, and numerical methods.  In 
Sec.~\ref{sec:numerical} we discuss numerical issues, including 
methods, boundary conditions, and convergence testing.  In 
Sec.~\ref{sec:flat} we treat dynamically sliced flat space models to 
demonstrate simple yet powerful code tests.  In Sec.~\ref{sec:waves} 
we focus on a series of weak gravitational wave spacetimes, 
replicating results from Ref.~\cite{Anninos94d} and extending their study 
to non-axisymmetric cases.  In Sec.~\ref{sec:bhole} we treat black 
hole spacetimes with a wide variety of slicings, and compare
with the analytical solution in the case of geodesically sliced black 
hole.  In all our test cases, we obtain rigorous self-consistent 
convergence and, in those cases published before, excellent agreement 
with known results.


\section{Theoretical Concepts}
\label{sec:basic}

In this section, we discuss some basic theoretical concepts and
introduce the choices that we have implemented.  We follow closely the
BM theoretical formulation of Ref.~\cite{Bona97a}.  Some
aspects of the ADM formulations are also discussed in others
papers~\cite{Anninos94c,Anninos94d,Bruegmann96}.

\subsection{The BM formulation}
The BM formulation of the Einstein equations is discussed in
detail in a previous paper in this series~\cite{Bona97a}.  For
completeness, here we write the basic equations, although the reader
is directed to Ref.~\cite{Bona97a} for further details and
discussions.  One of the fundamental advantages of this formulation is
that the whole system can be written in first order balance law form:
\begin{equation}
  \partial_t {\bf u} + \partial_k F^k_-{\bf u} = S_-{\bf u}
\label{3Dbalance}
\end{equation}
where the vector ${\bf u}$ displays the set of variables, and both
fluxes $F^k$ and sources $S$ are vector valued functions. We stress
that the fluxes $F^k$ and the sources $S$ do not contain any
derivative of the set of variables, which is crucial for analyzing the
causal structure of the system and for the application of appropriate
numerical methods.

The vector ${\bf u}$ has the following $37$ quantities:
\begin{equation}
  \left( g_{ij}, \alpha, K_{ij}, D_{kij}, A_k, V_k \right)
\end{equation}
where $g_{ij}$, $\alpha$, and $K_{ij}$ have their standard
definitions.  As we have introduced a first order system, the
following relations act as algebraic constraints imposed on the
initial slice only:
\begin{equation}
  A_k = \partial_k ln\,\alpha\, , \;\;\;\; D_{kij} = \frac{1}{2}\;\partial_k
  g_{ij}\;,
\label{Ds}
\end{equation}
and the special combination
\begin{equation}
  V_i = {D_{ir}}^{r}-{D^r}_{ri}\;,
\label{vector}
\end{equation}
is considered as an algebraic constraint which will hold if and only
if the momentum constraint is satisfied\cite{Bona97a}.  We define
$D^{k}_{ij} = g^{km} D_{kij}$, i.e., we use the three-metric
$g_{ij}$ to raise and lower indices on objects, even if they do
not transform as tensors.  This is just a notational convenience.
We also note that the shift vector $\beta^i$ is not in this dynamical set, as
it is considered a given arbitrary function whose spatial derivatives
$B_k^{\;i} = \frac{1}{2}\;\partial_k \beta^{i}$ are known at any time.

The fluxes in the set of Eqs.~(\ref{3Dbalance}) are:
\begin{eqnarray}
\label{fluxes}
  F^k_-g_{ij} & = & 0 \;, \label{fluxgamma} \\
  F^k_-\alpha      & = & 0 \;, \label{fluxlapse} \\
  F^k_-K_{ij}      & = & -\beta^k\,K_{ij} + \alpha\;[\; D^k_{ij}
                         - n/2\;V^k\;g_{ij}
                         \\ \nonumber
                   &   & + 1/2\;\delta^k_i\;(A_j+2\,V_j-D_{jr}^{\;\;r})
                         \label{fluxK} \\
                   &   & + 1/2\;\delta^k_j\;(A_i+2\,V_i-D_{ir}^{\;\;r})\;]
                         \;, \nonumber \\
  F^k_-D_{kij}     & = & -\beta^r D_{rij} + \alpha\;(K_{ij}-s_{ij})
                         \;, \label{fluxD} \\
  F^k_-A_k         & = & -\beta^r A_r + \alpha\;Q
                         \;, \label{fluxA} \\
  F^k_-V_i         & = & -\beta^k V_i + B^k_{\;i} - B_i^{\;k}
                         \;. \label{fluxV}
\end{eqnarray}

The sources for these equations are:
\begin{eqnarray}
  S_-g_{ij} &=& - 2\;\alpha\;(K_{ij}-s_{ij}) + 2\beta^r\,D_{rij}
                       \;,\label{sourcegamma} \\
  S_-\alpha      &=& - \alpha^2\;Q + \alpha\beta^r\,A_r
                       \;,\label{sourcelapse} \\
  S_-K_{ij}      &=&  2(K_{ir}B_j^{\;r}+K_{jr}B_i^{\;r}-K_{ij}B_r^{\;r})
                       \nonumber \\
      & & + \alpha\; [  \; -^{(4)}R_{ij}
                     - 2K_i^{\;k}K_{kj}+tr\,K\;K_{ij} 
                       \nonumber  \\
                 & & \;\;\;\; - \Gamma^k_{\;ri}\Gamma^r_{\;kj}
                     + 2D_{ik}^{\;\;r}D_{rj}^{\;\;k}
                     + 2D_{jk}^{\;\;r}D_{ri}^{\;\;k}
                     + \Gamma^k_{\;kr}\Gamma^r_{\;ij}
                       \nonumber \\
            & & \;\;\;\;-(2\,D_{kr}^{\;\;k}-A_r)(D_{ij}^{\;\;r}+D_{ji}^{\;\;r})
                       \label{sourceK}  \\
  & & \;\;\;\; + A_i(V_j-1/2\;D_{jk}^{\;\;k}) + A_j(V_i-1/2\;D_{ik}^{\;\;k})
                       \nonumber \\
                 & & \;\;\;\; + A_j(V_i-1/2\;D_{ik}^{\;\;k})
                     - nV^kD_{kij} \;]
                       \nonumber  \\
  & & + n/4\;\alpha g_{ij}\;[\; -D_k^{\;rs}\Gamma^k_{\;rs}
                     + D_{kr}^{\;\;r}D^{ks}_{\;\;s} -2\,V^kA_k \nonumber  \\
                 & &  \;\;\;\; + K^{rs}K_{rs}-(tr\,K)^2
                     + 2\alpha^2\;G^{00} \;]
                       \;, \nonumber \\
  S_-D_{kij}     &=& 0
                       \;, \label{sourceD} \\
  S_-A_{k}       &=& 0
                       \;, \label{sourceA} \\
  S_-V_i         &=&   \alpha\;[\alpha\;G^0_{\;i}
                     + A_r\;(K^r_{\;i}-tr\,K\;\delta^r_i)
                       \nonumber \\
                 & & + K^r_{\;s}(D_{ir}^{\;\;s}-2D_{ri}^{\;\;s})
                     - K^r_{\;i}(D_{rs}^{\;\;s}-2D_{sr}^{\;\;s})]
                       \label{sourceV} \\
                 & & + 2(B_i^{\;r} - \delta_i^r\;tr\,B)\;V_r
                     + 2(D_{ri}^{\;\;s}-\delta^s_i\;D^j_{\;jr})B^r_{\;s}
                       \;. \nonumber
\end{eqnarray}

We have used the shorthand
\begin{equation}
  s_{ij} = (B_{ij}+B_{ji})/\alpha,
\end{equation}
and we stress again that for notational convenience, we raise and
lower indices with the three-metric, so for instance we have written
$B_{ij} = g_{ik}B^{\;k}_{j}$, even though $B_{ij}$ is not a
tensor quantity. 

The free parameter $n$ allows one to select a
specific evolution system (it is zero for the ``Ricci'' system and one for
the ``Einstein'' system), as discussed in Ref.~\cite{Bona97a}. 

As in this paper we do not explore methods based on the the
diagonalization of the system (i.e., based on the characteristic
fields), we will not detail here the spectral decomposition. The
reader is directed to Ref.~\cite{Bona97a} for all the theoretical
foundations of hyperbolicity. Applications of advanced hyperbolic
methods to the eigensystem in one-dimensional problems can be found in
Ref.~\cite{Arbona98a}.

\subsection{ADM Formulation}
As explained below, the Cactus code is written in a modular 
``plug-in'' way to allow for any number of formulations of the 
evolution system.  For example, in addition to evolving the 
BM system, the Cactus code has a straightforward ADM 
integrator subroutine (what in Cactus language we call a ``thorn''), 
which solves 
the ADM system using a full leapfrog scheme described in 
~\cite{Press86} and similar to that used for evolutions in 
\cite{Bruegmann96}.  The current implementation of the ADM system 
assumes a zero shift vector, and can perform conformal differencing, 
as described below.  We use this independent code for comparisons 
between the BM system and the ADM system.  In this way all 
code infrastructure used to generate results is the same; only the 
formulation of the equations differs, permitting a clean comparison of 
results.

The standard ADM equations are\cite{York79}:
\begin{equation}
\label{metric evolution}
\partial_{t}g_{ij}=-2\alpha K_{ij}+
D_{i}\beta_{i}+D_{j}\beta_{i},
\end{equation}
\begin{eqnarray}
\label{excurv evolution}
\partial_{t}K_{ij} & = & -D_{i}D_{j}\alpha  \nonumber \\
                   & & + \alpha \;[\; 
 -^{(4)}R_{ij} + R_{ij} + tr\,K K_{ij} - 2K_{ik}K^{k}{}_{j} \;]\; \nonumber\\
 & &+\beta^{k}D_{k}K_{ij}+ K_{ik}D_{j}\beta^{k}+K_{kj}D_{i}\beta^{k}.
\end{eqnarray}
Here $R_{ij}$ is the Ricci tensor, $R$ the scalar curvature, and 
$D_{i}$ the covariant derivative associated with three-dimensional 
metric $g_{ij}$.  Note that these equations look much simpler 
than the BM Eqs.~presented above, but this is deceptive, as 
the expansion of the Ricci tensor and the covariant derivatives brings 
a large number of terms already expanded in the BM system.  
In fact, apart from the fact that the BM system introduces 
the $V_k$ to achieve hyperbolicity, the BM and ADM systems 
only differ by the introduction of first order quantities and by the 
use of flux conservative form.  It is useful to notice that 
substitution of the definition of $V_k$ (Eq.~(\ref{vector})) into all the 
fluxes and sources detailed above allows a flux-conservative, but not 
necessarily diagonalizable, treatment 
of the ADM system as a first order system.

\subsection{The Constraints}

The $3+1$ decomposition of the Einstein equations result in the
evolution equations, Eqs. (\ref{metric evolution}) and (\ref{excurv
  evolution}), and additional constraint equations.  These are the
energy or hamiltonian constraint,
\begin{equation}
\label{Hamiltonian constraint}
R+({\mathrm tr} K)^{2}-K^{ij}K_{ij}= 2\alpha^2\;G^{00}\;,
\end{equation}
and the momentum constraint,
\begin{equation}
D_{k}(K^{ij}-g^{ij}{\mathrm tr} K)= \alpha\;G^0_k\;,
\label{momentum_constraint}
\end{equation}
both written here in this standard ADM form. 

Using the BM variables, we can write a more natural way to 
measure the constraints for the BM formulation.  The Ricci 
scalar term in the hamiltonian constraint (Eq.(\ref{Hamiltonian 
constraint})) can be computed using
\begin{equation}
\label{eqn:bmhamcon}
R = -2\partial_k V^k + D_k^{\;rs}\Gamma^k_{\;rs}
              - D_{kr}^{\;\;r}D^{ks}_{\;\;s} 
\end{equation}
The treatment of momentum constraint is more subtle. In generating the
equation for the evolution of $V_i$ in the BM formulation, the
momentum constraint Eq.~(\ref{momentum_constraint}) is factored in.
Thus, the algebraic constraint Eq. (\ref{vector}) measures the time
integral of momentum constraint violation, since $\partial_t V_i -
\partial_t ({D^r}_{ri} - {D_{ir}}^r) = $ the momentum constraint.
Therefore, rather than measure the momentum constraint directly, we
measure the algebraic constraint Eq.~(\ref{vector}) in its place.

\subsection{Coordinate Systems}
\label{sec:coords}

We choose a 3D cartesian coordinate system with a general metric,
general extrinsic curvature, and an arbitrary 3D shift
vector.  In this way any slicing or shift condition may be imposed as
needed.  The use of cartesian coordinates avoids the introduction of
any coordinate singularities, and enables the treatment of many
problems in 3D, regardless of their geometry.

We also allow for a (time independent) conformal rescaling of the
three-metric, which can be useful in increasing accuracy in spacetimes
where the conformal factor is known analytically, or perhaps
numerically through a solution of the constraint
equations\cite{Anninos94c}.  The key point is that the derivatives of
the conformal factor, provided in the initial data, can be known with
much greater accuracy than is achieved via finite differencing on the
grid used for evolution, and exploiting this knowledge can improve the
accuracy of the evolution.  In this case we write the metric as
\begin{equation}
\hat g_{ij} = \psi^4 g_{ij}.
\label{eqn:conformal}
\end{equation}
This leads to a relationship between the physical variables, denoted
only here with a hat (i.e. $\hat g_{ij}$), and
conformal variables,
\begin{eqnarray}
\hat D_{kij} &=& \psi^4 (D_{kij} + 2\frac{\partial_k \psi}{\psi}
g_{ij}) \\
\hat V_i &=& V_i + 4 \frac{\partial_i \psi}{\psi}.
\end{eqnarray}

We use these relationships to move the conformal factor and its
derivatives out of the flux terms and into source terms, allowing us
to evolve the system without having to take numerical derivatives of
the conformal factor, while still maintaining a first order flux
conservative form. The complete transformed equations are given in
Appendix A. The usage of conformal rescaling is an optional
parameter in Cactus, and we only use it in the black hole spacetime
tests of Sec.~\ref{sec:bhole}.

\subsection{Gauge Choices}
\label{sec:gauge}

Buried in the above system of equations is the slicing condition.
Normally considered as a supplemental condition in the ADM evolution
system, it is an integral part of the evolution system here, which
for clarity we repeat here:
\begin{equation}
\partial_t \alpha  = - \alpha^2\;Q + \alpha\beta^r\,A_r.
\label{eqn:lapseevol}
\end{equation}
It is important to realize that one does not need to use this
evolution equation for the lapse, as the BM formulation as presented
above allows any arbitrary choice of lapse and shift.  In principle,
if one is not concerned about the hyperbolicity of the system, it is
possible to use any choice and even dynamical choices that involve
dependencies on the spacetime metric or the extrinsic curvature are
allowed. However, given that in the future we are particularly
interested in exploiting the hyperbolicity of the system, we will
concentrate our studies in the the family of slicings introduced in
Ref.~\cite{Bona94b,Bona97a}. Namely, we admit lapses
with the following gauge source function:
\begin{equation}
  Q = f(\alpha) {\mathrm tr}K,
\label{eqn:lapsevol}
\end{equation}
where the most common choices for $f$ will be the following: $f=0$,
which implies geodesic slicing, $f=1$, which implies harmonic slicing,
$f=1/\alpha$, which gives rise to the so-called ``$1+\log(g)$''
slicing.  As discussed in Ref.~\cite{Bona97a}, all choices with $f>0$
are singularity avoiding and permit a hyperbolic
system.  

Recent work\cite{Alcubierre97a,Alcubierre97b} has shown the
potential danger of hyperbolic gauges in numerical relativity, as
blow-up along characteristics may occur depending on the choices for
the initial data and gauge condition.  This occurs independently of
the formulation of the equations. It is even possible in simple
electrodynamics with a nonlinear choice of gauge. More research is
necessary to characterize the initial data and gauge choices that are
``safe'' from gauge pathologies. Until then, the time-honored usage of
elliptic conditions remains the safest alternative. Maximal slicing
corresponds to the limit of divergent $f$. We implement it in our code
by not evolving Eq.~(\ref{eqn:lapseevol}), but rather by setting $f$
to zero through the update step, and solving the elliptic gauge
condition
\begin{equation}
  \Delta \alpha = K^{ij} K_{ij} \alpha
\label{eqn:maximal_def}
\end{equation}
after the update stage. The variables $A_k$ related to the derivatives
of the lapse are then computed using centered finite differencing.

We also allow a non-zero shift vector.  The choice of appropriate
shift vector in 3D is still an open research area, and so here we
demonstrate simple tests of the shift terms, but we do not use the
shift to enforce any physically motivated coordinate conditions (e.g.,
minimal distortion).  We will treat the shift as a ``given'' arbitrary
function of spacetime whose derivatives are known at all time, which
we instantaneously update every $\Delta t$.  See Ref.~\cite{Bona97a}
for a full discussion of general shifts and special subtleties in
their implementation for hyperbolic formulations (See also
\cite{Abrahams95a,Abrahams96a} for discussions of treatment of a
general shift in another hyperbolic framework.)

\section{Numerical and Computational Concepts}
\label{sec:numerical}

\subsection{The Cactus Code and Computational Science}

As well as solving the Einstein equations, the Cactus code endeavors
to address several difficult problems in computational science.
Although these are addressed in detail elsewhere
\cite{Masso98b,Walker98b,Clune98a}, we review the basic ideas briefly
here.

Cactus is a parallel code, and is parallelized using the
standard MPI message passing interface \cite{MPI}. This allows high
performance portable parallelism using a distributed memory model. All
major high performance parallel architectures, including the SGI/Cray
Origin 2000, SGI/Cray T3E, HP/Convex Exemplar and IBM SP-2 support
this programming model. Moreover, using MPI allows computing on
clusters of workstations using any of several free implementations of
MPI. The parallelism software we developed in Cactus, described in
\cite{Walker98b}, is a generic domain decomposition package for
distributing uniform grid functions on various processors and
providing ghost-zone based communications with a variety of stencil
widths and grid staggerings. The code can also compile without MPI,
allowing one source code to be used for single processor workstation
development and for massively parallel high performance computing
simulations. Our parallelism software is similar in spirit to Parashar
and Browne's DAGH system \cite{Parashar95,Parashar96}, with the crucial
difference being that it does not support fully adaptive meshes, and
therefore has a much lower degree of computational complexity.
However, the system does support the creation of multiple grids which
are distributed across all processors. This feature is used to provide
automatic convergence testing, the importance of which is stressed
below. The support of multiple grid hierarchies also allows multigrid
solvers and fixed mesh refinement hyperbolic solvers to be built upon
this parallel software. We are presently collaborating with several
groups and colleagues to implement this and many other computational
features which will be reported elsewhere.

The implementation of the Bona Mass\'o and ADM evolution equations in
Cactus has been strongly optimized for high single processor
performance on cache based architectures. The code very effectively
utilizes the many-tiered memory structures of modern high performance
computing architectures, through a variety of techniques described in
Ref.\cite{Clune98a}. The combination of portable parallelism with high
single processor performance has led to a very well performing code.
In recent performance studies, the Cactus code
evolution system attained better than 66 GFlop/s performance on a 512
processor T3E-900, experiencing a speedup of more than 500 fold over 1
processor on the 512 processor system.

In addition to this performance related technology, the Cactus code
attempts to be a {\em usable} code in a {\em collaborative} setting.
The code has a clearly defined ``plug-in'' coding style, by which
users developing code to extend cactus do not modify the central code,
but rather place their subroutines in a ``thorn'' which has a well
defined calling structure. There are several positive benefits to this
software engineering decision, as managing and maintaining the code
becomes a distributed task.  Each ``thorn'' and the central code are
managed as separate modules using versioning software, and each small
chunk has a clearly defined maintainer. Experimentation by a user will
not disrupt the work of all other users, since other users will not
be required to use new and unstable ``thorns.'' With the thorn system,
we are able to maintain a {\em single} {\em central} version of the
Cactus code which all users of the code extend in a non-intrusive
manner.

\subsection{Boundary Conditions}
\label{sec:boundaries}
\begin{figure}
  \incpsfig{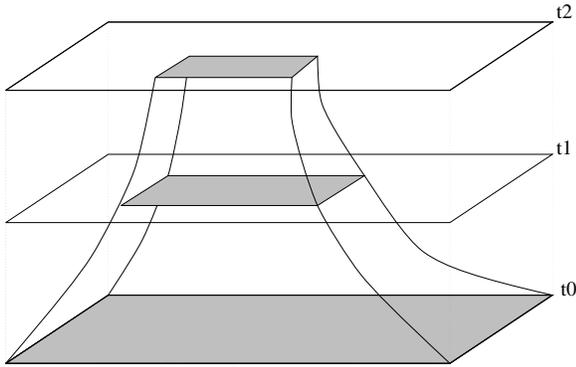}
\caption{Even with an inaccurate, but stable, outer boundary, the region which is
  causally disconnected from the boundary has finite size after some
  evolution. We note that in general relativity the coordinate speed
  of propagation is not one, so it is possible that causal connection
  to the boundaries does not move at a linear speed, as indicated by
  the curved edges of the boundary region here.}
\label{fig:domaindep}
\end{figure}

As discussed in the introduction, boundary conditions are a major open
research problem in numerical relativity.  It is beyond the scope of
this paper to formulate an adequately general outgoing boundary
condition.  We opt here to use very simple boundary conditions and
concentrate on our evolution in the interior.  We will demonstrate
that, although poor boundary conditions can lead to loss of
convergence in the interior of any numerically generated spacetime,
one can still find accurate solutions to the Einstein equations for a
finite time.  In the worst scenario, the interior solution should
always be valid for the Cauchy domain of dependence shown in
Fig.~\ref{fig:domaindep}, but generally one fares better than this with
reasonable conditions, such as those we use.

The boundary condition we use is a simple copying boundary (sometimes
called zero order extrapolation). That is, for each point on the
physical outer boundary, we copy all the variables from the point
nearest the inside. In practice, this condition will prove very
effective in several scenarios. It has the effect of canceling the
flux difference in the exterior (as the finite differences of the last
points will be zero). This is a valid approximation to ``outgoing''
boundary conditions when the boundary is close to linear perturbation
around flat spacetime. In this special case, the sources of the BM
system are close to zero and the system approximates a set of linear
wave equations, so canceling the exterior flux effectively prevents
any incoming information from outside the domain. 

Following Ref.\cite{Anninos94d}, we allow octant boundary conditions,
which are appropriate for spacetimes with rotational and
equatorial plane symmetry.  This allows us to simulate black hole
spacetimes with symmetries as full 3D problems, while using one eighth
the computational resources necessary when evolving on a full grid.
Many interesting problems, including Schwarzschild\cite{Anninos94d},
axisymmetric black hole collisions\cite{Anninos96a} and distorted
black holes\cite{Camarda97b}, and even some full 3D data sets with
certain dependence on the azimuthal angle\cite{Camarda97a,Allen98a},
can be treated with this symmetry, allowing a great savings in
computational resources.  Of course, our code can run without this
boundary condition also, and as demonstrated through comparisons
running full and octant grids \cite{Anninos94d}, the use of octant
symmetry does not affect results.

There are many other boundary conditions which are applicable to
three-dimensional numerical relativity, which we do not consider here.
However, they are worth mentioning.  The apparent horizon boundary
condition (AHBC)\cite{Seidel92a,Cook97a} adds a boundary at the causal
interior of a black hole spacetime.  Recent progress on the outer
boundary treatment, such as matching schemes to
perturbative\cite{Abrahams97a} or characteristic\cite{Bishop97a}
evolution schemes, look very promising, and could be ultimately used
by Cactus.  Another promising boundary treatment involves moving the
outer boundary to infinity\cite{Huebner98} by conformally rescaling
the metric, as per Friedrich's hyperbolic system\cite{Friedrich96}.
This has proven very successful in one-dimensional
calculations\cite{Huebner96,Frauendiener97} and higher dimensional calculations with
this method should be available soon.  Finally, we reiterate that
boundary conditions are a major motivation for hyperbolic treatments
of the Einstein equations.  Through study of the eigenfields and
eigenvalues of the transport system they provide more information
about the flow of information at the boundaries, which can be
exploited in numerical methods\cite{Arbona98a}.

The interior of black holes is usually handled with an isometry
condition, which identifies the interior of a black hole with the
isometric exterior via inversion through the sphere.  This has been
crucial in numerous black hole evolutions published to date (see,
e.g., \cite{Abrahams92a,Anninos94b,Anninos94d,Camarda97b,Allen98a}.
We do not use a three-dimensional isometry condition, as is described
in ~\cite{Anninos94d,Bruegmann96}, since we must transform not only
the metric and curvature tensor, but also the first order quantities
(such as derivatives of the metric and the vector $V_k$) which do not
transform as a tensor.  An isometry could be implemented for the BM formulation in
principle, but as we are looking to move to more general methods to
treat the black hole, such as an AHBC, we have chosen not to do so at
present.  Furthermore, as shown in Ref.\cite{Anninos94d}, with certain
slicing conditions like maximal slicing, even without AHBC the
isometry condition can be ignored and both regions inside and outside
the horizon can be evolved, as long as the lapse collapses
sufficiently quickly in the vicinity of the singularity.  We will
make use of this property of maximal slicing in tests presented below.
Recent work proposes an alternative to isometry conditions by
``stuffing'' with matter the interior of black holes\cite{Arbona97}
(``stuffed black holes'') and we are currently investigating this
approach for 3D spacetimes.

\subsection{Evolution Schemes}

\subsubsection{The Strang Splitting}

Following the numerical discussion of the BM system in
Ref.~\cite{Bona97a}, we will split Eq. (\ref{3Dbalance}) into two
separate processes. The transport part is given by the flux terms
\begin{equation}
  \partial_t {\bf u} + \partial_k F^k_-{\bf u} = 0 \;\; .
\label{3Dtransport}
\end{equation}
The source contribution is given by the following system of {\em
ordinary} differential equations
\begin{equation}
  \partial_t {\bf u} = S_-{\bf u} \;\; .
\label{3Dsources}
\end{equation}
Numerically, this splitting is performed by a combination of both flux
and source operators. Denoting by $E(\Delta t)$ the numerical evolution
operator for system (\ref{3Dbalance}) in a single timestep, we
implement the following combination sequence of subevolution steps:
\begin{equation}
  E(\Delta t) = S(\Delta t/2)\;T(\Delta t)\;S(\Delta t/2)
\label{splitting}
\end{equation}
where $T$, $S$ are the numerical evolution operators for systems
(\ref{3Dtransport}) and (\ref{3Dsources}), respectively.  This is
known as ``Strang splitting''~\cite{Press86}.  As long as both
operators $T$ and $S$ are second order accurate in $\Delta t$, the
overall step of operator $E$ is also second order accurate in time.

This choice of splitting allows easy implementation of different
numerical treatments of the principal part of the system without
having to worry about the sources of the equations.  Additionally,
there are numerous computational advantages to this technique, as
discussed in \cite{Clune98a}. Theoretical and practical advantages for
general relativistic hydrodynamics, where the source step couples the
equations for the whole system of Einstein plus matter equations, will
be detailed elsewhere.

\subsubsection{The Source update method}
Currently, we treat the source integration with a second order
predictor-corrector method\cite{Press86}.  During this step, we
only need to evolve the 16 quantities which have a source
($g_{ij}$,$K_{ij}$,$V_k$ and $\alpha$).

We use standard finite difference notation here.  Subscripts denote
grid index, and superscripts denote time index.  For instance,
$u^{n}_{i,j,k}$ is the value of field $u$ at spatial grid point
$i,j,k$ and time level $n$.  We use the special upper indices $p$ and
$c$ to denote the predicted and corrected values during an update
cycle, as we define below.
In order to update a variable $u$ (running through the 16 quantities 
with source) at time level $n$ to the future time level $n+1$, we 
first compute the ``predicted value'' $u^p_{i,j,k}$ at every point 
$i,j,k$ of our computational grid.
\begin{equation}
  u^p_{i,j,k} =  u^n_{i,j,k}  +  \Delta t\, S(u^n_{i,j,k})\;,
  \label{sopredict}
\end{equation}
where $u^n_{i,j,k}$ is $u$ at current time step $n$ and grid point
$i,j,k$, and $\Delta t$ is the time discretization interval.
With this predicted value of $u^p$, we compute the predicted sources 
and take a corrector step:
\begin{equation}
  u^c_{i,j,k} =  u^p_{i,j,k}   +   \Delta t \, S(u^p_{i,j,k}) \;.
  \label{socorrect}
\end{equation}
Finally, the evolved value of $u$ at the next time step $n+1$ is the average of
the value at time step $n$ and the correction:
\begin{equation}
  u^{n+1}_{i,j,k} = (u^{n}_{i,j,k} + u^{c}_{i,j,k})/2
  \label{soaverage}
\end{equation}
In practice, the steps (\ref{socorrect}) and (\ref{soaverage}) can be
combined into one. Note that this is a completely local operation at
every grid point, which allows a high degree of
optimization\cite{Clune98a}.  Higher order methods for source
integration can be easily implemented, but this will not improve the
overall order of accuracy.  However, in special cases where the
evolution is largely source driven\cite{Masso92}, it may be important
to use higher order source operators, and this method allows such
generalizations.

\subsubsection{The Flux Update Methods}

The implementation of numerical methods for the flux operator is much
more involved, and we have many choices at our disposal,
ranging from standard choices to advanced shock capturing
methods\cite{Leveque92,Bona96a,Bona97a}.  In this paper, we will limit
ourselves to two methods: the MacCormack method, which has proven to
be very robust in the computational fluid dynamics field (see, e.g.,
Ref.~\cite{Yee88} and references therein), and a directionally split
Lax-Wendroff method.  These schemes are fully second order in space
and time.  Although the Cactus code has a modular structure allowing
numerous numerical methods to be plugged in and applied to problems
for which they may be best suited, in this paper we restrict ourselves
to results with these two methods. Unless otherwise noted, results are
generated with the MacCormack method; use of the Lax-Wendroff solver
will be explicitly noted.

Following the previous notation we define our fluxes in
individual directions $x$, $y$, and $z$ as $FX$, $FY$, and $FZ$
respectively. 

The MacCormack method evolves a given quantity $u$, which now runs
through the 30 dynamical variables having fluxes
($K_{ij}$,$D_{kij}$,$V_k$,$A_k$; the $A_k$ and $D_{kij}$ only have fluxes in
one direction, which is explicitly exploited in our code) with the
following algorithm: First, in order to update the variable $u$ to the
time level $n+1$, we compute the ``predicted value''
$u^p_{i,j,k}$, with first order backward finite differences:
\begin{eqnarray}
  u^p_{i,j,k} =  u^n_{i,j,k} 
  & + &  {\Delta t\over \Delta x} (FX(u^n_{i,j,k}) - FX(u^n_{i-1,j,k}))
\nonumber\\
  & + &  {\Delta t\over \Delta y} (FY(u^n_{i,j,k}) - FY(u^n_{i,j-1,k}))
\label{pred}\\
  & + &  {\Delta t\over \Delta z} (FZ(u^n_{i,j,k}) - FZ(u^n_{i,j,k-1}))
\nonumber
\end{eqnarray}
where, in addition to the quantities defined above, $\Delta x$,
$\Delta y$, and $\Delta z$ are the spatial discretization
intervals.  Note that this predicted step can be done in a given
direction (say $x$), from grid points $2$ to $nx$ (total number of
grid points in that direction), as the first order
backward differencing only requires $i-1$.  With this predicted value
of $u^p$, we recompute predicted fluxes and sources and take a corrector
step with forward finite differencing:
\begin{eqnarray}
  u^c_{i,j,k} =  u^p_{i,j,k} 
  & + &  {\Delta t\over \Delta  x} (FX(u^p_{i+1,j,k}) - FX(u^p_{i,j,k}))
\nonumber \\
  & + &  {\Delta t\over \Delta  y} (FY(u^p_{i,j+1,k}) - FY(u^p_{i,j,k}))
\label{corr}\\
  & + &  {\Delta t\over \Delta  z} (FZ(u^p_{i,j,k+1}) - FZ(u^p_{i,j,k}))
\nonumber
\end{eqnarray}
Now we can correct the interior points of the domain from $2$ to
$nx-1$, as we have a prediction for the last plane at $nx$.  Finally,
the evolved value of $u$ at the next time step $n+1$ is the average of
the value at time step n and the correction:
\begin{equation}
  u^{n+1}_{i,j,k} = (u^{n}_{i,j,k} + u^{c}_{i,j,k})/2.
\end{equation}

A similar method could be obtained interchanging the order or backward
and forward derivatives in the predictor and corrector steps.  We note
that both methods can introduce certain spatial asymmetries in a
numerical evolution, due to the preferred order of finite difference
operations in the predictor and corrector steps. These asymmetries
converge away to second order, as we will discuss below when
presenting results.

The directionally split Lax-Wendroff method uses a series of one
dimensional Lax-Wendroff integrations to complete a full three
dimensional integration step. In one dimension, the Lax-Wendroff
scheme is
\begin{eqnarray}
u^{n+1/2}_{i+1/2} &=& \frac{1}{2}(u^n_{i+1}+u^n_i) +
\frac{\Delta t}{2 \Delta x} (F(u^n_{i+1}) -F(u^n_i)) \\
u^{n+1}_i &=& u^n_i + \frac{\Delta t}{\Delta x}
(F(u^{n+1/2}_{i+1/2}) -F(u^{n+1/2}_{i-1/2})).
\end{eqnarray}

Several options exist to turn Lax-Wendroff into a three dimensional
scheme. Here we choose directional splitting\cite{Press86}. Defining
$X(\Delta t)$ to be a one dimensional Lax-Wendroff in the
$x-$direction, $Y(\Delta t)$ in the $y-$direction and $Z(\Delta t)$ in
the $z-$direction we define a full flux time step as $X(\Delta t)
Y(\Delta t) Z(\Delta t)$ on the first step, $Y(\Delta t) Z(\Delta t)
X(\Delta t)$ on the second step, $Z(\Delta t) X(\Delta t) Y(\Delta t)$
on the third step, and then repeat the prescription. This permutation
leads empirically to a second order in space and time scheme, as we shall
demonstrate below. The advantage of this directionally split
Lax-Wendroff is that, by turning the problem into a set of one
dimensional PDEs, implementation of a simple inner (apparent horizon)
boundary condition becomes easier, as will be reported
elsewhere\cite{Walker98a}. 

\subsection{Convergence}
\label{sec:convergence}

Since the pioneering work of Choptuik\cite{Choptuik91}, the usage of 
convergence tests in numerical relativity is slowly becoming standard 
practice\cite{Anninos94d,Abrahams97a,Gomez98a,Alcubierre97b}.  The 
recent discovery and characterization of gauge 
pathologies\cite{Alcubierre97b} stresses the importance of careful 
convergence analysis, especially in 3D numerical relativity, as 
simulations may hide solutions that ``look'' reasonable for a given 
resolution but do not satisfy the constraints.  For completeness, here 
we review the basis of convergence tests.  We will discuss the case of 
numerical discretization of PDE's with finite differences.  Similar 
arguments can be developed for other approaches.

Assuming that we have well behaved solutions which allow a expansion
in Taylor series, we can relate the numerical solution $\tilde S$ to
the analytical solution $S$ in the following way:
\begin{equation}
\tilde S = S + O(\Delta^\sigma),
\end{equation}
where $\Delta$ is the grid
spacing. Consistent numerical simulations must demonstrate that some
form of this relation is obeyed, as the refinement of the grid should
always improve the solution. In many cases, it is actually possible to
measure the convergence rate $\sigma$.  This analysis is crucial if
one is to understand how close a given numerical solution is to the
true analytic solution, which is generally not known.

Given three discretized solutions,
$\tilde S(\Delta)$, $\tilde S(\Delta/q)$ and $\tilde S(\Delta/q^2)$ we
find that
\begin{eqnarray}
\label{eq:noanalyticsolution}
  L &\equiv& \tilde S(\Delta/q) - \tilde S(\Delta) = O\left
((\Delta/q)^\sigma - \Delta^\sigma\right) \\
  M &\equiv& \tilde S(\Delta/q^2) - \tilde S(\Delta/q) =
  O\left((\Delta/q^2)^\sigma - (\Delta/q)^\sigma\right).
\end{eqnarray}
We define precisely the intuitively clear ``-'' operator below.  
Dividing and canceling $\Delta^\sigma$ we find
\begin{equation}
  \frac{L}{M} = \frac{q^{-\sigma} - 1} {q^{-2\sigma} - q^{-\sigma}} =
  q^\sigma,
\end{equation}
so solving for $\sigma$,
\begin{equation}
\label{eqn:sigmadef}
\sigma = \frac{ \log\left(\frac{L}{M}\right)}{\log q}.
\end{equation}
Eq.~(\ref{eqn:sigmadef}) is the principal definition of the
convergence rate $\sigma$ that we will use.  In practice, we use $q =
2$ for our convergence tests, that is, we double or halve our grid
resolution in a sequence of simulations when determining $\sigma$.

The definition of the ``-'' operator used to form $L$ and $M$ is very
important. If the points of $S(\Delta/q)$ and $S(\Delta)$ are
coincident on the $\Delta$ grid, then simply pointwise subtraction
followed by a norm of the difference can generate the ``-''
operator. We can schematically represent this operation as
\begin{equation}
  \tilde S(\Delta/q) - \tilde S(\Delta) \equiv |\tilde S(\Delta/q)_{qi,qj,qk} -
  \tilde S(\Delta)_{i,j,k}|,
\end{equation}
although often more complicated index juggling that simply $i
\rightarrow qi$ is required.  We note that $|x|$ denotes some norm
over the $i,j,k$ space (e.g., maximum, ${\cal L}_1$, ${\cal L}_2$).

If the points are not coincident, a possibility 
is to use some norm over the solutions and then define ``-'' as the
difference of those norms. That is
\begin{equation}
  \tilde S(\Delta/q) - \tilde S(\Delta) \equiv |\tilde S(\Delta/q)| -
  |\tilde S(\Delta)|.
\end{equation}
We call this convergence in the norm.  This method has the advantage
that it is very easy to calculate during runtime of a parallel code,
but is often susceptible to large amounts of noise.  If we have an
interpolation operator $I^{\Delta/q}_{\Delta}$ which interpolates a
solution from a grid with resolution $\Delta/q$ to one with resolution
$\Delta$, we can define
\begin{equation}
  \tilde S(\Delta/q) - \tilde S(\Delta) \equiv |I^{\Delta/q}_{\Delta}
  \tilde S(\Delta/q) - \tilde S(\Delta)|.
\end{equation}
Generally, an interpolator of at least order $\sigma$ is required to
do this style of convergence testing.

Finally, when an exact solution is known, we only require two
numerical solutions to the equations to measure $\sigma$.  That is,
given a discretized solution at $\tilde S(\Delta)$ and $\tilde
S(\Delta/q)$ and an exact solution $S$, we can form two differences
pointwise,
\begin{eqnarray}
\label{eq:analyticsolution}
  L &=& S - \tilde S (\Delta)  = {\cal O} (\Delta^\sigma)\\
  M &=& S - \tilde S (\Delta/q) = {\cal O} (q^{-\sigma} \Delta^\sigma)
\label{eqn:lmdef2}
\end{eqnarray}
and therefore find the relationship
\begin{equation}
\label{eqn:knownsol}
  L = q^{\sigma} M,
\end{equation}
and again we recover $\sigma$ from Eq. (\ref{eqn:sigmadef}).
Simply said, for a second order method, the error should be four times
larger on the coarser grid than the finer grid. This method will prove
valuable for calculating convergence against known solutions,
convergence of constraints, and convergence of fictitious numerical
errors, such as asymmetries.

As before, we are faced with the problem of computing the quotient 
$L/M$ accurately, especially in the case of fields which go to zero.  
Once again, we can solve this problem by interpolating $M$ onto the 
grid which $L$ inhabits and forming the quotient pointwise.

When the convergence rate is expected to be second order, with this
technique we can also measure $\sigma$ graphically at all points.
That is, if we have a known solution $S$, we can plot, $\tilde
S(\Delta) - S$, $(\tilde S(2 \Delta) - S)/4$, $(\tilde
S(4 \Delta)-S)/16$ and so forth.  If the points agree, then we have
second order convergence.  This method has the advantage that point to
point noise present in calculating $\sigma$ can be eliminated ``by
eye,'' and we shall use this method often below.

Convergence testing is an essential component of a battery of code
tests.  Demonstrating that an evolution scheme has the appropriate
convergence order shows that boundary treatments, methods, and
infrastructure are coded properly.  Studying convergence properties
can help diagnose and track subtle errors in a code.  However, showing
that, for example the metric function $g_{xx}$ converges does not
imply that one is solving the Einstein equations; it merely means that
one is solving the coded evolution equation to second order.  Thus,
convergence testing against known solutions is important.  In a few
rare cases, notably a geodesically sliced black hole, there are exact
solutions to the non-linear dynamical Einstein equations.  In this
case one can show not only that numerical results converge to
something (that is, we find $\sigma = 2$ using definitions
(\ref{eqn:sigmadef})), but also that they converge to the right
thing (that is, we get order $\sigma = 2$ when comparing against the
known solution, using the definition (\ref{eqn:knownsol})).

With the Einstein equations, however, we play on favorable
ground, as we always have an analytic solution at our disposal: the
vanishing of the
constraints.  That is, all correct solutions to the fully nonlinear
Einstein equations have the property that the hamiltonian and momentum
constraints must identically vanish.  Regardless of the relativistic
system being simulated, if the initial data satisfies the constraints,
then so must all subsequent time steps.  We assume the behavior of the
hamiltonian constraint, $H$, is
\begin{equation}
  H(\Delta x) = 0 + E(\Delta^\sigma),
\label{eqn:hamcon}
\end{equation}
where $E(\Delta^\sigma)$ is the error due to finite differencing with
a spatial step $\Delta$.
Choptuik has investigated this point at great length in
Ref.\cite{Choptuik91}, where he shows that for a consistent finite
differencing of the free evolution of the Einstein equations, the
constraints have the same order error as the evolution scheme.
Choptuik demonstrated this in spherical symmetry, and here we
demonstrate this in full three dimensional numerical relativity.  With
relation Eq.~(\ref{eqn:hamcon}) in hand, forming $L$ and $M$ in the
language of Eq.~(\ref{eqn:lmdef2}) simply amounts to looking at the
value of the constraint.  If we double the resolution, and the
numerical code is solving the Einstein equations, our constraints must
drop by a factor of four (for a second order scheme) everywhere.  (We
note that one may use the constraints to eliminate one of the
evolution equations, and with this approach it would be reasonable to
expect that the code could demonstrate an independent construction of
the eliminated evolution equation converging to zero, rather than the
constraint.)

Having discussed the convergence techniques we use to study the
performance of the Cactus code, we describe briefly our philosophy of
their use before moving on to examples below.  An important point is
that a 3D code should exhibit {\em convergence}, even if the
resolution is too low to exhibit a high degree of {\em accuracy}.  For
instance, a given numerical result may differ from the true analytic
solution by a large factor.  This is not necessarily a major concern,
as long as doubling the resolution can quarter the error.  By running
simulations at different resolutions, one can then estimate how close
the numerical solution is to the analytic solution, understand the
behavior of the truncation error, and estimate the resolution required
to obtain a solution to the desired accuracy.  We regard this as a
crucial requisite of a code, which as we show below, our code has.

We note that it is possible that resolution can be too low to allow an
evolution beyond a certain point.  For instance, a geodesically sliced
black hole with very low resolution may crash before (or after) a
higher resolution simulation, since there is a physical singularity
present.  However, in this case convergence should show the region in
which the simulation is accurate.  When a solution starts failing to
converge, the evolution is probably about to fail.  We will see an
example of this in Sec.~\ref{sec:bhmaximal}.  Convergence at the
boundaries also offers useful information.  Using our simple ``copying''
boundary condition without any advanced treatment of the system, we
expect that condition to have a first order effect on phenomena which
interact with the boundary.  Finally, we regard second order
convergence to be desirable, but not necessary, in order to verify a
given numerical result.  For instance, often one does not have second
order convergence near boundaries.  But such an effect can be studied
and understood.  The key point is that one should know that a code
converges at or above the {\em expected} order, even if that order is one.

\section{Flat Space Tests}
\label{sec:flat}
\subsection{Dynamically sliced Flat space}

One of the crudest first tests of any 3D cartesian based code is, 
given geodesically sliced Minkowski space, does the code produce 1 and 
0 forever.  Of course Cactus does, but this test is almost useless, 
since one cannot measure convergence, and all constraints are 
trivially satisfied.  A more interesting and much more important test 
is that of a {\em dynamically} sliced flat spacetime.  That is, we 
choose an initial lapse in Minkowski space which is not unity 
everywhere, and then we evolve this system with a ``live'' slicing 
condition on the lapse $\alpha$, such as harmonic slicing.  Such an 
idea has been suggested in the past by York~\cite{York79} and also 
implemented and studied in detail by Mass\'o\cite{Masso92}.  We have 
examined three distinct cases, 1D periodic data (e.g., $\alpha$ is a 
periodic function of one coordinate only), 3D periodic data, and 3D 
data where $\alpha$ falls off to unity at large radii.  The first two 
allow us to study Cactus without boundary affects, and the last allows 
us to evaluate the quality of our boundary conditions.  In this 
section we present 3D simulations with ``copying'' boundary conditions 
and harmonic slicing, and in the following sections we also chose 
various shifts in both one and three dimensions.  Simulations here are 
performed with the Einstein system ($n=1$).

We note this problem is similar to the example used to study
coordinate conditions discussed in~\cite{Balakrishna96a}, but with
some important differences.  First, we evolve with harmonic slicing
throughout the entire evolution, rather than using the lapse to
generate a small ``bump'' in tr$K$, followed by maximal slicing, as
in Ref.~\cite{Balakrishna96a}.  Secondly, in harmonic slicing of flat
space, the lapse evolution equation becomes wavelike, so our initial
pulse travels off the grid as a wave pulse.

\begin{figure}
  \incpsfig{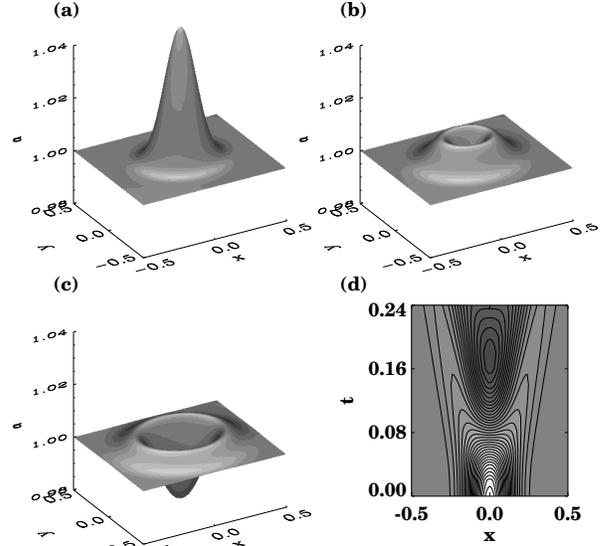} \caption{We show slices of the
    evolution of the lapse in the 3D dynamically sliced flat spacetime
    described in the text. Figures (a), (b), and (c) show slices in
    the $z=0$ plane at times $t=0.0$, $0.1$, and $0.2$. In Figure (d)
    we show the value of $\alpha$ on the $y=0$, $z=0$ line evolving in
    time as a colored contour map. We note that in this simulation,
    the majority of the pulse has not yet hit the boundary of the
    computational domain. $101^{3}$ grid points were used with a
    resolution of $\Delta x = 0.01$.}
\label{fig:flattylapse}
\end{figure}

In the 3D case, we choose an initial lapse with a
gaussian bump specified by
\begin{equation}
\label{eqn:flattydef}
\alpha = 1 + A \exp\left(\frac{r^2}{\sigma^2}\right)   .
\end{equation}
In the 1D case, we choose an identical form, simply replacing $r$ with 
$x$, $y$, or $z$.  In both cases, we see simple wave-like propagation 
in the lapse, as demonstrated below. For our first 
test, we evolve this dynamically sliced flat space system in 3D with a 
resolution $\Delta x = \Delta y = \Delta z = 0.01$ on a grid of 
$101^{3}$ centered around $r=0$.  We choose the parameters $A = 0.05$ 
and $\sigma^2 = 0.05$.  Two-dimensional slices in the $z-$plane of the 
evolution of this initial lapse in a harmonically sliced spacetime, 
with ``copying'' boundary conditions discussed in Sec.  
\ref{sec:boundaries} above, are shown in Fig.~\ref{fig:flattylapse}.  
Other metric functions, although initially taking a Minkowskian form, 
develop similar dynamics.

\begin{figure}
  \incpsfig{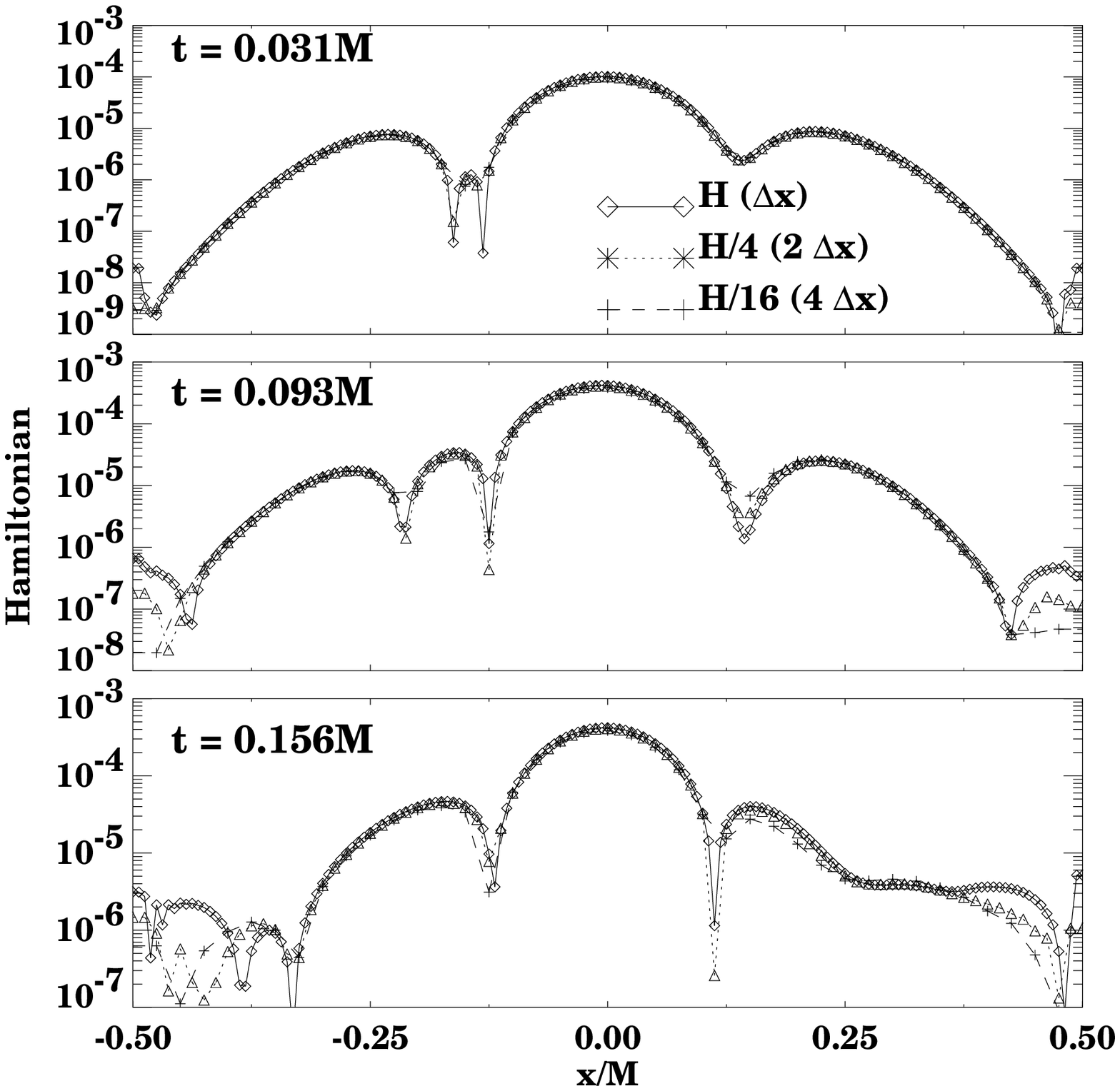} \caption{We demonstrate convergence
    of the hamiltonian constraint to zero in the interior of the 3D
    evolved dynamically sliced flat space.  Although the $x-$axis is
    shown here, other directions have similar results. The fact that
    the high resolution hamiltonian is equal to one-quarter the medium
    resolution, and this one is itself one-quarter the lower
    resolution, indicates that the hamiltonian converges rigorously to
    zero at second order in the interior.  However, the logarithmic
    scale reveals, at a very low level, a lack of second order
    convergence near the boundaries at later times.  This is caused by
    our outer boundary condition, which is not expected to be second
    order accurate.}
\label{fig:flattyhamconv}
\end{figure}

We demonstrate that the hamiltonian constraint converges at second
order in the interior in Fig.~\ref{fig:flattyhamconv}, where we show
the constraint at three different resolutions, with the appropriate
factors of four and sixteen.  The fact that the lines are coincident
demonstrates second order convergence.  The actual value of the
convergence exponent on the grid is above 1.9 for the entire
evolution, until the pulse interacts strongly with the boundary.

\begin{figure}
\incpsfig{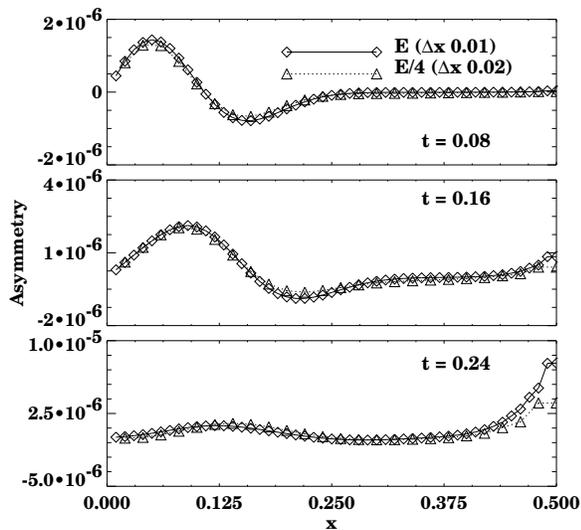} \caption{We show
symmetry violation due to the MacCormack predictor corrector method,
and how that converges away.  For the dynamically sliced flat space
model, we show $E = \alpha(x)-\alpha(-x)$ for $x > 0$ at the times
$t=0.8$, $t=0.16$ and $t=0.24$.  We show $E/4$ at a low resolution (dotted
line) and $E$ at twice the resolution (solid line).  The fact that
the high resolution error is less than or equal one-quarter the low
resolution error indicates that the method's asymmetry converges to
zero at second order.  An interesting feature of this figure is that
it demonstrates the first order nature of our boundary condition
clearly.  Since the lapse becomes dynamic on the boundary at later times,
convergence order drops from two (which the evolution system obeys) to
one (which the boundary condition obeys) as the wave propagates
towards and through the boundary.}
\label{fig:flattysymmetry}
\end{figure}

We noted above that due to the upwind/downwind nature of the
MacCormack predictor-corrector method we use, certain asymmetries in
the evolution are introduced.  In Fig.~\ref {fig:flattyhamconv} we see
that symmetry around the origin of the coordinate system is not
maintained except in the limit of a converged solution. (We note
rotational symmetries are obeyed. By this we mean that, given
symmetric data, our code will generate {\em identical} solutions along
an $x$-directed and $y$-directed slice of our data. However both of
these solutions will be (identically) asymmetric around the origin.)
This asymmetry is purely an artifact of our method having an
upwind/downwind nature, as shown in the finite difference
representation. As such, this asymmetry should
converge away at second order. In Fig.~\ref{fig:flattysymmetry} we
show that this asymmetry is an artifact of numerical error, and
consequently, converges to zero by measuring the asymmetry, $E =
\alpha(x) - \alpha(-x)$, for the evolved flat space case. Clearly this
should be zero in the converged limit, so the numerical solution
should obey $E(\Delta x / 2) =
E(\Delta x)/4$ if our method converges at second order. From
Fig.~\ref{fig:flattysymmetry} we see that this relationship is obeyed
except at the boundaries, where our boundary condition imposes a first
order asymmetry on the system at late times.

Figs.~\ref{fig:flattyhamconv} and \ref{fig:flattysymmetry} also give
an interesting indication of our boundary conditions when dynamics are
present at the boundaries.  As shown in Fig.~\ref{fig:flattylapse}, the
traveling pulse in the lapse is approaching the boundary by late times
in our simulation. Once the dynamics reach the boundary, convergence
drops from second order towards first order there. This is indicated
in Fig.~\ref{fig:flattysymmetry} by the high resolution case (solid) having
more than one-quarter the error of the low resolution case (dotted
line), and by non-second-order convergent (although small) errors in
the hamiltonian constraint, as shown in Fig.~\ref{fig:flattyhamconv}.
That is, the solid line is above the dotted line.

\begin{figure}
  \incpsfig{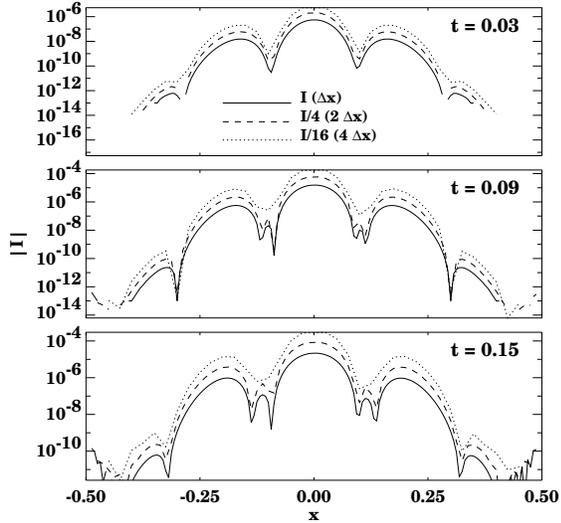} \caption{We show the Riemann 
  invariant $|I|$ along the $x$-line in the dynamical flat spacetime at 
  various resolutions. We show the invariant at three resolutions
  with appropriate factors of $1/4$ and $1/16$, and note that the
  invariant is converging towards zero {\em faster} than second
  order. In this case, the convergence exponent is very close to four.}
\label{fig:flattyInormEvol}
\end{figure}

So far, we have only measured convergence of metric functions and
constraints.  We can also examine the physical properties of our
underlying spacetime.  In this spacetime, we can demonstrate that we
are evolving Minkowski space by measuring the Riemann invariants $I$
and $J$, computed using a $3+1$ method\cite{Gunnarsen95}.
These should be identically zero, but they will not be due
to finite differencing errors.  However, we can test how they behave
with varying resolution.  In Fig.~\ref{fig:flattyInormEvol} we show
$|I|$ at three different resolutions for the distorted flat space case
considered here.  We note that, firstly, $|I|$ is small, and also that
it decreases {\em faster} than second order with grid resolution
towards zero. In fact, in this case the convergence exponent for $|I|$
is very close to four. Clearly boundary effects are evident,
driving the system away from the underlying flat space.

\subsection{Testing the Shift}

We now introduce the shift vector $\beta^{i}$ to test its effect on
the solution.  The dynamically sliced flat spacetime is an excellent case
to test the shift terms in Cactus. We first examine a constant shift,
and then move to a spatially varying shift to test all terms related
to the shift vector.

\subsubsection{A Test of a Constant Shift}

As a first simple test, we chose the one dimensional periodic initial
lapse, and evolve this on an explicitly 1D grid with periodic boundary
conditions (that is, we use Cactus on a $(nx,1,1)$, $(1,ny,1)$ or
$(1,1,nz)$ sized grid).  The initial lapse is chosen the same as in
Eq.~(\ref{eqn:flattydef}), with $r$ replaced with $x$, $y$, or $z$
alone, with $\sigma^2 = 0.05$. In this harmonically sliced system with
a constant shift, the evolution equations become wavelike for the
lapse, with the propagation velocity being $1\pm \beta$.

\begin{figure}
  \incpsfig{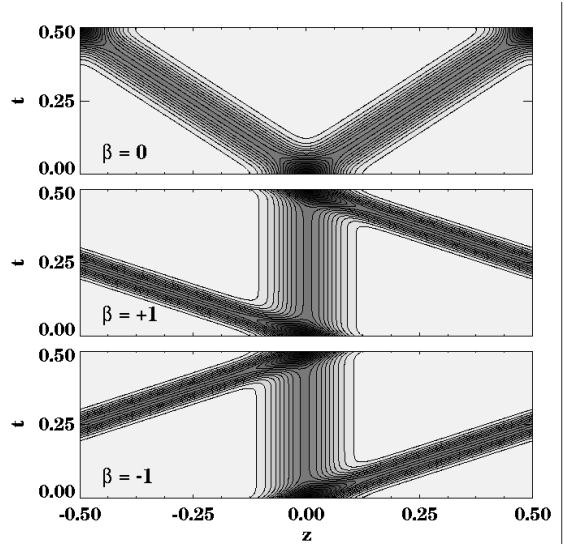}
\caption{We show propagation of the lapse in a dynamically
  (harmonically) sliced Minkowski space where the initial lapse is
  chosen to be a periodic gaussian function of $z$ alone. We see the
  expected affect of the shift. That is, shifts of $\pm 1$ force the
  propagation of the lapse to have zero coordinate speed in one
  direction.
}
\label{fig:onedflattyshift}
\end{figure}

In Fig.~\ref{fig:onedflattyshift} we see exactly this propagative
behavior. The lapse function $\alpha$ is shown for three cases, $\beta
= 0$ and $\beta \pm 1$.  For $\beta = 0$, the wave propagates with
speed $c=1$ in both directions.  For the shift chosen as $\pm 1$ we
see the speed of the waves to be two or zero, depending on the
direction of propagation and the sign of the shift. This can be
clearly read from the graph, where the propagation in the $t$
direction (vertically) is $0.5$ in all cases, and the propagation
distance in the $z$ direction is $0.5$ in the zero shift case, and
$1.0$ and $0.0$ in the $\pm 1$ shift case.  The other metric
functions, not shown, exhibit similar behavior.

\subsubsection{A Test of a Spatially Dependent Shift and an Important Lesson}
\label{sec:flattyeeconv}

We next turn to a spatially non constant shift as a test of our code,
\begin{equation}
  \beta^x = \beta^y = \beta^z = A e^{-(x^2+y^x+z^2)/{\sigma}^2}.
\label{eqn:shiftchoice}
\end{equation}
We here only consider the cases of $A < 1$, a sub-tachyonic shift.
The gaussian width $s$ is chosen so the shift is resolved but
effectively vanishes before the boundaries.  This choice of shift will
test all terms in our (non-conformal) evolution equations, since
it has derivatives of all shift terms in all directions.  The
following runs were performed with $\Delta x = \Delta y = \Delta z =
0.01$, $\sigma^2=0.02$, $A = 0.5$, and with $101$ grid zones in each direction.

Using this shift, we discovered an error in our code, which is worth
discussing. In an initial
version of our code, we had an error in the shift term for the sources
of the $V$ variables. Rather than the correct term,
\begin{equation}
  2({D_{ri}}^s - \delta_i^s {D^j}_{jr}){B^r}_s
\end{equation}
we had the different, although very similar,
\begin{equation}
  2 ({D_{ri}}^s - \delta_i^s {D^j}_{jr}) {B_s}^r.
\end{equation}
(Recall that $B$ is not symmetric.)  As we show now, by only
performing convergence tests we were able to diagnose and track down
the code error, without appealing to any analytic solutions beyond the
vanishing of the constraints.

\begin{figure}
  \incpsfig{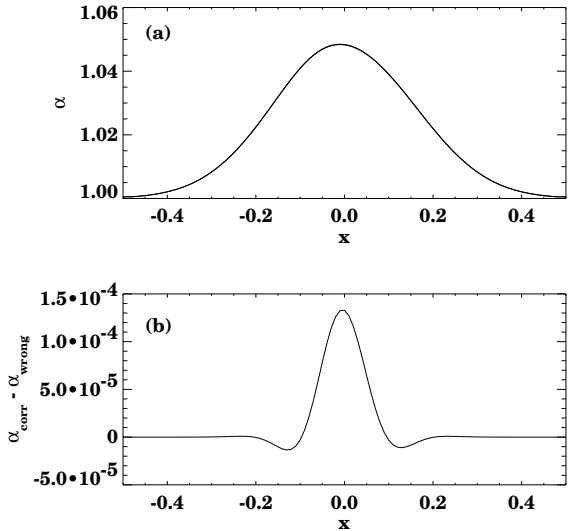} \caption{We show two evolutions of
    $\alpha$ in our distorted flat space model with a spatially
    dependent shift, using the Einstein equations in one case, and the
    equations with a small error in the second.  In (a) we show the
    numerical solution after eight iterations for the case with the
    correct shift terms with a solid line, and the results with an
    error in the shift with a dashed line.  At this level, the plots
    are indistinguishable.  In (b) we show the difference between the
    two evolutions, and notice the difference is negligible compared
    to the disturbance in the lapse.}
\label{fig:evdiffflat}
\end{figure}

In Fig.~\ref{fig:evdiffflat} we show the evolution after some time choosing the shift
in Eq.~(\ref{eqn:shiftchoice}), with and without the error above.
As is clear, the evolutions are very similar; in fact, had two
different codes given this result,
without further testing one would be tempted to say the results are the
``same'' and so the codes ``agree''.

\begin{figure}
  \incpsfig{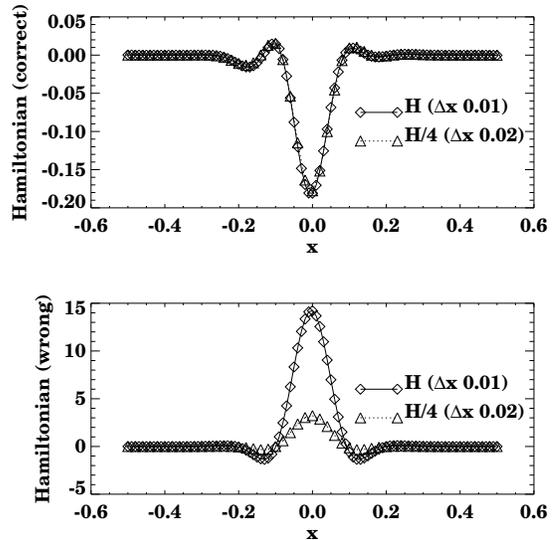}
\caption{We show the convergence of the hamiltonian constraint for the
  Einstein equations above and the non-convergence of the constraint
  for the Einstein equations with an error below.  We note that even
  though the error in our lapse evolution is very small, the
  convergence simply fails for the incorrect equation (note in the
  lower plot that the hamiltonian is the same for both resolutions,
  although the figure might mislead the reader because we introduce
  the factor of 4 that we would expect for convergent results).
  Again, this demonstrates second order convergence for the correct
  equations. We note that in both cases the hamiltonian constraint is
  ``large''; about 0.2 in the high resolution correct case (0.8 in
  the low resolution case) and about 15 in the incorrect case (being
  non convergent, stays the same for both grid resolutions). The only
  way to determine if the constraint is too ``large'' is to test its
  convergence towards zero, which is a feature of only the Einstein
  equations in this case.}
\label{fig:flevhamconv}
\end{figure}

However, in Fig.~\ref{fig:flevhamconv} we show that the hamiltonian
constraint, as defined by Eq.~(\ref{eqn:bmhamcon}), converges to zero for the
Einstein equations, and fails to do so for the system which is not.
The failure to converge is clear and large.  We note that even with fairly low
resolution we can demonstrate that our code is correct or
incorrect by showing merely the convergence of the constraints and we
did not need an exact solution for the spacetime (other than the
vanishing of the constraints). We feel that this
clearly demonstrates that convergence testing
constraints is an important and strong test of any code.

\section{Wave Spacetime Tests}
\label{sec:waves}

Although hyperbolic reformulations of the 3D Einstein equations have
not been used in a wide variety of spacetimes before this publication,
they have been applied to linearized wave
spacetimes\cite{Anninos94d,Abrahams97a}.  The current version of this
code owes much to the implementation of the ``H'' code described in
Ref.\cite{Anninos95d}.  As we reviewed in the introduction, this ``H''
code used a previous BM formulation of the equations that
required the exclusive use of harmonic slicing and zero shift
vector~\cite{Bona92}.  That code is now obsolete, although all the
tests of the ``H'' code described in Ref.~\cite{Anninos94d} can be
replicated successfully by this new and much more advanced version of
the code. All the tests presented here are run with the ``Ricci''
system ($n=0$), as this corresponds more closely to the simulations
performed with the ``H'' code.  Here we detail some of these
comparisons, evolving linear initial data that describe weak
gravitational waves.  The interesting transition from linear to
non-linear effects described in Ref.~\cite{Anninos96b} will not be
studied here, although it is possible to reproduce those effects with
the two formulations (BM and ADM) implemented in Cactus.

Further studies of stronger gravitational wave interactions and their
possible collapse to a black hole are underway and will be described in a
future publication in this series, where appropriate slicing conditions
for wave spacetimes will be considered in detail. In this section, we
will focus on two cases, colliding plane waves and quadrupolar
waves, and limit our gauge to harmonic slicing.

\subsection{Plane Waves}

We consider linearized plane wave solutions, following the test in
section III of Ref.~\cite{Anninos94d}. The line element is written
\begin{equation}
ds^2 = -dt^2 + (1+f(t,z))dx^2 + (1-f(t,z))dy^2 + dz^2.
\end{equation}
For small $f$, the linearized Hamiltonian constraint is satisfied, and
the evolution of the spacetime is governed by the linear wave equation
\begin{equation}
\partial_t^2 f(t,z) = \partial_z^2 f(t,z),
\end{equation}
that describes plane waves propagating in the $z$ direction. We use
the Gaussian-shaped packet:
\begin{eqnarray}
f(t,z) &=& 
                  A_{R}  e^{-(2\pi(t-z-a)/\sigma)^2}
                  \cos\left(\frac{2\pi}{\lambda}(z-t)\right)
                \nonumber \\
       & & + A_{L}  e^{-(2\pi(t+z-a)/\sigma)^2}
                    \cos\left(\frac{2\pi}{\lambda}(z+t)\right)
         \;,
\end{eqnarray}
The amplitudes $A_R$ and $A_L$ represent the amplitudes of waves
traveling to the right and left, respectively, with a Gaussian shape
of width $\sigma$ and centered at $z=\pm a$ at $t=0$. $\lambda$ is the
wavelength of the Gaussian-modulated oscillations.

\begin{figure}
  \incpsfig{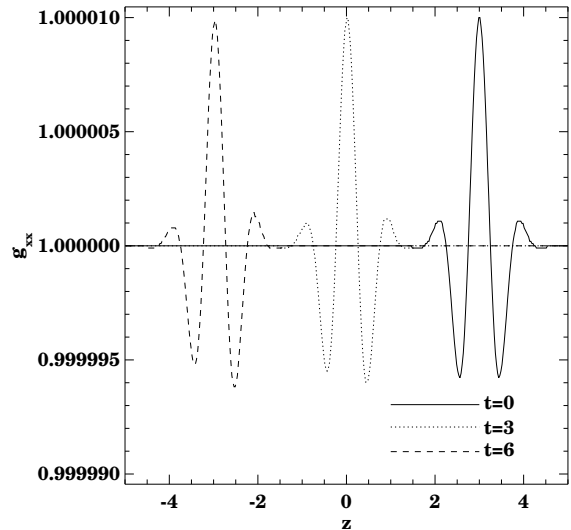}
\caption{We show the evolution of $g_{xx}$ for a single plane
  wave moving in the $-z$ direction at times $t=0$, $t=3$ and $t=6$
  for a gaussian wave packet. 
  This figure replicates Fig.~1(a)
  of Ref.~\protect\cite{Anninos94d}. The dispersive nature of the
  MacCormack method can be appreciated in the non-symmetric
  propagation of the gaussian packet.}
\label{fig:singlewave}
\end{figure}

In Fig.~\ref{fig:singlewave} we show the evolution of the metric
component $g_{xx}$ for a single wave moving in the $-z$ direction.  We
have chosen the shape parameters $\sigma=2$,$\lambda=1$,$A_L=0.00001$,
$A_R = 0$ and $a=3$, with $\Delta z = 0.025$.  This figure replicates
Fig.1(a) of Ref.~\cite{Anninos94d}, which used an ADM code with an
staggered-leapfrog algorithm.  We notice that the wave is transported
with a small loss of amplitude, due to dispersive effects in the
MacCormack predictor corrector scheme.  The measured convergence
$\sigma$ is very close to 2.  This is one of the simplest tests of a
numerical code designed to evolve waves and the results obtained are
in agreement with the well-known numerical properties of our standard
methods applied to the linear wave equation.

\begin{figure}
  \incpsfig{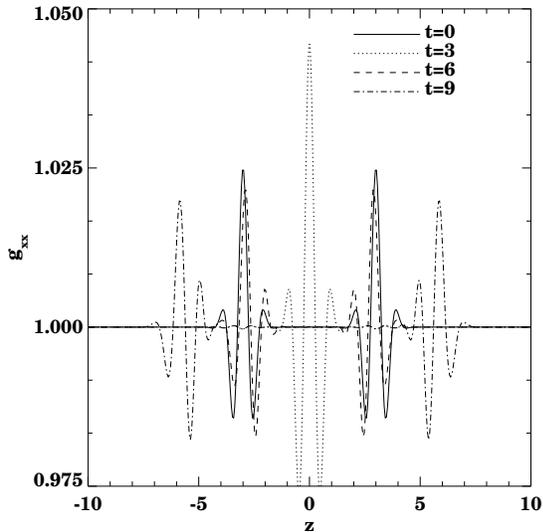}
\caption{We show the evolution of $g_{xx}$ for two colliding plane wave
  packets. At $t=0$ the two packets are centered at $z=\pm 3$, they
  collide and superimpose at $t=3$. At $t=6$ the left and right
  packets have interchanged their positions and should be coincident
  with their shapes at the initial time. The difference is due to
  numerical dispersion. Continuing the evolution, the packets are more
  dispersed at $t=9$. This figure replicates Fig.~6(d) of
  Ref.~\protect\cite{Anninos94d}. The MacCormack method used here does
  not exhibit the drifting after the collision exhibited by the
  staggered-leapfrog method in Fig.~6(a) of that reference.}
\label{fig:collidewave}
\end{figure}

A more involved test results from colliding plane waves.  Unlike the
previous test, in this case we deal with nontrivial spacetimes:
theoretically, it is known that such spacetimes will develop a
singularity in the future (in the non-linear
regime)~\cite{Yurtsever88a,Yurtsever88b}; numerically, coupled
nonlinear and finite differencing effects can lead to spurious
numerical evolution~\cite{Anninos94d}.  Hence, they provide a stronger
test of a numerical code.  In Fig.~\ref{fig:collidewave} we show the
evolution of a colliding wave system.  Two wave packets originally
start, moving inwards, centered at $z=\pm 3$. We choose the same
parameters as the single wave packet except for the amplitudes
$A_R=A_L=0.025$.  The packets collide at the center at time $t=3$ and
then continue on.  Once again, dispersion is visible when the waves
return to their original images at $t=6$.  This figure replicates Fig.
6(d) of Ref.\cite{Anninos94d}. There it was shown that the
staggered-leapfrog method was prone to a large secular drifting after the
packets collided, which does not occur with our MacCormack method.

\subsection{Pure Quadrupolar waves}

The numerical simulation of quadrupolar linearized wave solutions to
the Einstein equations has been established as an standard test of 3D
numerical codes\cite{Abrahams97a,Anninos94d,Shibata95}. One of the
reasons of their appeal is the existence of a family of analytic
solutions for both even- and odd-parity and the independent azimuthal
modes\cite{Teukolsky82}. But more importantly, we also need to model
their evolution accurately, as quadrupolar modes are a dominant signal
in the late time evolution of black hole spacetimes. In this section
we compare evolutions of quadrupolar waves in Cactus with previous
results, following again the extensive tests and discussions of
Ref.\cite{Anninos94d}. Due to the length of the analytical
expressions, we do not write the solutions here and refer to the
reader to Ref.\cite{Teukolsky82} or Ref.\cite{Evans86}. 

\begin{figure}
  \incpsfig{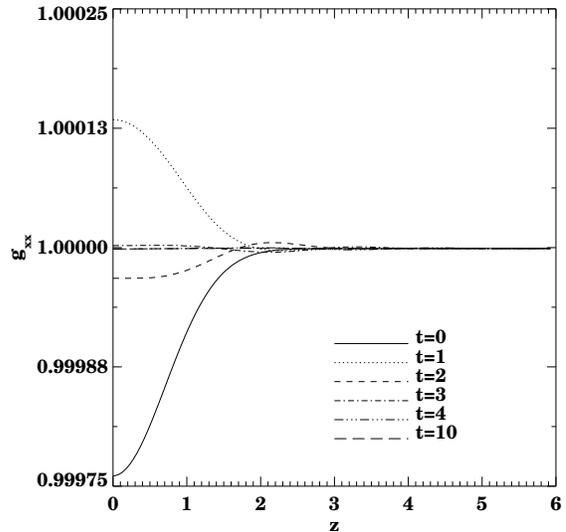}
\caption{The evolution of metric function $g_{xx}$ 
  along the z line is shown for linear quadrupolar waves with $l=2$,
  $m=0$ and a low amplitude packet, which corresponds to a
  perturbation of 0.025\% in the metric functions.  The wave expands
  outward as time increases, returning to a flat profile after $t=4$.
  This replicates Fig.9(c) of Ref.~\protect\cite{Anninos94d}.}
\label{fig:tw0low}
\end{figure}

We start by evolving even-parity waves with an amplitude of $10^{-5}$ 
and quadrupole numbers $l=2$ and $m=0$.  The details of this setup are 
given in section VI of Ref.\cite{Anninos94d}.  In 
Fig.~\ref{fig:tw0low} we show the evolution of $g_{xx}$ along the 
$z$--axis performed on a grid of $120^3$ points with $\Delta x = 
\Delta y = \Delta z = 0.05$.  This replicates Fig.~9(c) of 
Ref.\cite{Anninos94d}.  We can see how an initially moderate wave 
packet near the center of the grid oscillates and propagates off the 
grid, as expected.  

\begin{figure}
  \incpsfig{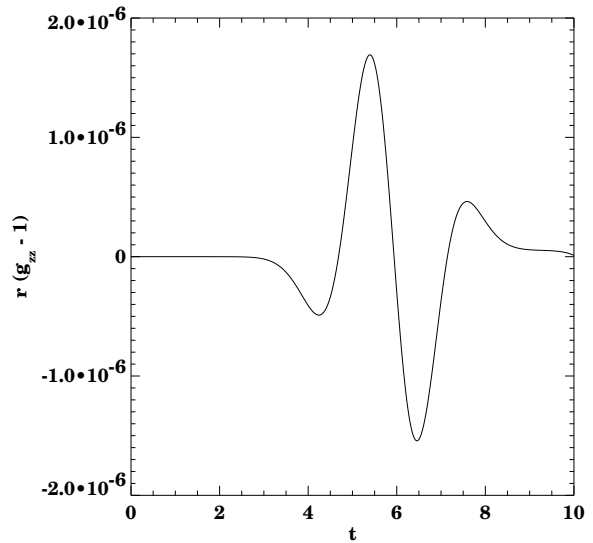}
\caption{We show the time evolution of the wave-like quantity 
 $r (g_{xx}-1)$ measured at the outer
  boundary for the simulation shown in the previous figure. The wave
  pulse arrives to the boundary at around $t=4$, oscillates and leaves
  the computational grid. This serves as indicator of the outgoing
  condition provided by our simple copying boundaries}
\label{fig:tw0low_extract}
\end{figure}

In Fig.~\ref{fig:tw0low_extract} we show the time evolution of the
quantity $r (g_{xx}-1)$ at the outer boundary of our grid, as an 
indicator of the clean outgoing condition provided by our simple 
copying boundaries.  This measure of the wave simply separates the 
perturbation from the background Minkowski metric, and corrects for 
the $1/r$ falloff.  It is not a gauge-invariant measure of waves, such 
as that used in Ref.~\cite{Allen98a}.  Detailed studies extending these 
results beyond linear wave regimes are under way and will be published 
elsewhere.

\begin{figure}
  \incpsfig{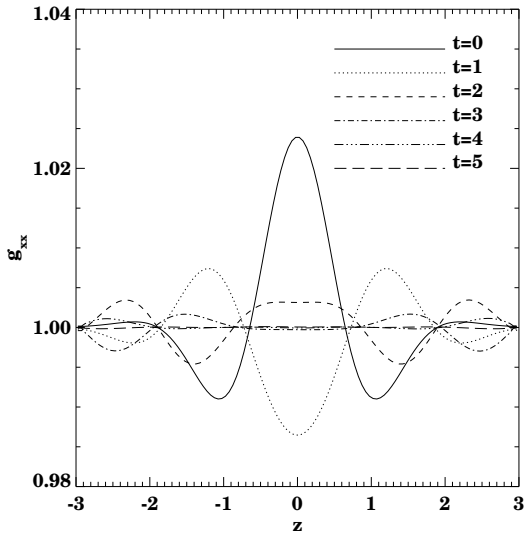}
\caption{We show the evolution of $g_{xx}$ for the $l=2$, $m=2$
  stronger amplitude quadrupolar wave
  packet along the $z$--axis. The perturbation on the metric components
  is around 2.5\% for this higher amplitude. Although this packet is fully 3D
  and can not be evolved using an octant of the spacetime, the metric
  component $g_{xx}$ is symmetric around the origin.}
\label{fig:tw2}
\end{figure}

\begin{figure}
  \incpsfig{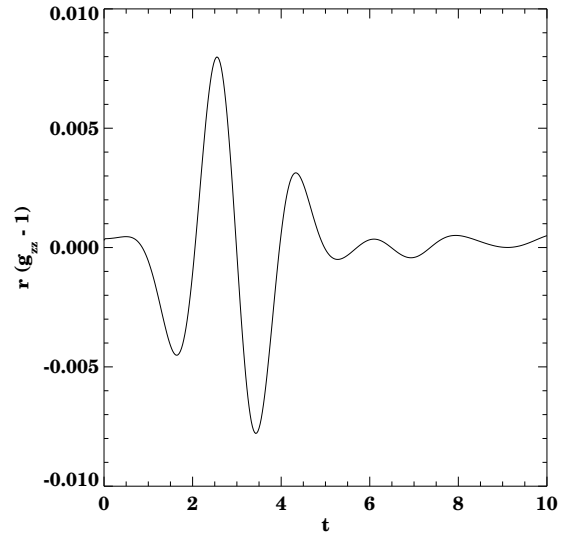}
\caption{We again show the time evolution of $r(g_{xx} - 1)$ at the
  outer boundary for the simulation shown
  in the previous figure. Again, although our simple copying 
boundary condition, coupled with the MacCormack method, does a 
reasonably good job of allowing the wave to propagate through the 
boundary, at late times boundary effects are evident. Note that 
the outer boundary, at $z=\pm3$, is now closer to the origin.}
\label{fig:tw2_extract}
\end{figure}

Ironically, the extensive work of Ref.\cite{Anninos94d} does not
include results with any truly 3D spacetime, as the cases studied for
quadrupolar waves correspond to axisymmetric waves of azimuthal number
$m=0$. In this paper we will extend the results of that reference by
setting up a slightly more realistic scenario, tuning the parameters to
mimic what we expect from late time ringdown of black hole
simulations. Therefore, we will evolve non-axisymmetric quadrupolar
waves with $l=2$, $m=2$ and a stronger amplitude wave, with $A=0.001$,
corresponding to a perturbation of 3\% in the metric components. In
this full 3D case, we do not use an octant of the spacetime, but rather 
set up a full grid with the origin in the center.  Again, the grid has 
$120^3$ points with $\Delta x=\Delta y=\Delta z = 0.05$.  As we have a 
full grid, the outer boundary is now closer.  In Fig.~\ref{fig:tw2} we 
see the evolution of the now stronger initial packet propagate 
outwards, as expected.  In Fig.~\ref{fig:tw2_extract}, we again show 
the ``waveform'' measured directly by the function $r(g_{xx} - 1)$ at 
the outer boundary, which is allowing the wave to cleanly propagate 
off the grid.  At late times boundary effects become visible.  See 
Ref.\cite{Abrahams97a} for an excellent discussion of the problem of 
outgoing conditions in this scenario and possible solutions with 
perturbative techniques.

\begin{figure}
  \incpsfig{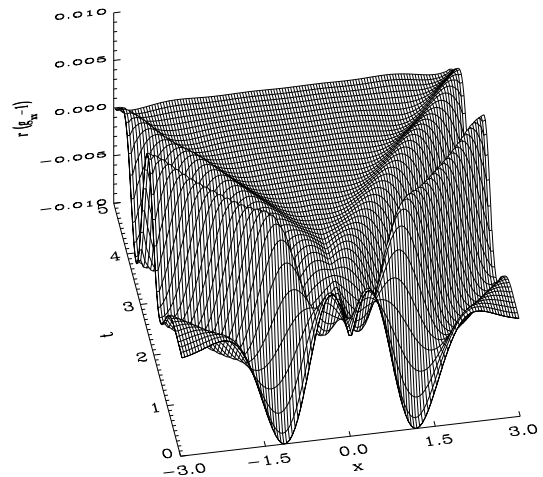}
\caption{We show the time evolution of the ``extraction'' function
  $r(g_{xx}-1)$ along the $x$ line. The surface plot has time along
  the $y$--axis. The $r$ factor corrects for the $1/r$ fall-off, so we
  can see that the wave propagates from the center and off the
  boundaries. The previous figure corresponds to the $y$--axis (i.e.,
  time) boundary of this plot.}
\label{fig:tw2_surfrad}
\end{figure}

To better visualize the temporal evolution of this wave, in
Fig.~\ref{fig:tw2_surfrad} we show the value of $r (g_{xx}-1)$ along the
$z-$axis evolving in time as a surface. We can see that the wave
propagates cleanly away from the center and off the boundaries, as
expected.

\begin{figure}
  \incpsfig{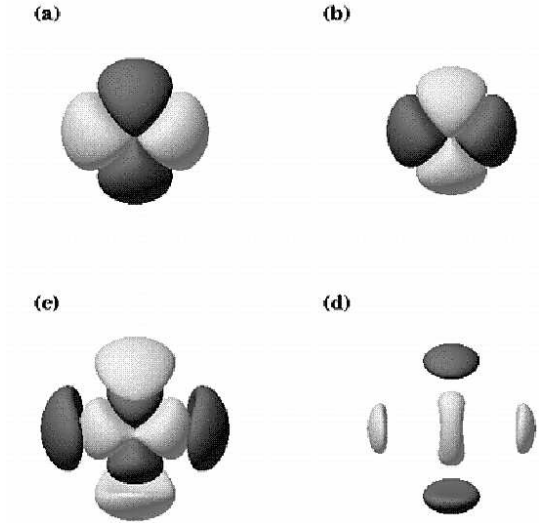}
\caption{We show four time snapshots of the evolution of the packet
  presented in the last figures. Two isosurface values of the
  spherical metric component $g_{rr}$, reconstructed from the evolved
  cartesian components, are shown at times $t=0$ (a), $t=1$ (b), $t=2$
  (c) and $t=3$ (d). The dark and light colored isosurfaces correspond
  to the values $0.9997$ and $1.0003$ respectively. They oscillate
  around the center and propagate outwards. }
\label{fig:tw_full3d}
\end{figure}

The best way to visualize the full 3D nature of these waves
and their propagation would be to show a movie, which obviously we can
not do in printed form.  In Fig.~\ref{fig:tw_full3d} we show four
snapshots of such a movie, showing two isosurface values of the metric
component $g_{rr}$, constructed from the cartesian metric. 

\begin{figure}
  \incpsfig{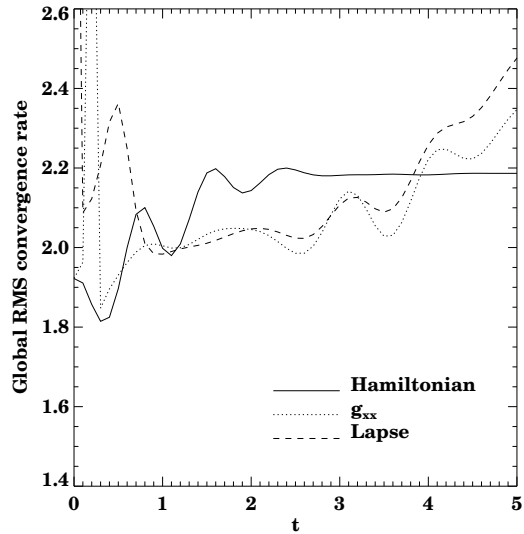}
\caption{We show the time evolution of the global 
  convergence rate (computed in the ${\cal L}_2$ norm) of the
  hamiltonian constraint, $g_{xx}$ and the lapse function. All
  quantities converge at second order. In particular, 
  the hamiltonian constraint converges also at second order, although
  it does not converge to zero, since it is only satisfied to linear
  order.}
\label{fig:tw_conv}
\end{figure}

All the wave tests presented in this section converge as expected. In
Fig.~\ref{fig:tw_conv} we show the time evolution of the convergence
rate $\sigma$ obtained using the ${\cal L}_2$ (i.e., RMS) norm over
the entire grid. We measure convergence for the hamiltonian
constraint, the metric component $g_{xx}$, and the lapse, and note
that all converge at or above second order, as expected. We do not need to
resort to the linear solution to measure convergence. In this special
case, we do not measure the convergence of the constraints to zero, as
the initial data is linear and only satisfies the constraints to
linear order. Thus, we measure the convergence of the hamiltonian as
we would any other quantity, using
three different resolutions to create the convergence exponent.

\section{Black Hole Tests}
\label{sec:bhole}

Black hole spacetimes are currently one of the major motivations for
developing 3D numerical relativity.  The waveforms emitted by
inspiraling colliding black holes are expected to be one of the most
likely candidates for early detection by laser
interferometers\cite{Flanagan97a,Flanagan97b}, and hence are urgently
in need of general 3D simulations.  Thus, black hole spacetimes are
important tests of our code, and we will follow the work of
Ref.~\cite{Anninos94c} in these code tests of Schwarzschild black
holes.  More dynamic black hole studies, including the simulations of
3D excitation and ringdown of the quasinormal modes of distorted black
holes~\cite{Allen97a,Allen98a,Camarda97b,Camarda97c}, and of black
hole collisions~\cite{Anninos96c}, are in progress and will reported
and compared against published results in a future paper in this
series.

Black hole spacetimes are in many ways similar to other spacetimes.  
An initial metric evolves with some slicing conditions, and the 
constraints should converge as in any spacetime. However, 
special difficulties are encountered due to the presence of 
singularities.  Thus, as discussed in the introduction, present 
Cauchy evolutions of general 3D black hole spacetimes do not allow a 3D code 
to run forever, as they can when propagating disturbances in flat 
space or low amplitude waves.  At some point, a time slice may hit a 
singularity and crash, or stretch the grid so much that the simulation 
will no longer be able to continue.  At this point, we will see ``blow 
ups'' on our grid, convergence will fail (starting, usually, at the 
lower resolution grids), and we will have to stop our code.  Thus 
evolving black holes for many tens of $M$, where $M$ is the ADM mass, 
with a demonstration of convergence is still ``state of the art'' in 
numerical relativity.

In this section, we test Cactus using a single black hole with the
Einstein-Rosen bridge topology with an isotropic radial coordinate
$r$.  That is, the spatial line element takes the form
\begin{equation}
  ds^2 = \Psi^4(dr^2 + r^2 d\Omega^2)
\end{equation}
with
\begin{equation}
\Psi = 1 + \frac {M}{2r}.
\label{eq:bhconformal}
\end{equation}
We satisfy the constraint equations with this metric and initial
$K_{ij} = 0$.  For more detail, see Ref.~\cite{Anninos94c}. For all
the work which follows, we choose $M=1$.

This data is isometric in inversion through the sphere, or throat,
located at $r = \frac{M}{2}$. The singularity
at $r = 0$ is also related to the remapping of a second universe
on the other side of the bridge to the origin in our flat space.
However, rather than evolve the Einstein-Rosen bridge black hole
spacetime with the natural $S^2 \times R$ topology (as used in
axisymmetric simulations such as \cite{Bernstein93b,Brandt94b}), we
evolve it on an $R^3$ manifold which contains a point where the
conformal factor is infinite. This was one of the techniques used in
Ref.~\cite{Anninos94c}, and has recently been generalized to generate
full 3D, binary black hole data with spin and momenta\cite{Brandt97b}.

As in Ref.\cite{Anninos94c,Bruegmann96}, we handle the infinity in the 
conformal factor numerically using two tricks.  First, we do not place 
a grid point at $r = 0$, but rather we stagger the origin, with 
grid points at $\Delta x / 2$ and $-\Delta x / 2$.  Secondly, we 
exploit knowledge of the conformal factor and its derivatives in our 
finite differencing.  This allows us to factor out the infinity from 
the evolved quantities as known derivatives in the source terms, and 
evolve fields which are unity everywhere.  This approach to computing 
``conformal derivatives'' is quite general, and can be used with a 
numerically generated initial data set as well.  Note that this 
conformal rescaling of the equations, as discussed in 
Sec.~\ref{sec:coords} and Appendix A. is different from the conformal 
rescaling done in typical ADM codes, including the Cactus ADM thorn, 
where the Ricci tensor is formed directly with conformal derivatives 
of the system.  For our first order system, we do not form the Ricci 
tensor, and therefore we must treat the conformal rescaling differently 
in order to preserve a first order system, and still allow only 
conformal variables to appear in the fluxes.

Here we consider various slicings of a single black hole spacetime.  
We do not discuss or demonstrate multiple black hole or distorted 
black hole spacetimes here, since we wish only to show code tests at 
this time.  However, preliminary tests show that the results presented 
here carry over into more dynamical black hole spacetimes.  This is a 
major and active research area in 3D numerical relativity in which we 
are presently engaged.  In the final part of this section, we also 
perform tests of the Schwarzschild black hole system with the ADM 
equations in Cactus, and compare with the results from the 
BM formulation.  All simulations in this section are done 
with $\Delta x = \Delta y = \Delta z$, $nx = ny = nz$, and with the 
conformal rescaling of the BM system, or conformal 
differencing in the ADM system.  For the BM system, all 
simulations were performed with the Einstein system.

\subsection{Geodesic Slicing}

A black hole spacetime evolved with geodesic slicing ($\alpha = 1$,
$\beta^i = 0$) can only be evolved until points initially on the
throat hit the singularity unless points are excised from the grid, as
shown in Refs.~\cite{Anninos94c,Bruegmann96}.  At that point any code
evolving this system will crash.  We know that
observers initially at rest in the Schwarzschild spacetime that this
crash must come at $t=\pi M$.  The crash will appear as an infinity or
undefined value at a point on the numerical grid.

Despite this critical failing, the geodesically sliced Schwarzschild
spacetime is useful as an analytic solution for the three-metric
exists, the Novikov solution~\cite{Misner73}. This solution expresses
the metric in terms of cyclic infall times for initially non-moving
observers. Expressions for these solutions in isotropic radius are
given in~\cite{Bruegmann96}, although the final term is
missing a square root, thus for completeness we give expressions
here~\cite{BruegmannPrivateComm}. We use a slightly different notation
than Ref.~\cite{Bruegmann96}: $r$ is our isotropic radius, $r_{\rm{a}}$ is
the areal radius, $r_{\rm{max}}$ is the maximum (areal) radius for an
observer during the cyclic infall (and is therefore the initial areal
radius, so $r_{\rm{a}} = r_{\rm{max}}$ at $\tau = 0$), $\tau$ is the proper
time of an observer (and therefore the coordinate grid time, as
$\alpha = 1$), and $g_{rr}$ and $g_{\theta \theta}$ are the
conformal isotropic metric components. The relevant expressions for an
$M=1$ black hole are
\begin{mathletters}
\label{eqn:novikov}
\begin{eqnarray}
  r_{\rm{max}}(r) &=& \frac{(1 + 2 r)^2}{4 r}, \\
  \tau(r_{\rm{a}},r_{\rm{max}}) &=& r_{\max} \left(\frac{r_{\rm{a}}}{2} \left(1 -
      \frac{r_{\rm{a}}}{r_{\rm{max}}}\right)\right)^{\frac{1}{2}} \nonumber\\
   &&+ 2
  \left(\frac{r_{\rm{max}}}{2}\right)^{\frac{3}{2}}\rm{arccos}
  \left(\left(\frac{r_{\rm{a}}}{r_{\rm{max}}}\right)^{\frac{1}{2}}\right),
  \label{eqn:tauofr}  \\
  \frac{\partial r_{\rm{a}}}{\partial r_{\rm{max}}} &=& \frac{3}{2} -
  \frac{r_{\rm{a}}}{2 r_{\rm{max}}}\nonumber \\
  &&+
  \frac{3}{2}\left(\frac{r_{\rm{max}}}{r_{\rm{a}}}-1\right)^\frac{1}{2}
  \rm{arccos}\left(\left(\frac{r_{\rm{a}}}{r_{\rm{max}}}\right)^{\frac{1}{2}}\right),\\
  g_{rr} &=& \left(\frac{\partial r_{\rm{a}}}{\partial
r_{\rm{max}}}\right)^2,\\
  \text{and}\nonumber\\
  g_{\theta \theta} &=& \left(\frac{r_{\rm{a}}}{r_{\rm{max}}}\right)^2.
\end{eqnarray}
\label{eqn:geodesic}
\end{mathletters}
To construct the metric, we must numerically invert
relation Eq.~(\ref{eqn:tauofr}) to find $r(\tau, r_{\rm{max}})$.
Simple bisection solves this problem. Aside from this minor
complication, constructing the solution is straightforward.

\begin{figure}
  \incpsfig{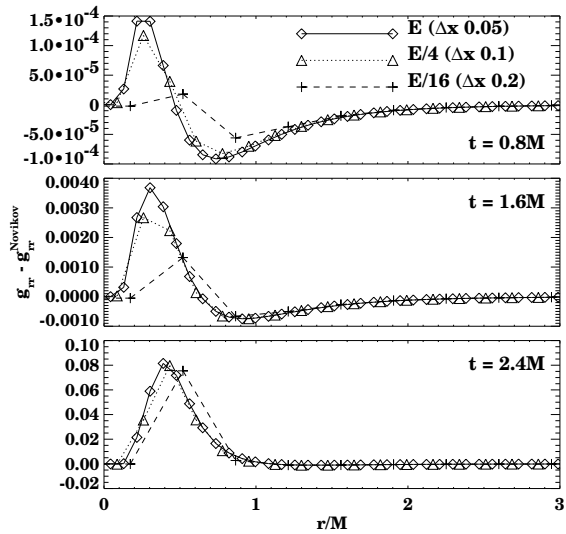}
\caption{We show the difference of the radial metric between the
  analytic Novikov solution and the full three-dimensional numerical
  evolution. Data is extracted along a diagonal line. We define $E$ as
  the difference between the analytic solution and the numerical
  solution. We show $E/16$ for $\Delta x = 0.2$, $E/4$ for $\Delta x =
  0.1$ and $E$ for $\Delta x = 0.05$.  We note that the data points
  are practically identical, showing second order
  convergence.}
\label{fig:novcomp}
\end{figure}

We present here two demonstrations that our code is in fact creating 
the correct solution for the geodesically sliced black hole spacetime.  
In Fig.~\ref{fig:novcomp} we show the difference between the $g_{rr}$ 
produced by the code (which is constructed from the full 
evolved cartesian three metric) and the analytic expression in 
Eq.~(\ref{eqn:geodesic}).  We extract the data along a diagonal line.  
We show the difference at three different resolutions, adjusting the 
lower resolution differences by factors of 1/4 and 1/16, respectively.  We 
note that the points (shown as crosses, diamonds, and triangles) are, 
for all practical purposes, identical in this figure, strongly 
indicating second order convergence at every point on the grid.

\begin{figure}
  \incpsfig{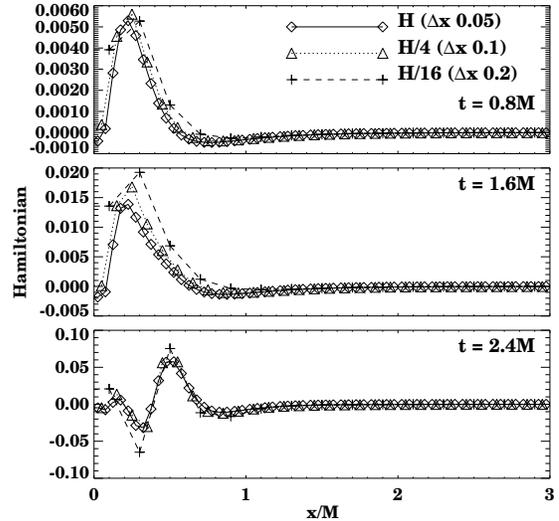}
\caption{
  We show the hamiltonian constraint, $H$, for the geodesically sliced
  black hole at three different resolutions.  We show $H/16$ for $\Delta
  x = 0.2$, $H/4$ for $\Delta x = 0.1$ and $H$ for $\Delta x = 0.05$.
  We note that the lines are identical, indicating second order
  convergence. }
\label{fig:geohamcon}
\end{figure}

In Fig.~\ref{fig:geohamcon}, we show similar plots for the hamiltonian 
constraint.  We show the lower resolution constraints divided by 4 and 
16 respectively.  Once again, the fact that these lines are visually 
coincident (although not completely identical) strongly demonstrates 
that our code is converging at second order.  Note that the error is 
largest near the throat, located at $r = 0.5M$, which is well 
inside the horizon at late times as it rushes towards the singularity.

From these two diagrams we can calculate four values of the 
convergence exponent everywhere, since we have two quantities to 
measure against exact solutions, at three resolutions.  Doing this 
analysis gives a convergence exponent between about 1.8 and 2.1 
(oscillatory in time), once again demonstrating second order 
convergence to an exact solution.

\begin{figure}
  \incpsfig{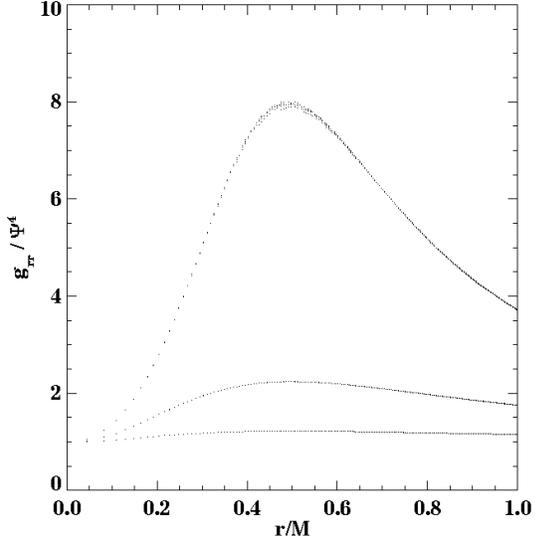}
\caption{
We use a scatter plot to show $g_{rr}$ vs. $r$ for all points
  in a region of the cartesian grid for a geodesically sliced black
  hole. This scatter plot allows one to see how well spherical
  symmetry is maintained by eye.  From this plot, it is clear that
  deviations from sphericity occur near the peak at late times, and
  are fairly small.  This figure was generated with $\Delta x =
  0.05M$, and the slices are shown at $t=0.9M$, $1.8M$, and $2.7M$.}
\label{fig:grrscat}
\end{figure}

The black hole spacetime also provides a strong test of a code's
ability to preserve the appropriate spherical and rotational
symmetries inherent in the initial data set.  Especially near the
singularity, there is a rapid growth of strong gradients surrounding a
black hole, which must be computed in the separate cartesian metric
functions on a cartesian grid.  These individual functions do not
exhibit the underlying symmetries of the black hole, so it can be
difficult to model spherical or axisymmetric phenomena without
introducing spurious effects due to resolution and coordinate
geometry.  As noted, the MacCormack method 
exactly obeys rotational symmetries, with $g_{xx}$ along an $x$-line
through the origin and $g_{yy}$ along a $y$-line through the origin
being the same to machine precision, but has no such property for
spherical symmetry.  Therefore we must test whether spherical symmetry
is preserved. We can make this test visually, by displaying all the
points on our grid versus their radial position for some spherically
symmetric quantity, such as $g_{rr}$ in a ``scatter'' plot. In
Fig.~\ref{fig:grrscat} we do exactly this for the high resolution
geodesically sliced black hole above. The ``width'' of the scatter
plot at late times indicates that deviations from sphericity are
becoming larger as the solution evolves towards the singularity.

\begin{figure}
\incpsfig{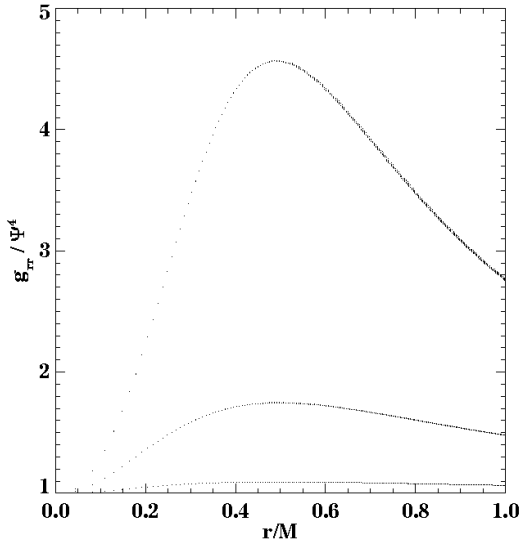}
\caption{We use a scatter plot to show $g_{rr}$ vs $r$ for all points
  in a region of our cartesian grid for a geodesically sliced black hole. This
  simulation uses the directionally split Lax-Wendroff solver. This
  figure was generated with $\Delta x = 0.05M$ and slices are shown at
  $t=0.6M$, $1.5M$, and $2.4M$. We note that even though the method is
  explicitly split in cartesian directions it maintains excellent
  spherical symmetry.}
\label{fig:grrlwscat}
\end{figure}

We repeat these tests using our Lax-Wendroff directionally split
update method. In Fig.~\ref{fig:grrlwscat} we show a scatter plot of
the conformal radial metric function in the neighborhood of its peak
at $r=M/2$. We notice that spherical symmetry is obeyed very well
despite the fact that this method is a manifestly cartesian method.

\begin{figure}
\incpsfig{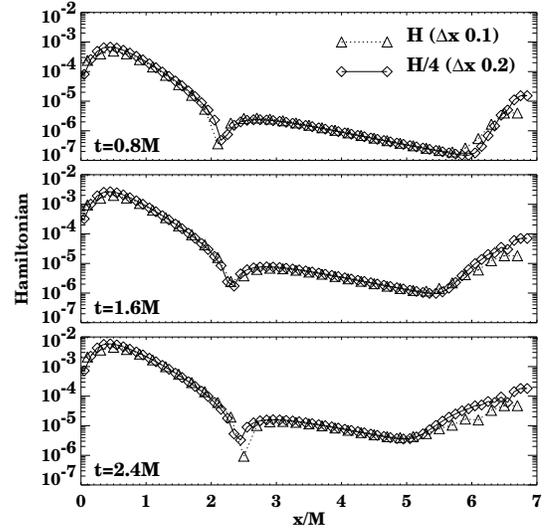}
\caption{We show the convergence of the hamiltonian constraint for a
  geodesically sliced black hole with the flux evolution solved with
  the directionally split Lax-Wendroff method. We note that
  convergence is excellent, and points away from the boundary are
  visually coincident, demonstrating second order convergence of the
  constraints towards zero away from the boundaries.}
\label{fig:lwhamcon}
\end{figure}

In Fig.~\ref{fig:lwhamcon} we show the solution is indeed converging
at second order. We measure the convergence order of the hamiltonian
constraint and find it is converging at or above $\sigma = 2$ during
the entire evolution.

\subsection{Algebraic Slicings}

Algebraic slicing conditions have been used for three dimensional 
black hole evolutions in the past with a relatively high degree of 
success, as shown in Refs.~\cite{Anninos94c,Camarda97a,Camarda97b}.  
Such slicings typically use Eq.~(\ref {eqn:lapseevol}) to provide a 
condition on the lapse.  Here we examine the use of such slicings in 
3D black hole spacetimes in the BM formulation.  We note 
that such slicings also have been shown under certain conditions to 
develop coordinate pathologies~\cite{Alcubierre97b}, but we will not 
investigate those issues here.  The main purpose of this section is to 
compare results of Cactus with previously published results on 
Schwarzschild black hole evolutions.  

The simulations in Refs.~\cite{Anninos94c,Camarda97a,Camarda97b} used 
both a diffusion term added to the lapse evolution equation to achieve
stability, and an enforced isometry condition, mapping the highly 
resolved region exterior to the throat into the poorly resolved region 
interior to the throat.  As detailed in Ref.~\cite{Anninos94c}, 
explicit enforcement of this isometry was very important in obtaining 
accurate long time evolutions of the system, as it allows one to avoid 
numerical evolution in the coarsely resolved region near the 
singularity: one simply maps the accurately computed exterior into 
this region before proceeding to the next time step.  Although the 
algebraic slicing conditions studied actually do obey the isometry 
operation, and will attempt to preserve it numerically during an 
evolution, without an explicit isometry operator in the code, large 
errors will develop inside the throat, causing a code to crash.

An isometry condition could be applied within Cactus, but with the 
BM system this leads to an additional difficulty in that the 
isometry conditions on the $D_{ijk}$, $A_{i}$ and $V_{i}$ variables is 
non-trivial, since these are not tensor quantities.  Due to this 
complication, and due to the promise of alternative techniques such as 
apparent horizon boundary conditions which do not require isometry 
conditions, we have currently chosen not to implement an isometry 
condition in Cactus.  Under these conditions it is difficult to 
achieve the same accuracy and long runtime that were available to an 
isometry based code, when algebraic slicings are used.  We stress that 
this not a limitation of the code or the formulation of the equations, 
but merely a sensitivity of such slicings in black hole simulations without 
an explicit isometry operator.  Similar results are obtained, for 
example, with the ``G'' code used to generate results in 
Refs.~\cite{Anninos94c,Camarda97a,Camarda97b}.  Furthermore, we will 
see in the next section that maximal slicing, which as shown in 
Ref.~\cite{Anninos94c} does not require the isometry operator, works 
very well in Cactus.

\begin{figure}
\incpsfig{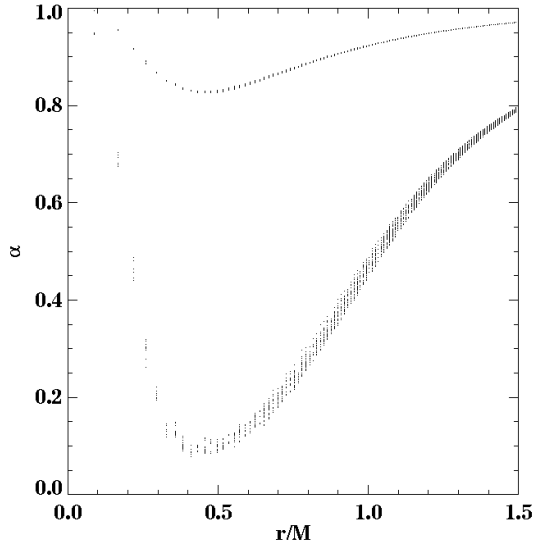}
\caption{We use a scatter plot to show the collapse of the lapse at
  $t=2.1M$ and $t=3.5M$ for a ``1+log'' sliced black hole. We note
  that with this dynamical local slicing, spherical symmetry is not
  preserved to such a high degree as in the geodesically sliced
  case, especially near the point of large gradient in the lapse. This
  inaccuracy in the dipping of the lapse will cause the code to crash
  shortly after this plot. This data was produced with $\Delta x = 0.1M$.}
\label{fig:algalpscatter}
\end{figure}

With those remarks in mind, in Fig.~\ref{fig:algalpscatter} we show 
the lapse profile in a scatter plot at $t=2.1M$ and $3.5M$ for a ``$1+\log$'' 
($f=1/\alpha$ in Eq.~(\ref{eqn:lapseevol})) sliced black hole.  
Clearly the failure of the lapse to collapse in the center, combined 
with poor resolution of the consequent gradient in the lapse, will not 
allow an accurate evolution in this region.  Convergence testing this 
solution at three resolutions, $\Delta x=0.1M$, $\Delta x = 0.2M$, and 
$\Delta x = 0.4M$, in this case on a full grid rather than an octant 
grid, we see that the simulation on the medium resolution grid crashes 
first, at $t=4.2M$.  We note that the lapse is collapsing most quickly 
at $r=M/2$, indicating that our system is trying to preserve the 
underlying isometry present in the initial data, as it should.

\begin{figure}
  \incpsfig{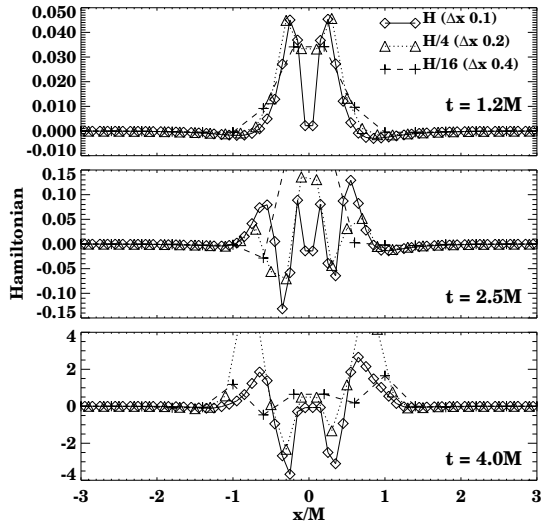}
\caption{We show the convergence of the hamiltonian constraint for the
  ``$1+\log$'' sliced spacetime. We show 1, 4, and 16 times the
  hamiltonian constraint at $\Delta x = 0.4M$, $0.2M$, and $0.1M$
  respectively. We note that the low resolution grid does {\em not}
  converge, and the entire system fails to converge at late
  times. The medium resolution grid simulation will crash at
  $t=4.2M$.}
\label{fig:alghamconv}
\end{figure}

We can understand the nature of the algebraically sliced spacetime
without isometry or diffusion by studying its convergence
properties. In Fig.~\ref{fig:alghamconv} we show the convergence of the
hamiltonian constraint towards zero at three different
resolutions. Several important points in this figure should be
noted. Firstly, the very low resolution simulation ($\Delta x = 0.4M$,
almost the radius of the throat) does not converge from the second
slice, $t=2.5M$. As
we shall see below, this is because this very low resolution
simulation ``misses'' the isometry, and the lapse collapses, leading
to the longest time evolution of the three simulations. We note,
however, that at the first two displayed times, the medium and high
resolution simulations are converging appropriately, as indicated by
the almost coincident lines in Fig.~\ref{fig:alghamconv}. As the medium
resolution grid nears its crash time at $t=4.2M$, however, there is no strong
evidence of second order convergence in the system near the hole.

\begin{figure}
  \incpsfig{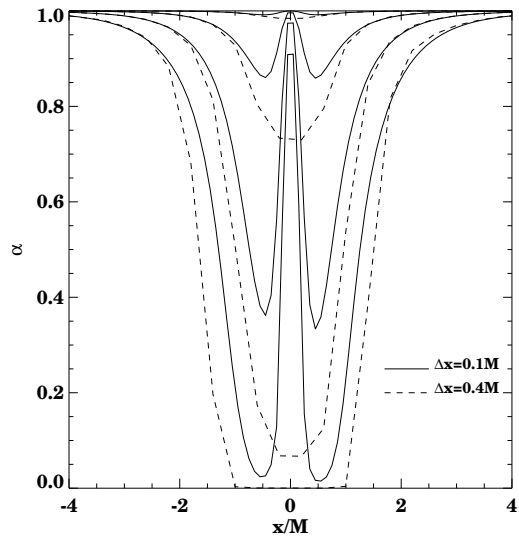}
\caption{We note that on a low enough resolution grid, the isometry
  condition inherent in the Schwarzschild spacetime, which maintains
  $\alpha = 1$ at the center of the grid, is ``missed'' so the lapse will
  collapse everywhere in the center of the grid. Ironically, this
  under-resolution will allow the
  low resolution simulation to run considerably longer than the high
  resolution simulation, but the solution will not converge, of
  course. Lines are shown with $\Delta x = 0.1M$ (solid line) and
  $0.4M$ (dashed line) at $t=0,1,2,3,$ and $4M$.}
\label{fig:lowhighcoll}
\end{figure}

Returning to the mystery of the lowest resolution grid, we show in 
Fig.~\ref{fig:lowhighcoll} the evolution of the lapse on the low 
($\Delta x = 0.4M$) and high ($\Delta x = 0.1M$) resolution grids.  On 
the lowest resolution grid the system simply has too few points to 
obey the isometry, and the lapse collapses uniformly around $r=0$, 
allowing a long time evolution.  The highest resolution grid clearly 
attempts to obey the isometry, but is destined to fail, due to the 
small number of points covering the region in $r < M/2$.  Thus the two 
evolutions do not converge towards the same slicing of the spacetime.  
This is clearly a dangerous feature running a simulation with very 
poor resolution: it can produce a solution which misses features, but 
still creates a reasonable looking (and, in this case, longer running) 
solution than a higher resolution run.

To summarize this subsection, algebraic slicings are convenient and 
inexpensive singularity avoiding slicing conditions.  We have shown 
that in Cactus they behave as expected, and converge at second order 
as they should.  But they must be used with care.  As already shown in 
Ref.~\cite{Anninos94c}, in a spacetime containing a singularity, they 
can still be very useful if an isometry operator is used to avoid 
evolving the region near the singularity.  Without it, evolution in 
this region inside the black hole throat is almost impossible in 3D.
These slicings
should still find use in a number of other circumstances, including 
use on black holes if precautions are taken near a singularity, if it 
exists on the grid. 

\subsection{Maximal Slicing}
\label{sec:bhmaximal}

Maximal slicing has long been a favorite slicing condition for
numerical relativity. Alas, the maximal slicing condition,
Eq.(\ref{eqn:maximal_def}), is an elliptic condition for the lapse,
which brings with it both a breaking of the hyperbolic system for the
lapse and its derivatives and a very large degree of computational
complexity. 

Solving three-dimensional elliptic equations is far more
difficult than solving three-dimensional hyperbolic ones, using much
more memory, and taking much more time. For the work in this paper, we
use an elliptic solver based on the freely available PETSc software
\cite{PETSc}, which uses Krylov subspace based matrix methods, such as
conjugate gradient, to solve the elliptic conditions which are rewritten
as a matrix equation after being cast in finite difference form. The
Cactus code has several additional elliptic solvers with various
degrees of efficiency and functionality, including several relaxation
based solvers, and a parallel multigrid solver developed by B. 
Br{\"u}gmann, based on the solver
used in Refs.~\cite{Brandt97b,Bruegmann97}.

There are various boundary conditions we can apply to the lapse at the 
outer boundary when using maximal slicing.  For example, we can allow 
the boundary value of our lapse to change in time, applying the same 
boundary condition to both the lapse and its derivatives that we apply to 
all other fields.  The ``flat'' boundary condition used here has the 
effect of copying the lapse from one point in the interior to the 
exterior after the maximal equation solve, which causes the lapse to 
collapse slowly at the outer boundary.  We can also use the more 
traditional approach, which keeps the boundary fixed at some initial 
value for the entire run.  Experience has shown that the best approach 
in this case is to be initialize $\alpha$ to the (static) Schwarzschild value, and 
then call the maximal solver, to create the initial lapse profile with 
the correct spherical outer boundary, as discussed in 
\cite{Anninos94c}.  This has the added advantage of holding the lapse 
at the Schwarzschild value near the boundary, reducing evolution of 
the metric there for some time.

As shown in Ref.~\cite{Anninos94c}, unlike with algebraic slicings, 
one can handle the region inside the throat of the black hole simply 
by ignoring it.  The elliptic maximal lapse was found to collapse rapidly 
throughout this troublesome region, quickly halting the evolution 
there.  Hence no special precautions, and no isometry operator, are 
needed to handle this region.  We will see the same behavior in Cactus 
below.  Although maximal slicing could be enforced with an isometry 
condition, as in axisymmetric 
simulations~\cite{Abrahams92a,Bernstein93a,Bernstein93b,Anninos93c}, 
it is not necessary to do so, and we shall not do so here.

We show here that Cactus runs and converges using maximal slicing by
evolving a single black hole on a $100^3$ and a $50^3$ sized
computational grid to $15 M$. We look for convergence in the
constraints, which should converge to zero, and also in $\rm{tr}K$,
which the maximal slicing condition should force to zero.

\begin{figure}
  \incpsfig{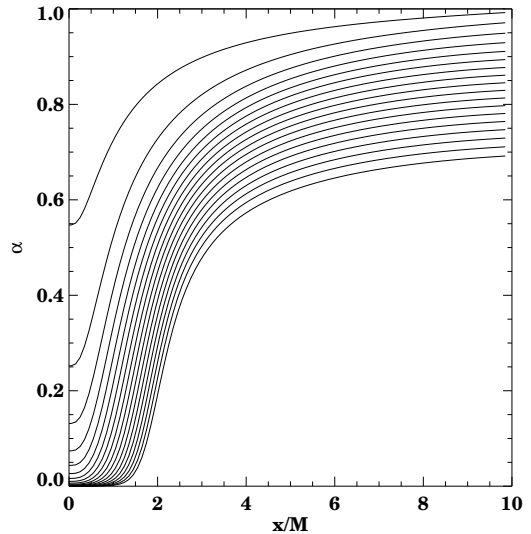}
\caption{We show the collapse of the lapse along the $x$--line for a
  maximally sliced black hole. We note the traditional collapse in the
  center. We also note that our outer boundary is not held static in
  this case, and thus the lapse collapses there. This collapse allows
  evolution with the outer boundary placed nearer the hole than in the
  static boundary case. This simulation has $\Delta x = 0.1M$ and the lapse is
  shown every $t=0.8M$ from $t=0M$ to $t=14.4M$.}
\label{fig:maxalp}
\end{figure}

\begin{figure}
  \incpsfig{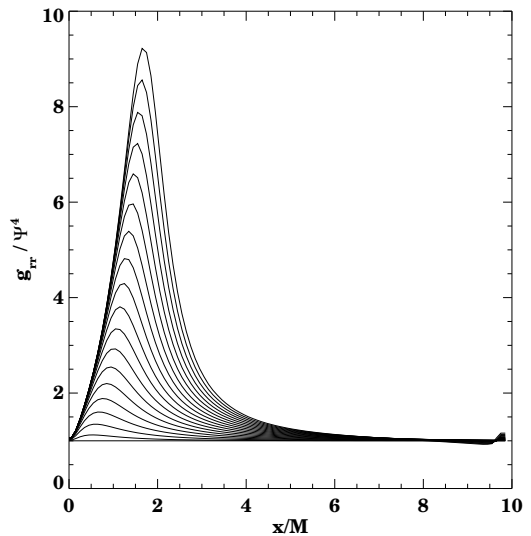}
\caption{We show the growth of the conformal 3-metric $g_{rr}/\psi^4$ along the
  $x$-axis in the maximal slicing case. This figure is the now infamous
  ``grid stretching'' figure, and demonstrates the problem which
  plagues all black hole simulations with singularity avoiding slicing
  without apparent horizon boundary conditions, namely the explosive
  growth of the radial metric function. Late time outer boundary
  problems are also evident in this plot. This simulation has $\Delta
  x = 0.1M$ and the metric is shown every $t=0.8M$ from $t=0M$ to
  $t=14.4M$.}
\label{fig:maxgrr}
\end{figure}

First we show the behavior of the solution. In Figs.~\ref{fig:maxalp}
and \ref{fig:maxgrr} we show the ``collapse of the lapse'' and ``peak
in $g_{rr}$'' which are familiar from maximally sliced black hole
evolutions in numerical relativity. Due to the singularity
avoiding coordinate slicing, we see the lapse collapses towards zero
at the center of our grid, which halts evolution, and leads to large
proper distances between coordinate points as the exterior evolves,
creating large gradients in the metric. We also note that the lapse
does collapse at the outer boundary in this simulation, and also that
at late times, outer boundary effects are noticeable in $g_{rr}/\psi^{4}$.

\begin{figure}
  \incpsfig{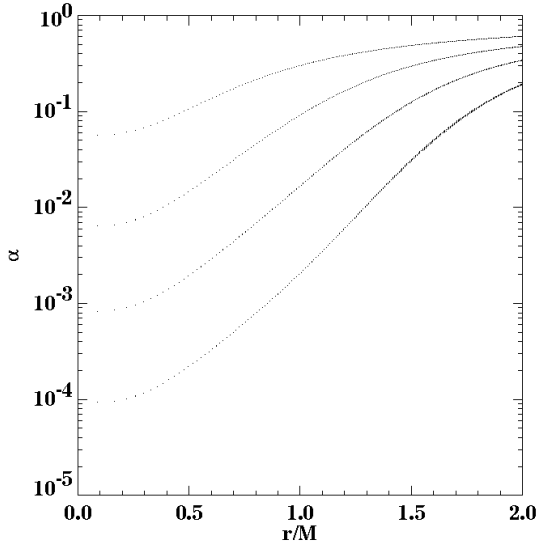}
\caption{We use a scatter plot to show the maintenance of spherical
  symmetry in the lapse in a maximally sliced black hole. We show the
  lapse on a log plot, and note that the collapse to very small lapse
  maintains spherical symmetry to a very high degree. Slices are shown
  at $t=3.6M$, $7.2M$, $10.8M$, and $14.4M$
  The resolution used is $\Delta x = 0.1M$.}
\label{fig:maxalp_scatter}
\end{figure}

In Fig.~\ref{fig:maxalp_scatter} we demonstrate that the system
maintains spherical symmetry using the initial lapse of one and
allowing the lapse to change at the outer boundary. We note that even
on this log scale, and at very small values of the lapse ($\alpha
\rightarrow 10^{-4}$) the system maintains excellent spherical symmetry.

\begin{figure}
\incpsfig{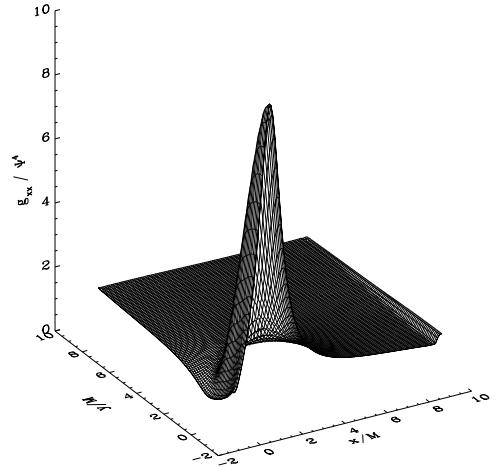}
\caption{We show the behavior of the cartesian metric functions 
  by showing a slice in the $x-y$ plane of $g_{xx}/\psi^4$ at
  $t=14.4M$ for the high resolution ($\Delta x = 0.1M$) simulation
  considered above. Note that this function behaves like $g_{rr}$
  along the $x-$axis, and $g_{\theta\theta}$ along the $y-$axis.}
\label{fig:gxx_surface}
\end{figure}

We emphasize that the growth in $g_{rr}$ is not something we simulate 
directly.  We do not evolve $g_{rr}$, but rather, we evolve cartesian 
metric functions.  These functions must not only display the growth in 
the radial metric function, but must also contain the decreasing
behavior of 
the angular metric functions.  That is $g_{xx}$ must behave like 
$g_{rr}$ along the $x-$line, but also like $g_{\theta\theta}$ along 
the $y-$ and $z-$ lines.  This leads to an even larger dynamic range 
in our cartesian metric functions than in the radial or angular metric 
functions alone.  In Fig.~\ref{fig:gxx_surface}, we demonstrate this by 
showing a slice in the $x-y$ plane of $g_{xx}/\psi^4$ at $t=14.4M$ for 
the high resolution ($\Delta x = 0.1M$) simulation considered above.  
It is clear from the figure that the function is growing along the 
$x-$ axis and dropping along the $y-$ axis, as expected.

\begin{figure}
  \incpsfig{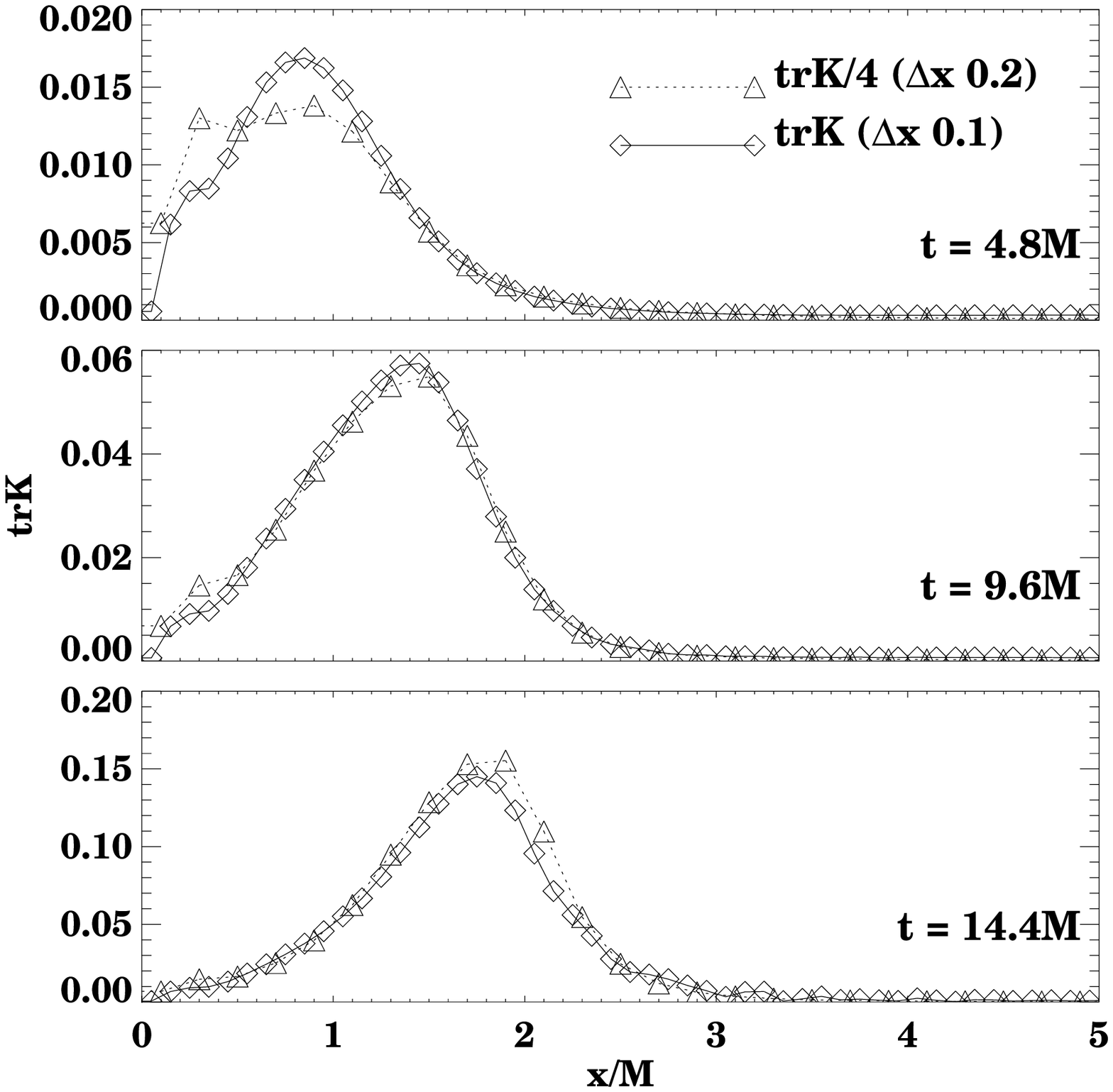}
\caption{We show the $trK$ at two different resolutions for the
  maximally sliced case. The dotted line is ${\rm tr}K/4$ calculated
  with $\Delta x = 0.2M$. The solid line is the ${\rm tr}K$
  calculated with $\Delta x = 0.1M$. Since the ${\rm tr}K=0$ should
  remain constant in maximal slicing, plotting these two quantities
  together demonstrates the second order convergence of our solution
  to the maximally sliced spacetime.}
\label{fig:maxtrKconv}
\end{figure}

One obvious question to ask of our simulation is whether or not our 
maximal slices are actually maximal, in that they keep the ${\rm tr}K$ 
zero.  This condition will be violated by our numerical simulation, 
but we can check whether the ${\rm tr}K$ converges towards zero.  In 
Fig.~\ref{fig:maxtrKconv} we do exactly this.  We show the ${\rm 
tr}K/4$ 
on the $50^3$ $\Delta x = 0.2$ grid and ${\rm tr}K$ on the $100^3$ 
$\Delta x = 0.1$ grid.  If the ${\rm tr}K$ converges to zero at second 
order we expect these lines to be identical.  From 
Fig.~\ref{fig:maxtrKconv} we can clearly see that our evolution 
converges to a maximal slice at second order.  

\begin{figure}
  \incpsfig{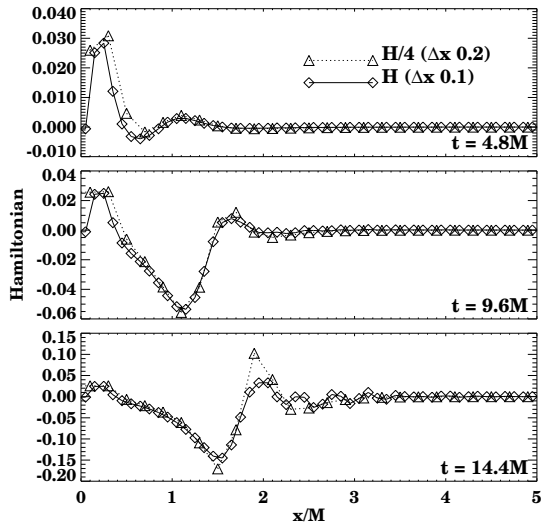}
\caption{We show second order convergence of the hamiltonian
  constraint in the maximally sliced black hole. We show $H$ at
  $\Delta x = 0.2M$ and $H/4$ at $\Delta x = 0.4M$. The fact that the
  points are visually coincident strongly indicates the system is
  converging towards a solution of the Einstein equations.}
\label{fig:maxhamconv}
\end{figure}

Similarly we can confirm that we are creating a solution to the 
Einstein equations in our maximally sliced spacetime by testing if the 
Hamiltonian constraint converges to zero.  In 
Fig.~\ref{fig:maxhamconv} we show one-quarter of the constraint on the $50^3$ $\Delta 
x = 0.2$ grid and the constraint on the $100^3$ $\Delta x = 
0.1$ grid.  Once again, we see the lines are close to identical 
visually, strongly indicating second order convergence.  

\begin{figure}
  \incpsfig{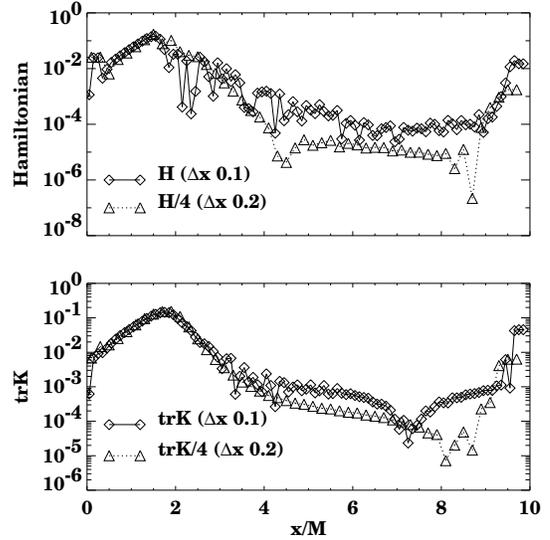}
\caption{We show the convergence of ${\rm tr}K$ and the hamiltonian
  constraint at $t=14.4M$ for the single black hole case considered in
  Fig.~\protect\ref{fig:maxtrKconv} and
  \protect\ref{fig:maxhamconv}. Here we use a logarithmic $y$ axis,
  which emphasizes that, at a very low level, the boundary introduces
  non second order convergent 
  features to the system. Since these 
  effects are several orders of magnitude below the dominant errors,
  they do not have an adverse effect on our solution (the area with
  large error is what crashes our code). However, it
  is clear that our boundaries lead to small, but non second order convergent,
  effects entering our grid in the less dynamic (spatial) regions.}
\label{fig:maxlogconv}
\end{figure}

Figs. \ref{fig:maxtrKconv} and \ref{fig:maxhamconv} demonstrate
fairly conclusively that in the regions where our error is
larger, our code is converging at second order. However, we note that
our simple boundary conditions do lead to small, but non second order convergent,
errors at low levels which are not visible in
Figs. \ref{fig:maxtrKconv} and \ref{fig:maxhamconv}. We can see these by
plotting the ${\rm tr}K$ and hamiltonian at late time (here $t=14.4M$,
the final time in Figs.  \ref{fig:maxtrKconv} and
\ref{fig:maxhamconv}) using a logarithmic $y$ axis, which we do in
Fig.~\ref{fig:maxlogconv}. Fig.~\ref{fig:maxlogconv} 
artificially inflates the non second order 
convergent features due to the boundary, but it
is instructive nonetheless. Since these features are on a very low level
(several orders of magnitude smaller than the dominant error) they
have no real adverse affect on our solution at this time.

\begin{figure}
  \incpsfig{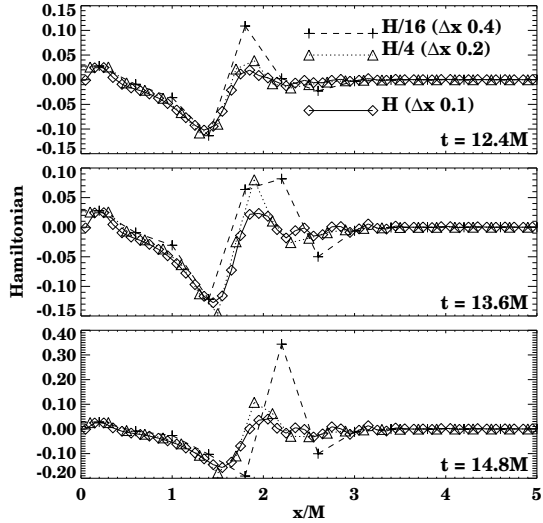}
\caption{We repeat the display in Fig.~\protect\ref{fig:maxhamconv},
  adding a third very low resolution simulation with $\Delta x =
  0.4M$. We see that the low resolution solution is not converging for
  $t>12M$, and it will crash soon after the final slice shown here.}
\label{fig:maxhamconvBad}
\end{figure}

For the sake of completeness, we also show the failing of convergence 
of these quantities when our grid is too poorly resolved.  In 
Fig.~\ref{fig:maxhamconvBad} we show the constraints for a $25^3$, 
$\Delta x = 0.4$ run.  Since only two points cover the entire initial 
horizon at this resolution, we cannot reasonably expect a converged 
answer, and we see that, even though the higher resolution simulations are 
converging at second order, the low resolution simulation has a worse 
convergence property.  At late times, this effect is mostly due to the 
lower resolution leading to an earlier crash time on the lowest 
resolution grid.

We repeat these tests with the outer boundary on the lapse held fixed,
which is the condition used previously in Ref.~\cite{Anninos94c}.
For this case, we must set the boundary farther away than in the
non-fixed case, setting it here initially at $t=15M$, rather than
$t=10M$. Additionally, we must use the maximal slicing solver to
generate an initial lapse which has the isotropic Schwarzschild form
\begin{equation}
\alpha = \frac{2r - M}{2r + M}
\end{equation}
at the outer boundary. This leads to an initial lapse other than one
everywhere, as discussed in Ref.\cite{Anninos94c}.  By holding the
lapse fixed, we avoid the collapse of the lapse near the boundaries,
and therefore evolve for a somewhat longer proper time in the outer
region. (We note that with a boundary very far away, as could be
provided by some form of adapted mesh structure, the two conditions
would be equivalent). Despite this difference, when we evolve the
maximally sliced black hole system with the two boundary conditions,
we see quantitatively the same behavior in the metric functions.

\begin{figure}
\incpsfig{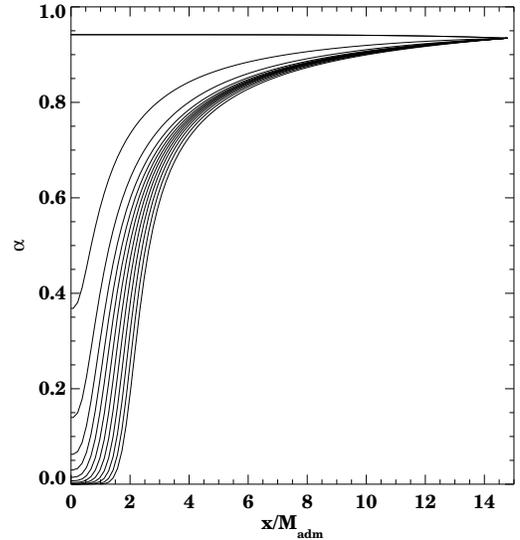}
\caption{We show the collapse of the lapse in the maximally sliced
  spacetime with the outer boundary held fixed. Note the initial lapse
  is {\em not} one, as it was in Fig.~\protect\ref{fig:maxalp}, but
  rather was the result of applying our maximal slicing solver to a
  lapse which obeys the Schwarzschild lapse outer boundary conditions.
  The resolution used here was $\Delta x = 0.15M$. Slices are shown
  every $1.2M$ between $t=0M$ and $t=14.4M$.}
\label{fig:fixed_alpcoll}
\end{figure}

In Fig.~\ref{fig:fixed_alpcoll} we show the lapse along the x line up
to $t=14.4M$, and note it remains fixed at its initial outer boundary
value, as expected. Comparing with Fig.~\ref{fig:maxalp}, we can
clearly see that this stops the lapse from collapsing over such a wide
portion of the grid, with the lapse at $x = 10M$ being around $0.9$ in
Fig.~\ref{fig:fixed_alpcoll}, and closer to $0.7$ in Fig.~\ref{fig:maxalp}.

\begin{figure}
\incpsfig{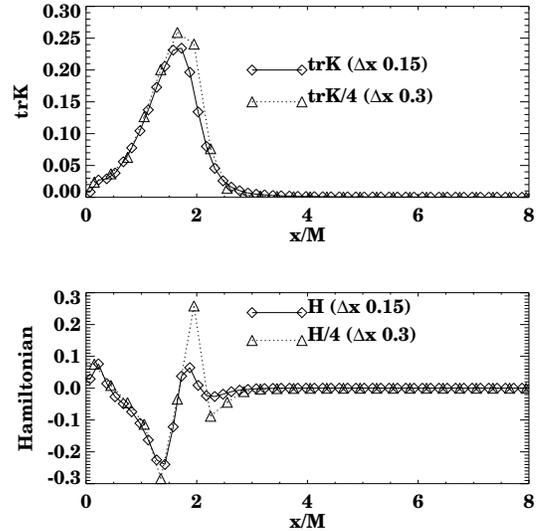}
\caption{We show the convergence of the ${\rm tr}K$ and the
  hamiltonian constraint towards zero at $t=12M$ for the maximally
  sliced spacetime with a fixed outer boundary. We note that, due to
  the lower resolution used in this simulation ($\Delta x = 0.15$
  rather than $0.1$) convergence of the constraints starts to fail
  near the peak earlier.}
\label{fig:fixed_conv}
\end{figure}

In Fig.~\ref{fig:fixed_conv} we show the convergence of the ${\rm tr}K$ and
hamiltonian constraint to zero at $t=12M$, and note we still achieve
second order convergence visually.

\subsection{Comparison with ADM Code}

In the above sections, we have demonstrated that the BM formalism can
generate convergent black hole spacetimes. In this section we confirm
that the Cactus ADM integrator is also second order convergent on
black hole spacetimes by repeating several of the above tests, and
compare ADM results with results from the BM formulation.

In general, we find that the BM and ADM systems generate comparable 
results, although, as shown below, the ADM system we have implemented 
will generally run some time longer than the BM system in maximally 
sliced black hole spacetimes, with large errors appearing first in the 
BM system.  The grid stretching problems ultimately ruins both 
calculations.  We emphasize that this is in not a shortcoming of the BM 
system; treating the system with advanced methods as described in, for 
instance, Ref.\cite{Arbona98a} will allow the BM system (in one 
dimension) to evolve for significantly longer times than the ADM 
system, showing a real advantage in using the first order system when combined with 
advanced numerical techniques.  Rather, this demonstrates that when 
evolving mathematically equivalent systems of equations, on problems 
such as these black holes that have large gradients, without using 
numerical methods designed to handle such features, both will fail 
when gradients become too steep to resolve.  The details of how the 
calculation fails can depend on many factors.  Thus for black holes, 
the present numerical methods applied to to BM system, which 
only moderately exploit the first order nature (in this case, ``flat'' 
boundaries, the Strang split, and the true MacCormack method) are not 
guaranteed to generate markedly better or longer numerical evolutions 
than, say, the ADM system with leapfrog.

\begin{figure}
\incpsfig{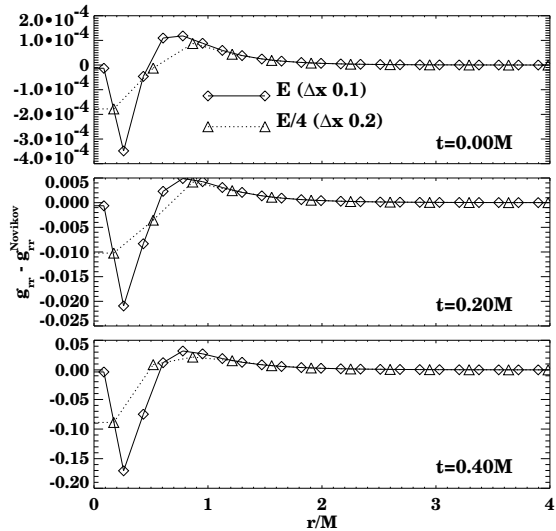}
\caption{We show that the ADM integrator in the Cactus code converges
  to second order against the analytic Novikov solution, repeating the
  test presented with the BM integrator in
  Fig.~\protect\ref{fig:novcomp}. We show the difference between the
  analytic solution and the computed solution at $\Delta x = 0.1$ and
  one quarter the error at $\Delta x = 0.2$. The fact that the points
  are visually coincident demonstrates second order convergence, which
  we see on our entire grid.}
\label{fig:adm_novikov}
\end{figure}

We return first to the geodesically sliced black hole. In
Fig.~\ref{fig:adm_novikov} we repeat the test from
Fig.~\ref{fig:novcomp}, by comparing $g_{rr}/\psi^4$ with the
analytic Novikov solution, Eq.(\ref{eqn:novikov}).  We show the
difference at a high resolution ($\Delta x = 0.1$) and $1/4$ the
difference at a lower resolution, ($\Delta x = 0.2$). The points are
visually coincident indicating second order convergence on our grid,
which we see in general.

The most interesting comparison is the maximally sliced black hole.
Studies of the three-dimensional maximally sliced black hole with the
ADM system have been undertaken in great detail in
Ref.\cite{Anninos94c}, so we only treat them briefly here, using the
cactus ADM integrator.

\begin{figure}
\incpsfig{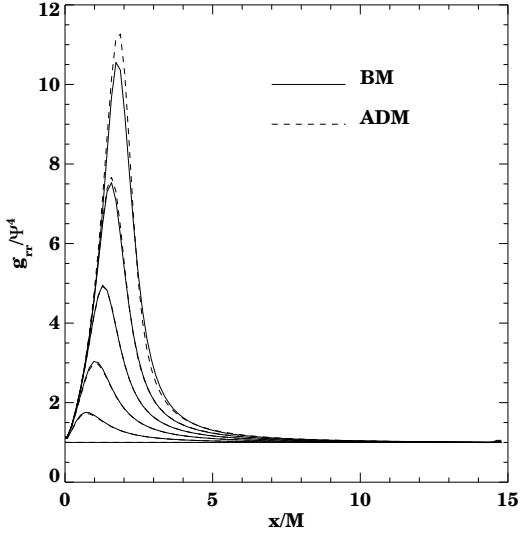}
\caption{We compare $g_{rr}/\psi^4$ for a maximally sliced single
  black hole spacetime evolved with $\Delta x = 0.15$ on a $100^3$
  grid with the ADM and BM integrators. In the ADM system all values
  are held fixed at the outer boundary, while in the BM system, only the
  lapse and its derivatives are held static, corresponding to the run
  in Fig.~\protect\ref{fig:fixed_alpcoll}. We show data every $3M$
  between $0$ and $15M$ along the $x-$line. We note that both systems
  exhibit qualitatively the same behavior.}
\label{fig:admbmsame}
\end{figure}

In Fig.~\ref{fig:admbmsame} we show that the BM and ADM system give
qualitatively the same behavior at a fixed resolution (since both
systems converge at second order, in the limit of infinite resolution
they would give identical results). We show the metric function
$g_{rr}/\psi^4$ every $3M$ between $0M$ and $15M$ with a resolution
$\Delta x = 0.15$ and a $100^3$ grid. We see clearly that the
behavior is the same in both cases, although at late times, the two
solutions are noticeably different near the peak in $g_{rr}$. 

\begin{figure}
\incpsfig{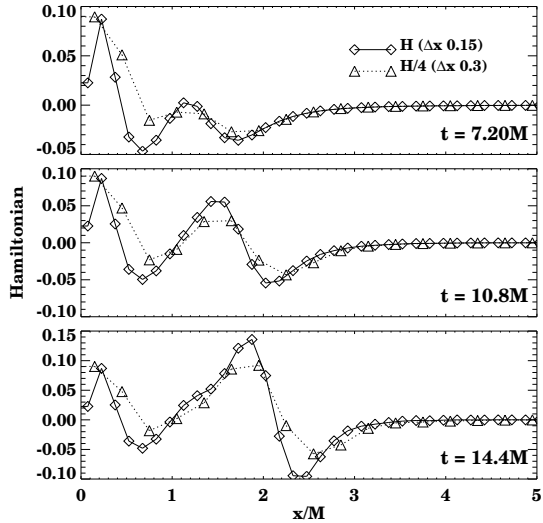}
\caption{We repeat the convergence test in
  Fig.\protect\ref{fig:maxhamconv} with the ADM integrator. We use
  parameters $\Delta x = 0.15$ on a $100^3$ grid and measure $H$ at
  $\Delta x = 0.15$ and $H/4$ at $\Delta x = 0.3$. We see almost second order
  convergence visually, and measure a convergence exponent around two
  over our entire grid. We note the the convergence order drops away
  from two as we approach the end of our simulation.}
\label{fig:adm_maxham}
\end{figure}

\begin{figure}
\incpsfig{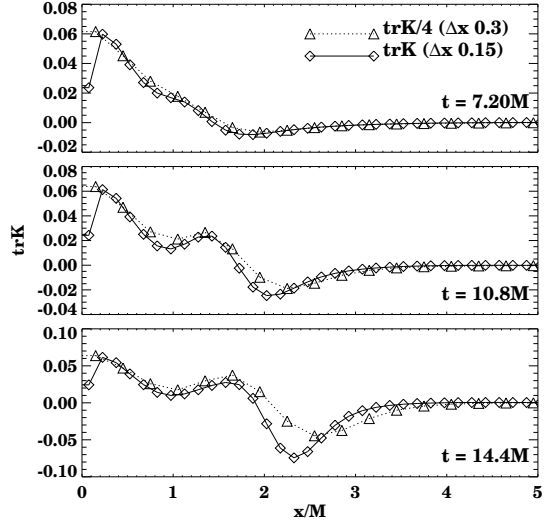}
\caption{We repeat the convergence test in
  Fig.\protect\ref{fig:maxtrKconv} with the ADM integrator. We use
  parameters $\Delta x = 0.15$ on a $100^3$ grid and measure ${\rm tr}K$ at
  $\Delta x = 0.15$ and ${\rm tr}K/4$ at $\Delta x = 0.3$. We see second order
  convergence visually, and measure a convergence exponent around two
  over our entire grid.}
\label{fig:adm_maxtrk}
\end{figure}

Even though the two solutions in Fig.~\ref{fig:admbmsame} are
different, both solutions are converging to second order, as shown in
Figs.~\ref{fig:adm_maxham} and \ref{fig:adm_maxtrk}. In these figures,
we repeat the tests of Figs.~\ref{fig:maxtrKconv} and \ref{fig:maxhamconv}
by convergence testing both the ${\rm tr}K$ and the hamiltonian
constraint against zero. We see converge close to second order
visually in the figures, and everywhere on the grid when measured
globally.

We finally directly compare the BM and ADM evolutions of the maximally
sliced black hole spacetime with parameter $\Delta x = 0.15$ on a
$100^3$ grid, with all fields held fixed at the outer boundary in the
ADM system, and the lapse held fixed with other fields having the
``copy'' boundary conditions in the BM system. We calculate the
hamiltonian constraint using Eq.~(\ref{eqn:bmhamcon}) in both the ADM
and BM system, constructing the BM $D_{ijk}$ and $V_i$ variables from
the ADM system with centered finite differences. 

\begin{figure}
\incpsfig{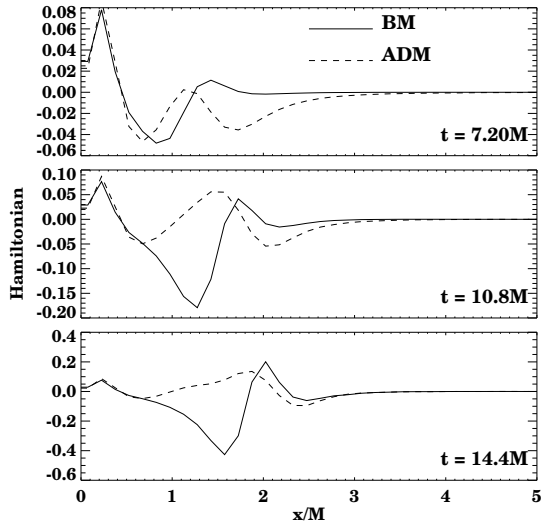}
\caption{We compare the hamiltonian constraint in the BM and ADM
  systems for the $\Delta x = 0.15$ $100^3$ maximally sliced black
  hole simulation. The hamiltonian constraint is evaluated by
  Eq.\protect\ref{eqn:bmhamcon} in both cases, with the BM $D_{ijk}$
  and $V_i$ variables constructed from the ADM simulation at every
  time step. We note that the errors in the constraint are
  comparable, but at late times, the errors in the BM system are
  larger near the maximum of the grid stretching. In the simulation
  shown here, the ADM code will run around $4-5M$ longer than the BM
  simulation (with crash times around $16M$ and $20M$ at this
  resolution). We note that the constraints converge to zero in both
  cases.}
\label{fig:bm_adm_ham_comp}
\end{figure}

In Fig.~\ref{fig:bm_adm_ham_comp} we can see that for a large part of
the run time, the hamiltonian constraint, although different, is of
the same order, around $0.1$. However as $t \rightarrow 15M$, the
hamiltonian constraint for the BM system around the peak in $g_{rr}$
drops to a larger (absolute) value than the ADM system. This dropping
continues, causing the BM system to crash about $4-5M$ before the ADM
system with the parameters chosen here.

\subsection{One-D AH Finder as a test of spherical spacetimes.}

Since the only black hole spacetimes we treat here are spherical, we
can use spherical expressions for the location of the apparent horizon
extracted along constant radial lines of the
spacetime. Here we choose diagonal lines. We assume the spherical
metric has the line element
\begin{equation}
  dl^2 = \psi^4 (g_{rr} dr^2 + g_{\theta \theta}
  d\Omega^2)\;,
\end{equation}
so the outgoing normal has the form
\begin{equation}
  s^a = \frac{1}{\psi^2 \sqrt{g_{rr}}} (1,0,0)\;.
\end{equation}
We can evaluate the expansion,
\begin{eqnarray}
  D_a s^a + K_{ab}s^a s^b - trK = \nonumber\\ 
 \frac{1}{\psi^2 \sqrt{g_{rr}}} 
   \left ( 4 \frac {\psi_{,r}}{\psi} + \frac {g_{\theta
        \theta,r}}{g_{\theta \theta}} + \frac {2} {r} \right ) - \frac
  {2 K_{\theta \theta}}{\psi^4 g_{\theta \theta}}
\label{onedahdef}
\end{eqnarray}
everywhere along this line. The point where the expansion crosses zero
defines the apparent horizon. By measuring $\psi^4 r^2
g_{\theta\theta} / 4 M^2$ there, we can monitor the horizon area,
which should be identically 1 using this normalization.

\begin{figure}
\incpsfig{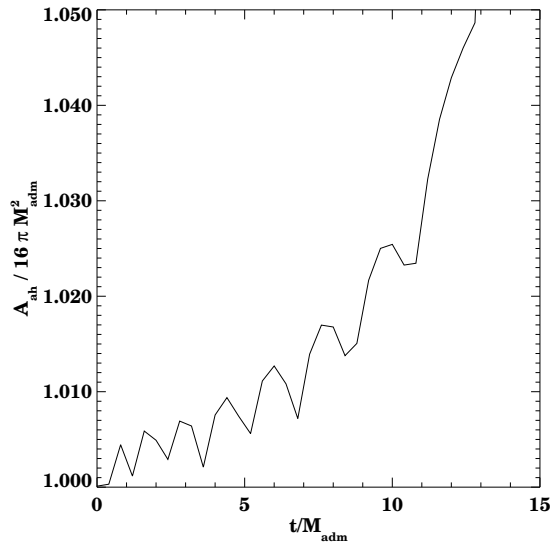}
\caption{We show the area of the apparent horizon for the $dx = 0.1M$
  maximally sliced black hole. The apparent horizon is extracted along
  the diagonal line. }
\label{fig:maximalAH}
\end{figure}

In Fig.~\ref{fig:maximalAH} we show the evolution of the apparent
horizon area up to $15M$ in the maximally sliced cases discussed
above.  We can make a crude estimate of how well our horizon is
converging by measuring the convergence exponent for its radial
location versus time.  Although this measure is plagued by
oscillations, we see that, on the whole, we have better than second
order convergence.  As well as a spherical AH finder, determined by
finding the zero of Eq.(\ref{onedahdef}), the Cactus code has a
parallelized implementation of Gundlach's Pseudo-spectral apparent
horizon finder \cite{Gundlach97a}.  Applications of this AH finder
during dynamic evolutions will be discussed elsewhere.

\section{Summary}

Hyperbolic formulations of Einstein's equations have been proposed by
a number of groups as a promising tool for numerical
relativity~\cite{Bona92,Bona94b,Bona97a,Abrahams95a,Abrahams96a,Abrahams97b}.
These reformulations of Einstein's equations have shown great strength
in 1D tests\cite{Bona94a,Bona94b}.  Early versions of the BM
hyperbolic formulation~\cite{Bona92} were developed into a full 3D
code and tested on dynamically sliced flat space~\cite{Masso92},
leading further to the development of the ``H'' code which was applied
to 3D gravitational wave studies~\cite{Anninos94d}.  A 3D version of
the Abrahams {et al.} hyperbolic formulation\cite{Abrahams95a} is also
currently under development~\cite{CookScheelPrivateComm}.  But these
3D codes have seen only limited development and application.

In this work we have performed the first systematic and detailed 
numerical exploration of a 3D hyperbolic formulation of Einstein's 
equations on a number of spacetimes of broad interest in physics and 
astronomy.  We developed and tested a full 3D numerical code, 
called Cactus, which implements the recent and more general 
BM hyperbolic formulation of Einstein's 
equations\cite{Bona94b,Bona97a}.  With this code, we showed on various 
dynamically sliced flat space, black hole, and gravitational wave 
spacetimes that this formulation allows for numerical treatment that 
is as stable and accurate as the traditional applications of the ADM 
formulation.

The Cactus code has a modular structure allowing for different 
formulations of the Einstein equations, including the ADM system, 
different numerical methods, and many different initial data, gauge, 
and analysis routines.  Cactus is developed on an advanced parallel 
computational infrastructure, achieving over 66GFlops/sec on a 512 
node Cray T3E supercomputer\cite{Clune98a,Walker98b}.  In this paper, 
within Cactus we compared Strang split MacCormack and Lax-Wendroff 
methods, applied to the Bona Mass\'o system, against a leapfrog 
implementation of the ADM system, and also against previous results 
obtained from two completely independent 3D codes (the ``G'' code, 
based on the ADM formulation, and the ``H'' code, described above).  
The numerical methods used were described in detail.

For the 3D black hole spacetimes, we studied (a) geodesically sliced 
black holes, and compared with the analytic solution of Novikov, (b) 
algebraic slicings, which have good singularity avoidance 
properties, (c) and maximal slicing, which is has traditionally been a
preferred choice for numerical black hole evolution. On all tests with 
both the ADM and BM formulation, the
code performed well, reproducing previous published results on 
spherical black hole evolution in 3D~\cite{Anninos94c}.

For 3D pure gravitational wave spacetimes, Cactus was tested on the 
evolution of linearized quadrupole and plane waves against previous 
results obtained with the ``G'' and ``H'' codes, again reproducing 
results of extensive studies published previously~\cite{Anninos94d}.  
Cactus was also tested with dynamically sliced Minkowski spacetimes, 
where quantities such as Riemann invariants were shown to converge to 
zero.

We also discussed the importance of convergence tests, and detailed a 
number of techniques we have developed to test convergence of the 
code.  We showed that Cactus is rigorously second order convergent, 
and we emphasized that convergence tests are important techniques for 
diagnosing code errors.

We emphasize that although this paper shows many successful 
applications of a 3D hyperbolic formulation of Einstein's equations, 
we have focussed on applying standard numerical methods for flux 
conservative systems, and on showing that they perform as well as 
standard methods applied to the ADM system.  We have not yet exploited 
the kinds of advanced numerical methods that can be applied to the 
eigenfields of a hyperbolic system.  Such numerical treatments are 
ultimately one 
of the major motivations for using hyperbolic systems in numerical 
relativity.  The application of numerical methods specifically 
designed for hyperbolic systems (e.g.  TVD 
methods~\cite{Leveque92,Bona96a}) has produced vast improvements in 1D 
studies of black holes, and their applications in 3D will are under 
development.  Advanced inner (e.g.  on a black hole horizon) and outer 
(e.g.  at the edge of a numerical grid) boundary treatments may also 
be possible through the use of the eigenfields.  The present Cactus 
code provides an advanced parallel tool for developing and testing 
such methods.

This paper is the first in an anticipated long series with many 
collaborators.  There are many directions in which research with this 
code is proceeding.  We are currently working on evolution of multiple 
black hole spacetimes, evolution of strong gravitational waves, 3D 
apparent horizon boundary conditions, self-gravitating scalar fields, 
advanced numerical treatments of the characteristic system, and full 
general relativistic hydrodynamics, among other projects.  We expect 
that future papers will build on this one, continuing to show careful 
comparisons with analytic solutions, demonstrating rigorous 
self-convergence, and discussing the effects of boundaries and 
numerical methods.

We plan to make the code used for the all calculations in this paper
publicly available at some point after the publication of
this paper, together with the parameter files used to produce the
figures and additional color figures and movies that provide more
details than it is possible to show in printed form. All this
information and instructions on uploading the code will be located at
the web address {\tt http://cactus.aei-potsdam.mpg.de/}.

\section{Acknowledgments}

The Cactus code has been developed in the Albert Einstein Institute in
Potsdam, Germany, with input from our many collaborators and friends
at Washington University in St.  Louis, the National Center for
Supercomputing Applications in Champaign, Illinois, the University of
the Balearic Islands in Mallorca, Spain, and elsewhere.  We would like
to thank Miguel Alcubierre, Gabrielle Allen, Miguel Angel Aloy, Pete
Anninos, Toni Arbona, Steve Brandt, Bernd Br\"ugmann, Dan Bullok,
Jaume Carot, Tom Clune, Ying Chen, Greg Daues, Thomas Dramlitsch, Ed
Evans, Toni Font, Ian Foster, Tom Goodale, Carsten Gundlach, Peter
H\"ubner, Jos\'e Mar\'{\i}a Ib{\'a}{\~n}ez, Sai Iyer, Klaus Ketelsen,
Szu-Wen Kuo, Gerd Lanferman, Lluis Mas, Richard Matzner, Wolfgang
Mertz, Mark Miller, Phillipos Papadopolous, Manuel Panea, Manish
Parashar, K.V. Rao, Sirpa Saarinen, Faisal Saied, Paul Saylor, Uli
Schwenn, Bernd Schmidt, Bernard Schutz, John Shalf, Hisa-aki Shinkai,
Barry Smith, Joan Stela, Wai-Mo Suen, Doug Swesty, Ryoji Takahashi,
Kip Thorne, Malcolm Tobias, John Towns, and Ed Wang for help with
testing, bug reports and fixes, encouragement, developing thorns, and
useful discussions.  Without help from all of these people, Cactus
would not be nearly as advanced and well tested as it is today.

We would especially like to thank Steve Brandt for providing an early
release of the thorn that computes the Riemann invariants.
Calculations and development were carried out at the Albert Einstein
Institute, the University of the Balearic Islands, and at the
Konrad-Zuse Computing Center. C.B. wishes to thank
Albert-Einstein-Institut for support by the AEI visitor program.  This
work is supported by the DGICyT of Spain under project PB-94-1177 and
also by NSF grant INT94-14185.  This work also benefitted greatly from
collaborations funded by NASA-NCCS5-153.

\section*{Appendix A: The Conformally Rescaled BM Equations}

Here we detail the equations modifications necessary 
to take into account a static conformal
factor for the metric. We will evolve a ``conformal'' metric
$g_{ij}$ related to the physical metric $\hat g_{ij}$ by the conformal
factor $\psi$:
\begin{equation}
{\hat g}_{ij} = \psi^4\;g_{ij} \;.
\end{equation}
We will keep the same formal definitions for the BM variables:
\begin{eqnarray}
D_{kij} &=& \frac{1}{2} \partial_k g_{ij}\;, \\
V_i     &=& {D_{ir}}^r - {D^r}_{ri}\;,
\end{eqnarray}
where indices are raised with $g^{ij}$ and lowered with $g_{ij}$.

With these definitions, the physical BM variables relate to the
conformal ones by
\begin{eqnarray}
{\hat D}_{kij} & = & \psi^4 (D_{kij} + 2 \psi_k g_{ij})\;, \\
{\hat V}_i     & = & V_i + 4 \psi_i \;.
\end{eqnarray}
We also introduce the following notation 
for the derivatives of the conformal factor
\begin{eqnarray}
\psi_i &=& \frac{\partial_i \psi}{\psi} \;, \\
\psi_{ij} &=& \frac{\partial_i \partial_j \psi}{\psi}\;.
\end{eqnarray} 
The Christoffel symbols relate by
\begin{equation}
\hat \Gamma^k_{\;ij} = \Gamma^k_{\;ij} + 2\;\delta^k_i\;{\psi_j
\over \psi} + 2\;\delta^k_j\;{\psi_i \over \psi} - 2\;{\psi^k \over
\psi}\;g_{ij}\;,
\end{equation}
and the Ricci tensor by
\begin{equation}
\hat R_{ij} = R_{ij} - Y_{ij} - Y_k^{\;k}g_{ij}\;,
\end{equation}
where we define
\begin{equation}
Y_{ij} = -\psi^2\;(\psi^{-2})_{;i;j} + 2\;{\psi^k\over \psi}
{\psi_k\over \psi}\;g_{ij}
\end{equation}
with covariant derivatives computed using $\Gamma^k_{\;ij}$. Thus,
\begin{equation}
Y_{ij} = 2 g_{ij} \psi^k \psi_k 
         + 2 ( \psi_{ij} - \psi_r {\Gamma^r}_{ij} - 3 \psi_i\psi_j ) \;.
\end{equation}

We can then derive a modified set of fluxes (note that the flux for the
extrinsic curvature does not change):
\begin{eqnarray}
  F^k_-g_{ij} & = & 0 \;, \\
  F^k_-\alpha      & = & 0 \;, \\
  F^k_-K_{ij}      & = & -\beta^k\,K_{ij} + \alpha\;[\; D^k_{ij}
                         - n/2\;V^k\;g_{ij}
                         \\ \nonumber
                   &   & + 1/2\;\delta^k_i\;(A_j+2\,V_j-D_{jr}^{\;\;r})
                         \\
                   &   & + 1/2\;\delta^k_j\;(A_i+2\,V_i-D_{ir}^{\;\;r})\;]
                         \;, \nonumber \\
  F^k_-D_{kij}     & = & -\beta^r (D_{rij} + 2\psi_r g_{ij})
                        + \alpha\;(K_{ij}/\psi^4-s_{ij})
                         \;, \\
  F^k_-A_k         & = & -\beta^r A_r + \alpha\;Q
                         \;,  \\
  F^k_-V_i         & = & -\beta^k (V_i + 4 \psi_i) + B^k_{\;i} - B_i^{\;k}
                         \;. 
\end{eqnarray}

The modified sources are:
\begin{eqnarray}
  S_-g_{ij} &=& - 2\;\alpha\;(K_{ij}/\psi^4-s_{ij}) \nonumber \\
                 & & +  2\beta^r\,( D_{rij}  + 2 \psi_r g_{ij} ) \;, \\
  S_-\alpha      &=& - \alpha^2\;Q + \alpha\beta^r\,A_r
                       \;,\\
  S_-K_{ij}      &=&  2(K_{ir}B_j^{\;r}+K_{jr}B_i^{\;r}-K_{ij}B_r^{\;r})
                       \nonumber \\
      & & + \alpha\; [  \; -^{(4)}R_{ij}
                     + (- 2K_i^{\;k}K_{kj}+tr\,K\;K_{ij})/\psi^4 
                       \nonumber  \\
                 & & \;\;\;\; - \Gamma^k_{\;ri}\Gamma^r_{\;kj}
                     + 2D_{ik}^{\;\;r}D_{rj}^{\;\;k}
                     + 2D_{jk}^{\;\;r}D_{ri}^{\;\;k}
                     + \Gamma^k_{\;kr}\Gamma^r_{\;ij}
                       \nonumber \\
   & &  \;\;\;\;-(2\,D_{kr}^{\;\;k}-A_r)(D_{ij}^{\;\;r}+D_{ji}^{\;\;r}) \\
  & & \;\;\;\; + A_i(V_j-1/2\;D_{jk}^{\;\;k}) + A_j(V_i-1/2\;D_{ik}^{\;\;k})
                       \nonumber \\
                 & & \;\;\;\; + A_j(V_i-1/2\;D_{ik}^{\;\;k})
                     - nV^kD_{kij} \;]
                       \nonumber  \\
  & & + n/4\;\alpha g_{ij}\;[\; -D_k^{\;rs}\Gamma^k_{\;rs}
                     + D_{kr}^{\;\;r}D^{ks}_{\;\;s} -2\,V^kA_k \nonumber  \\
                 & &  \;\;\;\; + (K^{rs}K_{rs}-(tr\,K)^2)/\psi^4
                     + 2\alpha^2\;G^{00} \;]\nonumber \\
  & &  -Y_{ij} + 2A_i\psi_j+2A_j\psi_i \nonumber\\
  & &  + g_{ij}\;[(n\!-\!1)Y^k_{\;k}-2A^k\psi_k]
                       \;, \nonumber \\
  S_-D_{kij}     &=& 0
                       \;, \\
  S_-A_{k}       &=& 0
                       \;, \\
  S_-V_i         &=&   \alpha^2\;G^0_{\;i}
                     + \alpha/\psi^4[ \;A_r(K^r_{\;i}-tr\,K\;\delta^r_i)
                       \nonumber \\
                 & & + K^r_{\;s}(D_{ir}^{\;\;s}-2D_{ri}^{\;\;s})
                     - K^r_{\;i}(D_{rs}^{\;\;s}-2D_{sr}^{\;\;s}) \nonumber\\
                 & & - 2\psi_r\;(3K^r_{\;k}-tr\,K\;\delta^r_k)\;] \\
                 & & + 2(B_i^{\;r} - \delta_i^r\;tr\,B)\;V_r
                     + 2(D_{ri}^{\;\;s}-\delta^s_i\;D^j_{\;jr})B^r_{\;s}
                     \nonumber \\ 
                 & & + 4 {B_i}^r \psi_r - 4 trB \psi_i
                       \;. \nonumber
\end{eqnarray}
Finally, our algebraic slicing condition becomes
\begin{equation}
Q = f(\alpha)\;tr\,K\;/\psi^4\;.
\end{equation}







\begin{thebibliography}{100}

\bibitem{Masso98b}
J. Mass\'o and P. Walker,   (1998), in preparation.

\bibitem{Bona92}
C. Bona and J. Mass\'{o}, Phys. Rev. Lett. {\bf 68},  1097  (1992).

\bibitem{Bona94b}
C. Bona, J. Mass\'o, E. Seidel, and J. Stela, Phys. Rev. Lett. {\bf 75},  600
  (1995).

\bibitem{Bona97a}
C. Bona, J. Mass\'o, E. Seidel, and J. Stela, Phys. Rev. D {\bf 56},  3405
  (1997).

\bibitem{Flanagan97b}
\'{E}anna \'{E}.~Flanagan and S.~A. Hughes, gr-qc/9710129,    (1997).

\bibitem{Flanagan97a}
\'{E}anna \'{E}.~Flanagan and S.~A. Hughes, gr-qc/9701039,    (1997).

\bibitem{Abrahams97a}
A.~M. Abrahams {\it et~al.}, Physical Review Letters {\bf 80},  1812  (1998),
  gr-qc/9709082.

\bibitem{Cook97a}
G.~B. Cook {\it et~al.},   (1997), gr-qc/9711078.

\bibitem{Gomez98a}
R. Gomez {\it et~al.},   (1998), gr-qc/9801069.

\bibitem{Mathews97}
G. Mathews and J. Wilson, Astrophysical Journal {\bf 482},  929  (1997).

\bibitem{Ruffert96a}
M. Ruffert, M. Rampp, and H.-T. Janka, Astron. Astrophys.  , to appear.
  astro-ph/9611056.

\bibitem{ruffert95}
M. Ruffert, H.-T. Janka, W. Keil, and G. Sch\"{a}fer,  in {\em Proc. of the
  17th Texas Symposium on Relativistic Astrophysics} (World Scientific,
  Singapore, 1995).

\bibitem{Oohara96}
K.-I. Oohara and T. Nakamura,  in {\em Relativistic Gravitation and
  Gravitational Radiation}, edited by J.-P. Lasota and J.-A. Marck (Cambridge
  University Press, Cambridge, England, 1997).

\bibitem{Choptuik93}
M. Choptuik, Phys. Rev. Lett. {\bf 70},  9  (1993).

\bibitem{Gundlach97d}
C. Gundlach, To appear in Adv. Theor. Math. Phys  (1998), gr-qc/9712084.

\bibitem{Abrahams93a}
A. Abrahams and C. Evans, Phys. Rev. Lett. {\bf 70},  2980  (1993).

\bibitem{Gundlach97b}
C. Gundlach,   (1998), gr-qc/9710066.

\bibitem{Anninos96c}
P. Anninos, J. Mass\'o, E. Seidel, and W.-M. Suen, Physics World {\bf 9},  43
  (1996).

\bibitem{Anninos94c}
P. Anninos, K. Camarda, J. Mass\'o, E. Seidel, W.-M. Suen, and J. Towns, Phys.
  Rev. D {\bf 52},  2059  (1995).

\bibitem{Bruegmann96}
B. Br\"ugmann, Phys. Rev. D {\bf 54},    (1996).

\bibitem{Anninos93b}
P. Anninos, D. Hobill, E. Seidel, L. Smarr, and W.-M. Suen, Phys. Rev. Lett.
  {\bf 71},  2851  (1993).

\bibitem{Anninos94b}
P. Anninos, D. Hobill, E. Seidel, L. Smarr, and W.-M. Suen, Phys. Rev. D {\bf
  52},  2044  (1995).

\bibitem{Camarda97a}
K. Camarda, Ph.D. thesis, University of Illinois at Urbana-Champaign, Urbana,
  Illinois, 1998.

\bibitem{Camarda97b}
K. Camarda and E. Seidel, Phys. Rev. D {\bf 57},  3204  (1998), gr-qc/9709075.

\bibitem{Allen97a}
G. Allen, K. Camarda, and E. Seidel,   (1998), in preparation.

\bibitem{Shapiro86}
S.~L. Shapiro and S.~A. Teukolsky,  in {\em Dynamical Spacetimes and Numerical
  Relativity}, edited by J.~M. Centrella (Cambridge University Press,
  Cambridge, England, 1986), pp.\ 74--100.

\bibitem{Price94a}
R.~H. Price and J. Pullin, Phys. Rev. Lett. {\bf 72},  3297  (1994).

\bibitem{Price94b}
P. Anninos, R.~H. Price, J. Pullin, E. Seidel, and W.-M. Suen, Phys. Rev. D
  {\bf 52},  4462  (1995).

\bibitem{Abrahams95c}
A. Abrahams and R. Price, Phys. Rev. D {\bf 53},  1972  (1996).

\bibitem{Baker96a}
J. Baker, A. Abrahams, P. Anninos, S. Brandt, R. Price, J. Pullin, and E.
  Seidel, Phys. Rev. D {\bf 55},  829  (1997).

\bibitem{Gleiser96b}
R.~J. Gleiser, C.~O. Nicasio, R.~H. Price, and J. Pullin, Physical Review
  Letters {\bf 77},  4483  (1996).

\bibitem{Rezzolla97a}
L. Rezzolla, A.~M. Abrahams, T.~W. Baumgarte, G.~B. Cook, M.~A. Scheel, S.~L.
  Shapiro, and S.~A. Teukolsky, Phys. Rev. D {\bf 57},  1084  (1998).

\bibitem{Berger84}
M. Berger and J. Oliger, Journal of Computational Physics {\bf 53},  484
  (1984).

\bibitem{Wild96}
L.~A. Wild, Ph.D. thesis, University of Wales, 1996.

\bibitem{Papadapoulos98a}
P. Papadapoulos, E. Seidel, and L. Wild, Physical Review D  (1998), submitted,
  gr-qc/9802069.

\bibitem{Seidel92a}
E. Seidel and W.-M. Suen, Phys. Rev. Lett. {\bf 69},  1845  (1992).

\bibitem{Anninos94e}
P. Anninos, G. Daues, J. Mass\'o, E. Seidel, and W.-M. Suen, Phys. Rev. D {\bf
  51},  5562  (1995).

\bibitem{Scheel94}
M.~A. Scheel, S.~L. Shapiro, and S.~A. Teukolsky, Phys. Rev. D {\bf 51},  4208
  (1995).

\bibitem{Marsa96}
R. Marsa and M. Choptuik, Phys Rev D {\bf 54},  4929  (1996).

\bibitem{Daues96a}
G.~E. Daues, Ph.D. thesis, Washington University, St. Louis, Missouri, 1996.

\bibitem{Anninos94d}
P. Anninos, J. Mass\'o, E. Seidel, W.-M. Suen, and M. Tobias, Phys. Rev. D {\bf
  56},  842  (1997).

\bibitem{Anninos96b}
P. Anninos, J. Mass\'o, E. Seidel, W.-M. Suen, and M. Tobias, Phys. Rev. D {\bf
  54},  6544  (1996).

\bibitem{Balakrishna96a}
J. Balakrishna, G. Daues, E. Seidel, W.-M. Suen, M. Tobias, and E. Wang, Class.
  Quant. Grav. {\bf 13},  L135  (1996).

\bibitem{Alcubierre97a}
M. Alcubierre, Phys. Rev. D {\bf 55},  5981  (1997).

\bibitem{Alcubierre97b}
M. Alcubierre and J. Mass\'o, Phys. Rev. D Rapid Comm., to
  appear. gr-qc/9709024.

\bibitem{Bruegmann97}
B. Br\"ugmann,   (1997), gr-qc/9708035.

\bibitem{Gomez97a}
R. Gomez, L. Lehner, R. Marsa, and J. Winicour,   (1997), gr-qc/9710138.

\bibitem{York79}
J. York,  in {\em Sources of Gravitational Radiation}, edited by L. Smarr
  (Cambridge University Press, Cambridge, England, 1979).

\bibitem{Friedrich96}
H. Friedrich, Class. Quant. Grav. {\bf 13},  1451  (1996).

\bibitem{Reula98a}
O. Reula, Living Reviews in Relativity {\bf 1},    (1998).

\bibitem{Leveque92}
R.~J. Leveque, {\em Numerical Methods for Conservation Laws} (Birkhauser
  Verlag, Basel, 1992).

\bibitem{Choquet83}
Y. Choquet-Bruhat and T. Ruggeri, Comm. Math. Phys {\bf 89},  269  (1983).

\bibitem{Friedrich85}
H. Friedrich, Comm. Math. Phys. {\bf 100},  1525  (1985).

\bibitem{Bona88}
C. Bona and J. Mass\'{o}, Phys. Rev. D {\bf 38},  2419  (1988).

\bibitem{Bona89}
C. Bona and J. Mass\'{o}, Phys. Rev. D {\bf 40},  1022  (1989).

\bibitem{Masso92}
J. Mass\'o, Ph.D. thesis, University of the Balearic Islands, 1992.

\bibitem{Bona92b}
C. Bona and J. Mass\'o,  in {\em Approaches to Numerical Relativity}, edited by
  R. D'Inverno (Cambridge University Press, Cambridge, England, 1992).

\bibitem{Gjertsen96}
R. Gjertsen, J. Mass\'o, M. Nardulli, E. Seidel, J. Shalf, and D. Weber,
  Forefronts {\bf 11},    (1996), cornell Theory Center.

\bibitem{Kuo97}
S. Kuo, M. Winslett, K. Seamons, Y. Chen, Y. Cho, and M. Subramaniam,  in {\em
  Proceedings of the SIAM Conference on Parallel Processing for Scientific
  Computing} (SIAM, Minneapolis, MN, 1997).

\bibitem{Bona94a}
C. Bona, J. Mass\'o, and J. Stela, Phys. Rev. D {\bf 51},  1639  (1995).

\bibitem{Arbona98a}
A. Arbona, C. Bona, J. Mass\'o, and J. Stela, In preparation  (1998).

\bibitem{Fritelli94}
S. Fritelli and O. Reula, Commun. Math. Phys. {\bf 166},  221  (1994).

\bibitem{Choquet95}
Y. Choquet-Bruhat and J. York, C. R. Acad. Sc. Paris {\bf 321},  1089  (1995).

\bibitem{Abrahams95a}
A. Abrahams, A. Anderson, Y. Choquet-Bruhat, and J. York, Phys. Rev. Lett. {\bf
  75},  3377  (1995).

\bibitem{Fritelli95}
S. Fritelli and O. Reula, Phys. Rev. Lett. {\bf 75},  4667  (1995).

\bibitem{MVP95}
M.~H. van Putten and D. Eardley, Phys. Rev. D {\bf 53},  3056  (1996).

\bibitem{Abrahams97b}
A. Abrahams, A. Anderson, Y. Choquet-Bruhat, and J. York, Class.Quant.Grav  A9
  (1997).

\bibitem{Scheel97}
M. Scheel, T. Baumgarte, G. Cook, S. Shapiro, and S. Teukolsky, Phys. Rev. D
  {\bf 56},  6320  (1997).

\bibitem{CookScheelPrivateComm}
G. Cook and M. Scheel, private communication.

\bibitem{Clune98a}
T. Clune, J. Mass\'o, M. Miller, and P. Walker, Technical report, National
  Center for Supercomputing Applications,  (unpublished), in preparation.

\bibitem{Walker98b}
P. Walker, Technical report, National Center for Supercomputing Applications,
  (unpublished), in preparation.

\bibitem{Press86}
W.~H. Press, B.~P. Flannery, S.~A. Teukolsky, and W.~T. Vetterling, {\em
  Numerical Recipes} (Cambridge University Press, Cambridge, England, 1986).

\bibitem{Abrahams96a}
A. Abrahams, A. Anderson, Y. Choquet-Bruhat, and J. York, C.R. Acad. Sci. Paris
   835  (1996).

\bibitem{MPI}
The Message Passing Interface (MPI) standard, http://www.ncs.anl.gov/mpi/.

\bibitem{Parashar95}
M. Parashar and J.~C. Browne, Proceedings of the International Conference for
  High Performance Computing  22  (1995).

\bibitem{Parashar96}
M. Parashar and J.~C. Browne, Proceedings of the $\rm 29^{th}$ Annual Hawaii
  International Conference on System Sciences  604  (1996).

\bibitem{Anninos96a}
P. Anninos, D. Hobill, E. Seidel, and W.-M. Suen, {\em Proceedings of the Sixth
  Canadian Conference on General Relativity and Relativistic Astrophysics}
  (Fields Institute Comunications, 1996), in press.

\bibitem{Allen98a}
G. Allen, K. Camarda, and E. Seidel, in preparation.

\bibitem{Bishop97a}
N. Bishop, R. Gomez, L. Lehner, and J. Winicour, Phys. Rev. D {\bf 52},
  (1997).

\bibitem{Huebner98}
P. H\"ubner, gr-qc/9804065 (1998).

\bibitem{Huebner96}
P. H\"ubner, Phys. Rev. D {\bf 53},  701  (1996).

\bibitem{Frauendiener97}
J. Frauendiener, gr-qc/971250 and gr-qc/9712052 (1997).

\bibitem{Abrahams92a}
A. Abrahams, D. Bernstein, D. Hobill, E. Seidel, and L. Smarr, Phys. Rev. D
  {\bf 45},  3544  (1992).

\bibitem{Arbona97}
A. Arbona, C. Bona, J. Carot, L. Mas, J. Mass\'o, and J. Stela, Phys. Rev. D
  {\bf 57},  2397  (1998).

\bibitem{Bona96a}
C. Bona,  in {\em Relativity and Scientific Computing}, edited by F. Hehl
  (Springer-Verlag, Berlin, 1996).

\bibitem{Yee88}
H.~C. Yee, {\em Computational Fluid Dynamics} (von Karman Institute for Fluid
  Dynamics, NASA Ames Research Center, CA, 1989).

\bibitem{Walker98a}
P. Walker, Ph.D. thesis, University of Illinois at Urbana-Champaign, Urbana,
  Illinois, 1998, in preparation.

\bibitem{Choptuik91}
M. Choptuik, Phys. Rev. D {\bf 44},  3124  (1991).

\bibitem{Gunnarsen95}
L. Gunnarsen, H. Shinkai, and K. Maeda, 
Class. Quant. Grav. {\bf 12},  133  (1995).

\bibitem{Anninos95d}
P. Anninos, K. Camarda, J. Mass\'o, E. Seidel, W.-M. Suen, M. Tobias, and J.
  Towns,  in {\em The Seventh Marcel Grossmann Meeting: On Recent Developments
  in Theoretical and Experimental General Relativity, Gravitation, and
  Relativistic Field Theories}, edited by R.~T. Jantzen, G.~M. Keiser, and R.
  Ruffini (World Scientific, Singapore, 1996), pp.\ 644--647.

\bibitem{Yurtsever88a}
U. Yurtsever, Phys. Rev. D {\bf 37},  2790  (1988).

\bibitem{Yurtsever88b}
U. Yurtsever, Phys. Rev. D {\bf 38},  1731  (1988).

\bibitem{Shibata95}
M. Shibata and T. Nakamura, Phys. Rev. D {\bf 52},  5428  (1995).

\bibitem{Teukolsky82}
S. Teukolsky, Phys. Rev. D {\bf 26},  745  (1982).

\bibitem{Evans86}
C. Evans,  in {\em Dynamical Spacetimes and Numerical Relativity}, edited by J.
  Centrella (Cambridge University Press, Cambridge, England, 1986), pp.\ 3--39.

\bibitem{Camarda97c}
K. Camarda and E. Seidel, in preparation.

\bibitem{Bernstein93b}
D. Bernstein, D. Hobill, E. Seidel, L. Smarr, and J. Towns, Phys. Rev. D {\bf
  50},  5000  (1994).

\bibitem{Brandt94b}
S. Brandt and E. Seidel, Phys. Rev. D {\bf 52},  856  (1995).

\bibitem{Brandt97b}
S. Brandt and B. Br\"ugmann, Phys. Rev. Lett. {\bf 78},  3606  (1997).

\bibitem{Misner73}
C.~W. Misner, K.~S. Thorne, and J.~A. Wheeler, {\em Gravitation} (W. H.
  Freeman, San Francisco, 1973).

\bibitem{BruegmannPrivateComm}
B. Br\"ugmann, private communication.

\bibitem{PETSc}
S. Balay, W. Gropp, L.~C. McInnes, and B. Smith, PETSc - The Portable,
  Extensible Toolkit for Scientific Compuation, 1998,
  http://www.mcs.anl.gov/petsc/.

\bibitem{Bernstein93a}
D. Bernstein, Ph.D. thesis, University of Illinois Urbana-Champaign, 1993.

\bibitem{Anninos93c}
P. Anninos, D. Bernstein, D. Hobill, E. Seidel, L. Smarr, and J. Towns,  in
  {\em Computational Astrophysics: Gas Dynamics and Particle Methods}, edited
  by W. Benz, J. Barnes, E. Muller, and M. Norman (Springer-Verlag, New York,
  1997), in press.

\bibitem{Gundlach97a}
C. Gundlach, Phys. Rev. D {\bf 57},  863  (1998).


\end{thebibliography}
\end{document}